\documentclass[twocolumn]{emulateapj}
\pdfoutput=1
\usepackage{apjfonts,amsmath}
\usepackage{graphicx,verbatim}
\newcommand{\beq}{\begin{equation}}
\newcommand{\eeq}{\end{equation}}
\newcommand{\bea}{\begin{eqnarray}}
\newcommand{\eea}{\end{eqnarray}}

\begin{document}	

\title{Simulations of Accretion Powered Supernovae in the Progenitors of Gamma Ray Bursts}
\author{
Christopher C. Lindner
\altaffilmark{1},
Milo\v s Milosavljevi\'c,
Rongfeng Shen
\altaffilmark{2},
and
Pawan Kumar
}
\affil{Department of Astronomy, University of Texas, 1 University Station C1400, Austin, TX 78712}
\altaffiltext{1}{NSF Graduate Research Fellow.}
\altaffiltext{2}{Department of Astronomy \& Astrophysics, University of Toronto, 50 St. George St., Toronto, Ontario M5S 3H4, Canada.}

\righthead{SIMULATIONS OF ACCRETION POWERED SUPERNOVAE}
\lefthead{LINDNER ET AL.}

\begin{abstract}
Observational evidence suggests a link between long duration gamma ray bursts (LGRBs) and Type Ic supernovae.  Here, we propose a potential mechanism for Type Ic supernovae in LGRB progenitors powered solely by accretion energy.  We present spherically-symmetric hydrodynamic simulations of the long-term accretion of a rotating gamma-ray burst progenitor star, a ``collapsar,'' onto the central compact object, which we take to be a black hole.  The simulations were carried out with the adaptive mesh refinement code FLASH in one spatial dimension and with rotation, an explicit shear viscosity, and convection in the mixing length theory approximation.  Once the accretion flow becomes rotationally supported outside of the black hole, an accretion shock forms and traverses the stellar envelope.  Energy is carried from the central geometrically thick accretion disk to the stellar envelope by convection.  Energy losses through neutrino emission and nuclear photodisintegration are calculated but do not seem important following the rapid early drop of the accretion rate following circularization.  We find that the shock velocity, energy, and unbound mass are sensitive to convective efficiency, effective viscosity, and initial stellar angular momentum.  Our simulations show that given the appropriate combinations of stellar and physical parameters, explosions with energies $\sim5 \times 10^{50} \,\textrm{ergs}$, velocities $\sim 3000 \,\textrm{km}\,\textrm{s}^{-1}$, and unbound material masses $\gtrsim 6\, M_\odot$ are possible in a rapidly rotating $16 \,M_\odot$ main sequence progenitor star.  Further work is needed to constrain the values of these parameters, to identify the likely outcomes in more plausible and massive LRGB progenitors, and to explore nucleosynthetic implications.

\keywords{ accretion, accretion disks --- black hole physics --- gamma rays: bursts --- stars: winds, outflows --- supernovae: general }

\end{abstract}

\section{Introduction}
\label{sec:intro}
\setcounter{footnote}{0}

A clear observational link has been established between long-duration gamma-ray bursts (LGRBs) and Type Ic supernovae \citep{Galama:98,Galama:00,Reichart:99,Bloom:02,DellaValle:03,DellaValle:06,Garnavich:03,Hjorth:03,Kawabata:03,Stanek:03,Matheson:03,Malesani:04,Campana:06,Mirabal:06,Modjaz:06,Pian:06,Chornock:10,Cobb:10,Starling:11}.  However, only a small percentage of Type Ic supernovae exhibit the late-time radio signatures of LGRBs \citep{Podsiadlowski:04,Soderberg:06b}.  LGRBs are believed to be manifestations of rotationally-powered ultrarelativistic outflows developing in the wake of the formation of black holes or neutron stars in rotating progenitor.  However, the exact mechanism for the production of LGRBs and their associated supernovae remains a subject of debate \citep[][and references therein]{Woosley:06b,Hjorth:11}.  At present, it is not clear whether the processes that give rise to LGRBs also drive a stellar explosions, or whether the explosions are driven independently, perhaps by the standard, neutrino-mediated mechanism.

In the standard supernova mechanism, an outward-moving shock forms after core-bounce.  This shock stalls, but may be reinvigorated by heating by neutrinos emitted during the neutronization near the proto-neutron star \citep{Bethe:85}, and in principle drive the star to explosion in the so-called ``delayed neutrino mechanism.''  Some simulations of this process in at least two spatial dimensions seem to produce successful explosions \citep[see, e.g.,][]{Buras:06b,Scheck:06,Mezzacappa:07,Murphy:08,Marek:09,Nordhaus:10}, although the success of two-dimensional and possibly three-dimensional simulations may be dependent upon the progenitor mass and the treatment of neutrinos \citep{Buras:06a,Nordhaus:10}.  Supernovae associated with LGRBs seem to be more energetic than the typical Type Ic supernovae \citep{Iwamoto:98,Woosley:99,Mazzali:03,Mazzali:06}, with large kinetic energies reaching $\sim 10^{52} \,\textrm{ergs}$.  Even if the neutrino mechanism can unbind the star, it still seems unclear whether it can deliver the energies found in supernovae associated with LGRBs.  An alternative or augmentative explosion mechanism may be required to explain the supernovae associated with LGRBs.  Alternatives to the neutrino mechanism call on the extraction of the rotational energy of the central compact object---a neutron star or a black hole---or on tapping the gravitational energy of the material accreting toward the compact object.  It remains to be determined which, if any, of the alternative pathways can deliver the large energies, and what are the resulting compact remnant masses.

If the post-bounce neutrino-mediated energy transfer is too weak to unbind all of the infalling stellar strata, some material may continue to fall onto the proto-neutron star and possibly take it to collapse further into a black hole \citep[e.g.,][]{Burrows:86, MacFadyen:01,Heger:03,Zhang:08,Sekiguchi:10,OConnor:10}.  This is especially relevant for rapidly rotating progenitors, as the progenitors with rapidly rotating cores may produce lower neutrino luminosities, decreasing the effectiveness of the neutrino-powered explosion mechanism \citep{Fujimoto:06,Lee:06}.

The infall or fallback of the stellar envelope should continue past the initial emergence of the event horizon, but then the structure of the accretion flow becomes sensitive to its angular momentum content.  Given sufficient angular momentum, the flow becomes rotationally supported.  Such a ``collapsar'' configuration has been proposed to naturally lead to the ultrarelativistic outflow in LRGBs \citep{Woosley:93}, as gamma rays can be produced in an ultrarelativistic jet launching from the magnetosphere of the black hole that forms in the aftermath of the collapse of the rotating progenitor. The jet is powered by a continuous infall and disklike accretion of the progenitor star's interior.  

It has long been hypothesized that a ``wind'' outflowing from a collapsar accretion disk could unbind the stellar envelope and synthesize sufficient $^{56}$Ni to produce an optically bright supernova \citep[e.g.,][]{MacFadyen:99,Pruet:03,Pruet:04,Kohri:05}.  The dynamics of the energy flow in such a system has yet to be elucidated.  In the present study, we utilize one dimensional hydrodynamic simulations with rotation (1.5D) to test the hypothesis that accretion power can drive an explosion of the star.  We do not simulate the core bounce, and simply posit that any prompt and neutrino-reinvigorated shock has failed and that the stellar atmosphere has not acquired outward motion and is free to accrete toward the black hole.

In \citet{Lindner:10}, we simulated the post-core-collapse hydrodynamical evolution of a rapidly rotating $14 \,M_\odot$ Wolf-Rayet stellar model 16TI of \citep{Woosley:06a} that has been proposed as a LGRB progenitor.  The rate at which the infalling stellar envelope was being accreted onto the black hole evolved through two distinct phases during the first $\sim 200\,\textrm{s}$ following the initial collapse of the stellar core.  First, the low specific angular momentum material of the inner layers of the star accreted quasispherically through the inner boundary, and is presumed to have accreted onto the black hole.  Then, the material that had sufficient angular momentum to become rotationally supported on the computational grid formed a thick accretion torus. Simultaneously, an accretion shock appeared at the innermost radii and traversed the star.  Most of the stellar envelope traversed by the shock was in radial hydrostatic equilibrium and convective; convection transported the energy dissipated at the smallest simulated radii toward the expanding shock.  The central accretion rate was nearly time-independent prior to rotating torus and shock formation, and dropped sharply afterwards.  The abrupt drop of the accretion rate closely resembled the prompt $\gamma$-ray and the early X-ray LGRB light curves measured with the NASA {\it Swift} satellite \citep{Tagliaferri:05,Nousek:06,OBrien:06}, adding weight to the hypothesis that the light curves are responding to an evolution of the central accretion rate \citep{Kumar:08a,Kumar:08b}. Because the innermost simulated radius was $500\,\textrm{km}$, much larger than the innermost stable circular orbit around the central black hole ($5-50\,\textrm{km}$), the accreted-mass-to-energy conversion efficiency was low and the shock acquired relatively low velocities, $\sim 1000\,\textrm{km}\,\textrm{s}^{-1}$, while in the interior of the star.  The star did not explode, but only lost mass to the thermally driven wind that set in after the shock had traversed the star.

In collapsars, a substantially larger accretion energy is dissipated at the radii left out from the \citet{Lindner:10} simulations, closer to the black hole, but only a fraction of this energy couples to the stellar envelope.  The rest may be lost to the emission of neutrinos and to the photodisintegration of hydrostatic elements into free nucleons as well as to advection into the black hole.  Crude analytical considerations \citep{Milosavljevic:12} suggest that following shock formation and the rapid accretion rate drop seen in \citet{Lindner:10}, neutrino losses are relatively small. Then, the amount of energy transferred onto the envelope is determined by the competition of the inward advective and the outward convective energy transport.  The advection arises from the inward drift of the fluid in response to magnetohydrodynamic (MHD) stresses; the convection arises from entropy gradients arising from the dissipation of MHD turbulence.  If convective transport is efficient, the amount of energy transferred from near the black hole to the shocked envelope can be sufficient to drive a fast shock with velocity $\gg 1000\,\textrm{km}\,\textrm{s}^{-1}$ and unbind the star.  The model of \citet{Milosavljevic:12} suggests that the parameters determining the viability and energy of such accretion-powered supernovae are the viscous stress-to-pressure ratio $\alpha$ and the convective mixing length  $\lambda_{\rm conv}$.   The model could not, of course, capture the consequences of the interplay of pressure and rotation at the critical radii where the two sources of radial support against gravity are comparable.  

In this work we show the results of a series of rotating one-dimensional simulations of the immediate aftermath of the collapse of a rapidly rotating LGRB progenitor star's core.  While one-dimensional, our simulations include rotation in a spherically-averaged sense and implement a modified $\alpha$-viscosity prescription.  One customarily refers to such simulations as ``1.5 dimensional.'' They also take into account optically thin cooling by neutrino emission, cooling and heating by nuclear processes, and energy and compositional transport by convection in the mixing length theory approximation.  This work is complementary to our rotating two-dimensional simulations (2.5D) of collapsar accretion \citep{Lindner:10}, in which we simulated only relatively large radii and did not incorporate nuclear and neutrino physics.  Here, we sacrifice in spatial dimensionality to make it possible to track rudimentary nuclear compositional transformation and simulate smaller radii ($r>25\,\textrm{km}$) over similarly extended time periods ($\sim 40-100\,\textrm{s}$).  In the presence of cooling by neutrino emission the rotating central torus may be geometrically thin \citep[e.g.,][]{MacFadyen:99,Popham:99,Kohri:05,Chen:07,Sekiguchi:10,Taylor:10}.  Therefore, we include corrections to approximate the effects of such flow.  The principal source of model uncertainty is the efficiency of convection which in the mixing length approximation can be parameterized with an effective value of the mixing length.  To our best knowledge, there has not been a systematic first principles study of convective efficiencies in the rapidly convecting regime.  Thus the mixing length $\lambda_{\rm conv}$, along with the viscous shear stress-to-pressure ratio $\alpha$, are the parameter dependences that we explore.

A magnetic outflow driven by a proto-neutron star may carry an energy similar to that of a supernova \citep[e.g.,][]{Bisnovatyi-Kogan:71,Wheeler:00,Thompson:04,Bucciantini:07,Burrows:07,Dessart:08}.  However, the outflow may be too axially collimated to produce a standard, quasi-spherical explosion \citep{Bucciantini:08,Bucciantini:09}.  Here, we assume that any explosion mechanism preceding the collapse into a black hole has failed.  Clearly, our one-dimensional model cannot capture the effects of the formation of a magnetized jet, after an accretion disk has formed.  Although this is an integral component to the collapsar model for LGRBs, we omit any treatment of the jet in the present work.

This work is organized as follows.  In Section \ref{sec:algorithm}, we discuss our numerical algorithm.  In Section \ref{sec:results}, we present the results our simulations.  In Section \ref{sec:discussion}, we identify the parameters critical to our model, and discuss their implications for real accretion powered supernovae. Finally, in Section \ref{sec:conclusions}, we summarize our conclusions.

\section{Numerical Algorithm}
\label{sec:algorithm}

The simulations were carried out with the piecewise-parabolic method (PPM) solver in the adaptive-mesh-refinement code FLASH \citep{Fryxell:00}, version 3.2,  in one spatial dimension. Although the rotating stellar collapse is inherently three-dimensional, we have chosen to approximate the key multidimensional effects, including angular momentum transport and convective energy and compositional transport, with a spherically-averaged transport scheme. In Section \ref{sec:angular_momentum}, we describe our implementation of angular momentum transport.  In Section \ref{sec:gravity}, we describe our calculation of the self-gravity of the fluid.  In Section \ref{sec:NSE}, we describe our modeling of the transition toward nuclear statistical equilibrium (NSE) in the hot inner accretion flow.  In Section \ref{sec:neutrino_cooling}, we discuss cooling by neutrino emission.  In Section \ref{sec:convection}, we describe our treatment of convective energy transport and compositional mixing.  In Section \ref{sec:thin_disk}, we describe the corrections that we apply in situations where, in the presence of cooling, the accretion flow is expected to be geometrically thin.  In Section \ref{sec:initial_model}, we describe our initial and boundary conditions.  In Section \ref{sec:tests}, we show the results from tests of the code.  Finally, in Section \ref{sec:limitations}, we briefly review the various limitations our method.

\subsection{Angular Momentum}
\label{sec:angular_momentum}

To include rotation and angular momentum transport in our one dimensional model, we track the specific angular momentum $\ell \equiv r v_\phi$, where $v_\phi$ is the azimuthal velocity, which we interpret as the mass-weighted spherical average of an underlying polar-angle-dependent angular momentum $\ell(r,\theta)$.  If, e.g., spherical shells rotate rigidly, $\ell(r,\theta)\propto \sin^2\theta$, and the fluid density is spherically symmetric, then the one-dimensional specific angular momentum is two-thirds of the midplane value, $\ell = \frac{2}{3}\ell_{\rm mid}$.  The azimuthal Navier-Stokes equation, combined with the equation of mass continuity, then implies the one-dimensional angular momentum transport equation \citep[see, e.g.,][]{Thompson:05}
\bea
\label{eq:ell_evolve}
& &\frac{\partial (\rho \ell)}{\partial t} + \frac{1}{r^2}\frac{\partial( r^2 v_r \rho \ell)}{\partial r} - \frac{1}{r^2}\frac{\partial}{\partial r}\left(r^3 \nu \rho \sigma_{r \phi} \right) = 0 ,
\eea
where $\nu$ is a shear viscosity and 
\beq
\sigma_{r \phi} = r \frac{\partial}{\partial r} \left(\frac{\ell}{r^2}\right)
\eeq
is the $r-\phi$ component of the shear tensor.  
The energy dissipated through shear viscosity was accounted for by including the specific heating rate \citep[see, e.g.,][]{Landau:59}
\beq
\label{eq:viscous_heating}
\dot \epsilon_{\rm visc} \equiv \frac{Q_{\rm visc}}{\rho}= \nu \sigma_{r \phi}^2  ,
\eeq 
where $Q_{\rm visc}$ denotes the volumetric viscous heating rate.
The dimensional reduction in equation (\ref{eq:ell_evolve}) is inaccurate in regions where the disk is geometrically thin. There, the mass-weighted spherical average closely approximates the midplane value, $\ell\sim \ell_{\rm mid}$.  We ignore this effect, but we do incorporate thermodynamic corrections addressing the transition to a thin disk in Section \ref{sec:thin_disk}. 

Our treatment of shear viscosity is similar to our methodology in \cite{Lindner:10}, and for completeness we reproduce our methodology here.  Since we do not simulate the magnetic field of the fluid, we utilize a local definition of the shear viscosity to emulate the magnetic stress arising from the nonlinear development of the magnetorotational instability (MRI; \citealt{Balbus:98} and references therein).  It should be kept in mind, however, that the effects of MRI are in some respects very different from those of the viscous stress. For example, the thick disk surrounding our collapsar black hole is convective; in unmagnetized accretion flows convection transports angular momentum inward, toward the center of rotation \citep{Ryu:92,Stone:96,Igumenshchev:00b}, whereas in magnetized flows, convection can also transport angular momentum outward \citep{Balbus:02,Igumenshchev:02,Igumenshchev:03,Christodoulou:03}.  Although we include treatment for convective energy flux and compositional mixing (see Section \ref{sec:convection}), we do not include angular momentum transport by convection.

Our definition of the local viscous stress emulating the MRI must be valid under rotationally supported, pressure supported, and freely falling conditions, and we proceed as in \citet{Lindner:10}.  \citet{Thompson:05} suggest that since the wavenumber of the fastest growing MRI mode, which is given by the dispersion relation $v_{\rm A} k\sim \Omega$ where $v_{\rm A}$ is the Alfv\'en velocity and $\Omega=v_\phi/r$ is the angular velocity, should in the saturated quasi-state state be about the gas pressure scale height, $k\propto H^{-1}$, the Maxwell $\rho v_{\rm A}^2$ and viscous $\nu\rho \Omega$ stresses (up to factors in $|d\ln \Omega/d\ln r|$ that we neglect) can be equated if the viscosity is given by
\beq
\label{eq:viscosity_Thompson}
\nu_{\rm MRI} = \alpha H^2 \Omega ,
\eeq
where $\alpha$ is a dimensionless parameter. If the pressure scale height is defined locally, 
\beq
\label{eq:local_pressure_height}
H=|\mathbf{\nabla}\ln P|^{-1} ,
\eeq
the viscosity defined in equation (\ref{eq:viscosity_Thompson}) suffers from divergences at pressure extrema.  To alleviate this problem, as in \citet{Lindner:10}, we define a second viscosity according to the \citet{Shakura:73} prescription 
\beq
\label{eq:viscosity_Shakura}
\nu_{\rm SS} = \alpha \frac{P}{\rho} \Omega^{-1} .
\eeq
Shakura-Sunyaev viscosity overestimates the magnetic stress in stratified hydrostatic atmospheres. We thus set the viscosity in equations (\ref{eq:ell_evolve}) and (\ref{eq:viscous_heating}) to equal the harmonic mean of the above two viscosities
\beq
\label{eq:harmonic_mean}
\nu = \frac{2\,\nu_{\rm MRI}\,\nu_{\rm SS} }{\nu_{\rm MRI}+\nu_{\rm SS}} ,
\eeq
where the pressure gradient in equation (\ref{eq:local_pressure_height}) is calculated by the finite differencing of pressure in neighboring fluid cells.  Additionally, we have applied a Gaussian kernel smoothing to the radial dependence of $H$ to help filter short-wavelength numerical instabilities.  We describe this procedure in Section \ref{sec:convection}. 

In FLASH, we treat specific angular momentum as a conserved ``mass scalar'' that is being advected with the fluid, which makes $\rho\ell$ a conserved variable; the corresponding centrifugal force is then incorporated in the calculation of the gravitational acceleration, as we explain in Section \ref{sec:gravity} below. Then the third parabolic term in equation (\ref{eq:ell_evolve}) is computed explicitly through the inclusion of the radial $\rho\ell$-flux $- r \nu \rho \sigma_{r\phi}$ in the advection of $\ell$.  

Numerical stability of an explicit treatment of a parabolic term places a upper limit on the time step
\beq
\label{eq:tstep_diffuse}
\Delta t < \frac{\Delta r^2}{2\nu} ,
\eeq 
where $\Delta r$ is the grid resolution.  For $\alpha \gg 0.01$, the viscous time step in our simulations becomes significantly shorter than the Courant time step.  In our test integrations with a $\gamma$-law equation of state \citep{Lindner:10}, we find that, while not implying an outright instability, a choice of $\Delta t$ that saturates the limit in equation (\ref{eq:tstep_diffuse}) results in weak stationary staggered perturbations in the fluid variables.  We ignore this complication and allow our time step to be set by the limit in equation (\ref{eq:tstep_diffuse}) of the cell with the smallest viscous diffusion time across the cell.

\subsection{Gravity}
\label{sec:gravity}

We calculate contributions to the gravitational potential from a 
central point mass and a spherically-symmetric extended envelope.  
General relativistic effects become important at the innermost radius, which in some simulations is as small as $r_{\rm min} = 25 \,\textrm{km}$.  At radii $r\sim r_{\rm min}$, the black hole dominates the enclosed mass after about $0.5\,\textrm{s}$.  Thus, we describe the gravity of the black hole using the approximate, pseudo-Newtonian gravitational force for a rotating black hole proposed by \citet{Artemova:96}, which is a generalization of the \citet{Paczynski:80}  pseudopotential to rotating black holes. However, we continue to calculate the gravity of the fluid in the Newtonian limit.  The Artemova et al.\ gravitational acceleration in the equatorial plane of a rotating black hole is given by 
\beq
\label{eq:Artemova}
\mathbf{g}_{\rm BH} (r,\theta=\pi/2)= -\frac{G M_{\rm B H}}{r^{2-\beta}(r - r_{\rm H})^\beta} \mathbf{\hat{r}},
\eeq
where $r_{\rm H} = [1+(1-a^2)^{1/2}]GM_{\rm BH}/c^2$ is the radius of the event horizon expressed in terms of the dimensionless spin parameter $a$, and $\beta=r_{\rm ISCO}/r_{\rm H} -1$ is a dimensionless exponent with $r_{\rm ISCO}$ denoting radius of the innermost stable prograde equatorial circular orbit.  We assume a dimensionless spin parameter of $a=0.9$ in these calculations. Our treatment does not incorporate general relativistic corrections to the viscous stress and momentum equations \citep[see, e.g.,][]{Beloborodov:99}.

We adopt the form of  the gravitational acceleration in equation (\ref{eq:Artemova}), which was derived for the equatorial plane of the black hole, to represent the mass-weighted spherical average of the gravitational acceleration, by setting $g_{\rm BH} (r)=g_{\rm BH} (r,\theta=\pi/2)$.  This approximation is appropriate when the accreting mass is concentrated in the equatorial plane, especially when the innermost disk is geometrically thin, and is probably rather inaccurate for an accretion flow that is geometrically thick down to $r_{\rm ISCO}$.  Our simulations predict a geometrically thin disk at $r \lesssim 100\,\textrm{km}$ or greater radii after material has circularized in our simulation, so this assumption seems adequate.

For each zone, the gravitational acceleration due to fluid self gravity is calculated from
\bea
\mathbf{g}_{\rm self} (r_i) &=& -\frac{4 \pi}{3} \frac{G}{r^2} \left\{ \rho_i \left[r_i^3 - \left(r_i - \frac{\Delta r_i}{2}\right)^3 \right] \right. \nonumber\\ & &
+ \left. \sum_{r_k<r_i} \rho_k \left[ \left(r_k + \frac{\Delta r_k}{2}\right)^3 - \left(r_k - \frac{\Delta r_k}{2}\right)^3  \right] \right\} \mathbf{\hat{r}}
\eea
where $\Delta r_i$ and $\Delta r_k$ are the radial widths of the grid cells.
The net gravitational and inertial acceleration in our calculation is then given by
\begin{equation}
\mathbf{a}_{\rm {tot}} = \mathbf{g}_{\rm BH} + \mathbf{g}_{\rm self} + \mathbf{a}_{\rm cent} ,
\end{equation}
where 
\begin{equation}
\mathbf{a}_{\rm cent} = \frac{\ell^2}{r^3} \mathbf{\hat{r}} 
\end{equation}
is the centrifugal acceleration.

\subsection{Nuclear Processes and the Equation of State}
\label{sec:NSE}

To calculate the internal energy of the fluid, we use the Helmholtz equation of state (EOS) of \citet{Timmes:00} included with the FLASH distribution, which accounts for the contributions to pressure and other thermodynamic quantities from radiation, ions, electrons, positrons, and Coulomb corrections.  We track the abundances of 47 nuclear isotopes treated in the NSE calculations of \citet{Seitenzahl:08} and pass the local nuclear composition to the EOS as input.   Given density, temperature, and nuclear composition, the Helmholtz EOS provides the internal energy, density, pressure, entropy, specific heats, adiabatic indices, electron chemical potential, and various derivative thermodynamic quantities.  During the course of the thermodynamic update and the cooling update which is operator-split from the thermodynamic update, the temperature must be derived from the internal energy, and in the Helmholtz EOS, this is achieved by numerically solving for the implicit relation 
\beq
\label{eq:simple_eos}
\epsilon_{\rm EOS}(\rho,T,{\mathbf X}) = \epsilon
\eeq
for the temperature, where $\epsilon$ is the specific internal energy and $\mathbf{X}\equiv (X_1,...,X_{47})$ is the vector of isotopic mass fractions $X_i$.  

The fluid heats and cools in response to nuclear compositional transformation. We do not integrate a nuclear reaction network, but instead model the change of the nuclear composition as a gradual convergence to nuclear statistical equilibrium (NSE) in the part of the flow where the convergence time scale $\tau_{\rm NSE}$ is comparable to or shorter than the age of the system.  In this model, as we explain below, the nuclear composition responds instantaneously to a change of the temperature, implying that the dependence of the composition on the temperature must be taken into account, in a manner that conserves the combined specific internal and nuclear energy $\epsilon+\epsilon_{\rm nuc}$ when solving the EOS for temperature. Here,
$\epsilon_{\rm nuc}$ is the specific (negative) nuclear binding energy of the fluid
\begin{equation}
\label{eq:binding_energy}
\epsilon_{\rm nuc} = \sum_{i} \frac{X_i E_{{\rm B},i}}{A_i m_p} ,
\end{equation} 
while $E_{{\rm B},i}$ is the negative nuclear binding energy of the isotope and $A_i$ is the atomic mass of the isotope.

The time scale for convergence to NSE can be approximated via \citep[][see, also, \citealt{Calder:07}]{Khokhlov:91}
\begin{equation}
\tau_{\rm NSE}=\rho^{0.2} \exp \left(\frac{179.7}{T_9} - 40.5\right) \,\textrm{s},
\end{equation}
where $T=10^9 \,T_9 \,\textrm{K}$ and $\rho$ is the density in $\textrm{g}\,\textrm{cm}^{-3}$. At relevant densities, this time scale is of the order of one second for $T_{\rm NSE}\approx 4\times10^9\,\textrm{K}$.  We calculate the nuclear mass fractions using the publicly available solver of \citet{Seitenzahl:08} which solves for the NSE mass fractions $X_{{\rm NSE},i}$ of 47 nuclear isotopes as a function of density $\rho$, temperature $T$, and proton-to-nucleon ratio $Y_e=\sum_i Z_iX_i/A_i$, where $Z_i$ is the atomic number an isotope. At temperatures $T > 3 \times 10^9 \,\textrm{K}$ we model convergence to NSE via
\begin{equation}
\label{eq:X_convergence_NSE}
\left(\frac{\partial X_i}{\partial t}\right)_{\rm nuc} = \frac{X_{i,{\rm NSE}} (\rho,T_{\rm NSE},Y_e)- X_i}{\tau_{\rm NSE}(\rho,T)} ,
\end{equation}
where $T_{\rm NSE}$ is the temperature that the fluid element would have given enough time to relax into NSE while keeping the total specific energy $\epsilon+\epsilon_{\rm NSE}$ and proton-to-nucleon ratio $Y_e$ fixed. The temperature $T_{\rm NSE}$ is implicitly defined by the condition (cf.\ equation [\ref{eq:simple_eos}])
\bea
\label{eq:T_NSE}
& &\epsilon_{\rm EOS}[\rho,T_{\rm NSE},{\mathbf X}_{{\rm NSE}}(\rho,T_{\rm NSE},Y_e)] + \epsilon_{\rm nuc}[{\mathbf X}_{{\rm NSE}}(\rho,T_{\rm NSE},Y_e)]  \nonumber\\ 
& &= \epsilon(\rho,T,\mathbf{X})+\epsilon_{\rm nuc}(\mathbf{X}) .
\eea
This condition ensures that the sum of the internal and nuclear energy densities in NSE would equal the sum of the two energy densities in the model.
We solve equation (\ref{eq:T_NSE}) for $T_{\rm NSE}(\rho,Y_e,\epsilon,\epsilon_{\rm nuc})$ iteratively and then update the abundances by discretizing equation (\ref{eq:X_convergence_NSE}) with
\bea
 X_i (&t&+\Delta t) = X_{i,\rm{NSE}}(\rho,T_{\rm NSE},Y_e) \nonumber\\ & &
+ [ X_i(t) - X_{i,\rm{NSE}} (\rho,T_{\rm NSE},Y_e) ]\exp\left[-\frac{\Delta t}{ \tau_{\rm{NSE}}(\rho,T)}\right] .
\eea
Following the update of the nuclear mass fractions, we update the specific internal energy to account for heating or cooling due to any change in specific nuclear binding energy 
\beq
\label{eq:epsilon_update}
\epsilon(t+\Delta t) = \epsilon(t) + \epsilon_{\rm nuc}[\mathbf{X}(t)] - \epsilon_{\rm nuc}[\mathbf{X}(t+\Delta t)] ,
\eeq
and finally update the temperature from equation (\ref{eq:simple_eos}).

This prescription does not affect the proton-to-nucleon ratio $Y_e$; that latter is a conserved mass scalar in our simulations. Thus, the expected partial neutronization in the mildly degenerate innermost segment of the accretion flow is not calculated and our prescription cannot be used to accurately estimate the $^{56}$Ni fraction within the Fe-group elements synthesized in the simulation.

\subsection{Cooling}
\label{sec:neutrino_cooling}

The hot innermost accretion flow cools via neutrino emission. At the densities observed in our simulation, the disk and stellar atmosphere are transparent to neutrinos.   The two most significant neutrino-emission channels \citep[e.g.,][and references therein]{DiMatteo:02} are:

1. Pair capture on free nucleons (the Urca process): $p + e^- \rightarrow n + \nu$ and $n + e^+ \rightarrow p + \bar{\nu}$. The cooling rate is
\begin{equation}
\label{eq:neutrino_cooling1}
Q_{eN} = 9\times10^{33} \rho_{10} T_{11}^6 X_{\rm nuc} ~~{\rm ergs~cm^{-3}~s^{-1}} ,
\end{equation}
where $\rho = 10^{10} \rho_{10} \,{\rm g}\,{\rm cm}^{-3}$, $T=10^{11} \,T_{11} \,\textrm{K}$, and $X_{\rm nuc}=X_p+X_n$ is the mass fraction in free nucleons.

2. Pair annihilation ($e^- + e^+ \longrightarrow \nu + \bar{\nu}$).  The cooling rate is
\begin{equation}
\label{eq:neutrino_cooling2}
Q_{e^+e^-} = 1.5\times10^{33} T_{11}^9 ~~{\rm ergs~cm^{-3}~s^{-1}} .
\end{equation}
All three flavors, $e$, $\mu$, and $\tau$, of neutrinos are included.

We have included the above neutrino cooling rates in our calculations, where losses are computed via
\begin{equation}
\epsilon (t+\Delta t)= \epsilon(t) -  \frac{Q_\nu}{\rho}\Delta t ,
\end{equation}
where $Q_\nu=Q_{eN} + Q_{e^+e^-}$ is the total volumetric neutrino cooling rate.
The update of the internal energy due to cooling is operator split from the update due to nuclear compositional change.

\subsection{Convection}
\label{sec:convection}

We introduce convective energy transport and compositional mixing within the framework of mixing length theory \citep[e.g.,][]{Kuhfuss:86}.  In the calculation of the convective transport fluxes, we ignore the radial variation of the mean molecular weight as well as rotation, and the condition for instability is simply the Schwarzschild criterion, $\partial s/\partial r < 0$. Then, in unstable zones, the convective energy flux is
\begin{equation}
\label{eq:flux_with_dsdr}
F_{\rm conv} = -\frac{1}{2} c_P \rho v_{\rm conv} \lambda_{\rm conv}\left( \frac{\partial T}{\partial s} \right)_P \, \frac{\partial s}{\partial r} ,
\end{equation} 
where $c_P$ is the specific heat at constant pressure, $\lambda_{\rm conv}$ is the length over which convection occurs, $s$ is specific entropy, and $v_{\rm conv}$ is the convective velocity.  The convective velocity can be approximated by
\begin{equation}
\label{eq:v_convective}
v_{\rm conv}\sim \frac{1}{2}\lambda_{\rm conv} \left[-\frac{g}{\rho} \left(\frac{\partial\rho}{\partial T}\right)_P  \frac{T}{c_P} \frac{ds}{dr}\right]^{1/2}, 
\end{equation}
where $g<0$ is the gravitational acceleration in the local rest frame of the convectively unstable fluid
\begin{equation}
\label{eq:grav_accel_fluid_frame}
g = g_{\rm grav} - \frac{dv}{dt} =  \frac{1}{\rho} \frac{\partial P}{\partial r} = -\frac{P}{\rho H} ,
\end{equation}
and $g_{\rm grav}=g_{\rm BH}+g_{\rm self}$ is the net gravitational acceleration in the inertial frame, $v$ in the second step denotes the mass-weighted spherical average of the fluid velocity at radius $r$.  To parameterize our uncertainty regarding the value of the convective mixing length, we introduce a dimensionless parameter $\xi_{\rm conv}\sim {\cal O}(1)$ defined as 
\beq
\label{eq:lambda_C}
\xi_{\rm conv}\equiv \left(\frac{\lambda_{\rm conv}}{H}\right)^2 .
\eeq
Then, combining equations (\ref{eq:flux_with_dsdr}) -- (\ref{eq:lambda_C}), we obtain the standard expression
\begin{equation}
\label{eq:F_C}
F_{\rm conv} = \frac{1}{4} \xi_{\rm conv} H^2 c_P \left[ -\frac{P}{H} \left( \frac{\partial \rho}{\partial T} \right)_P  \right]^{1/2} \left(-\frac{T}{c_P}\frac{\partial s}{\partial r}  \right)^{3/2} ,
\end{equation}
which is appropriate even when the fluid is not in hydrostatic equilibrium and $v_r\neq 0$.

In evaluating the convective energy flux at a boundary (face) of a computational cell, we use face-centered linear interpolation of the density, temperature, and pressure.  The zone-centered values of the specific heat $c_{P}$, specific entropy $s$, and thermodynamic derivatives $(\partial P / \partial T )_\rho$, and $( \partial P / \partial \rho )_T$ are returned by the EOS routine, and the face-centered values are again computed by linear interpolation. Then,  $( \partial \rho/\partial T )_P$ is calculated from 
\begin{equation}
\left( \frac{\partial \rho}{\partial T} \right)_P = -\left( \frac{\partial P}{\partial T} \right)_\rho \left/ \left( \frac{\partial P}{\partial \rho} \right)_T \right. .
\end{equation}
The convective energy flux never exceeds
\begin{equation}
F_{\rm conv} \leq \rho \epsilon c_{\rm s} ,
\end{equation}
where $c_{\rm s}=(\gamma_c P/\rho)^{1/2}$ is the adiabatic sound speed, and $\gamma_c$ is the adiabatic index.

We anticipate that a local application of MLT, in which the expression for the convective energy flux contains a pressure derivative in the denominator, may contain an instability.  The instability is an artifact of modeling the intrinsically nonlocal convective energy transport with a local nonlinear differential operator.  To control---if not entirely prevent---undesirable outcomes of the instability, we filter short wavelength perturbations in the calculation of the pressure scale height $H$ that enters our estimates of the viscosity and the energy flux transported by convection by applying a Gaussian smoothing
\begin{equation}
P_{\rm smooth} (r) = \frac{\sum_i k_i (r) \,P_i}{\sum_i k_i(r)} ,
\end{equation}
where the summations are over all of the cells in the simulation, and the spherically-averaged smoothing kernel $k_i$ is given by
\begin{equation}
k_i (r)= \frac{1}{2\sqrt{2\pi}}\frac{\Delta r_i r_i}{r\sigma}  \left\{\exp \left[ -\frac{(r - r_i)^2}{2 \sigma^2} \right] -\exp \left[ -\frac{(r + r_i)^2}{2 \sigma^2} \right]  \right\} .
\end{equation}
Here, $\sigma$ is a radius-dependent smoothing length that we set to $\frac{1}{2}r$. Similarly, in the evaluation of the specific entropy derivative in equation (\ref{eq:F_C}), we smooth the specific entropy $s$ via 
\begin{equation}
s_{\rm smooth} (r) = \frac{\sum_i k_i (r) \,\rho_i \,s_i}{\sum_i k_i(r) \, \rho_i } .
\end{equation}
The filtering affects only the evaluation of $F_{\rm conv}$ and helps avoid breakdown of our transport scheme, but residual artificial non-propagating waves do develop, and saturate, on wavelengths comparable to the smoothing length. 

The accretion shock formally presents a negative entropy gradient but physically does not give rise to convection.  The upstream of the shockwave is marginally convectively stable as the shockwave traverses the progenitor's convective core, and becomes absolutely stable in the radiative envelope.  To prevent spurious convection across the shock transition, we modify the convective flux to decline to zero linearly near the shock
\begin{equation}
\label{eq:shock_limiter}
F_{{\rm conv},{\rm mod}}(r) = \begin{cases} (1-r/r_{\rm shock}) F_{\rm conv}(r), & r<r_{\rm shock} , \\
0, & r\geq r_{\rm shock} , \end{cases}
\end{equation}
where $r_{\rm shock}$ is the radius of the accretion shock front which we track during the simulation. 

\citet{Murphy:11} argue that on physical grounds, in quasi-stationary ``stalled'' shocks in the standard core-collapse context, the distance from the shock $r_{\rm shock}-r$ is the appropriate convective length scale near the shock, as convective eddies can grow to the largest size available to them.  If we had set the convective mixing length $\lambda_{\rm conv}$ proportional to the distance from the shock, which is the adaptation of MLT that Murphy \& Meakin suggest, equation (\ref{eq:shock_limiter}) would have contained a quadratic factor $(1-r/r_{\rm shock})^2$, instead of the linear factor $(1-r/r_{\rm shock})$ that we employ.  The physically motivated modification of $\lambda_{\rm conv}$ of Murphy \& Meakin, which we became aware of after the completion of this work, and our ad hoc version should give rise to similar dynamics, especially when the shock travels outward as in our simulations.

Convection also gives rise to compositional mixing in the convective region.  We model the mixing of nuclear species in the diffusion approximation \citep[e.g.,][]{Cloutman:76,Kuhfuss:86}
\beq
\label{eq:mixing}
\left[\frac{\partial (\rho X_i)}{\partial t}\right]_{\rm mix}  =  - \frac{1}{r^2} \frac{\partial}{\partial r} (r^2 {\cal F}_{{\rm mix},i}) ,
\eeq
where 
\beq
{\cal F}_{{\rm mix},i} = - \frac{1}{3} \nu_{\rm conv} \rho \,\frac{\partial X_i}{\partial r}
\eeq
is the mass flux of species $i$ transported by convection, while $\nu_{\rm conv}$ is the compositional diffusivity which
we take to be proportional to the convective velocity multiplied by the pressure scale height 
\beq
\nu_{\rm conv} = \xi_{\rm mix} v_{\rm conv} \lambda_{\rm conv} ,
\eeq
and $\xi_{\rm mix}\sim {\cal O}(1)$ is a dimensionless parameter.
We again apply the flux limitation behind the shock front in the form of the linear factor in equation (\ref{eq:shock_limiter}).  The compositional diffusion is also subject to the timestep limitation imposed in equation (\ref{eq:tstep_diffuse}).  It is worth noting that compositional diffusion implies a flux of nuclear energy given by
\beq
F_{\rm nuc,mix} = \sum_i \frac{E_{{\rm B},i}\,{\cal F}_{{\rm mix},i} }{A_i\, m_p} .
\eeq

The entropy transport equation implied by our algorithm is
\bea
\label{eq:internal_energy_conservation}
\rho T \frac{d s}{d t}+ \frac{1}{r^2} \frac{\partial }{\partial r} (r^2 F_{\rm conv, mod} ) = Q_{\rm visc} - Q_\nu + Q_{\rm nuc} ,
\eea
where $d/dt=\partial/\partial t+v_r\partial/\partial r$ and 
\beq
Q_{\rm nuc} = - \rho \sum_i \frac{E_{{\rm B},i}}{A_i m_p} \frac{\partial X_i}{\partial t}
\eeq
is the rate of heating or cooling associated with nuclear compositional transformation (see equations [\ref{eq:binding_energy}] and [\ref{eq:X_convergence_NSE}]).

\subsection{Thin Disk Corrections}
\label{sec:thin_disk}

We have thus far assumed a rotating, quasi-spherical accretion flow.  However, near the black hole, where cooling by neutrino emission and nuclear photodisintegration into nucleons is significant, the flow can become geometrically thin \citep[e.g.,][]{MacFadyen:99,Popham:99,Kohri:05,Chen:07}.  In this case, the quasi-spherical treatment underestimates the density, pressure, and temperature near the midplane of thin disk, where the bulk of the neutrino emission takes place.  We introduce a correction that adjusts the temperature of the flow to be closer to the physical, thin-disk value.

In what follows, the quantities applying to the thin disk will be marked with tilde.  Let $\tilde H_z$ denote the vertical half-thickness of the thin disk.  We assume that the vertical half-thickness of the quasi-spherical flow is $H_z\sim \frac{1}{2} r$.  Since the thin and the quasi-spherical flow must contain the same column density, $\tilde H_z \tilde\rho \sim \frac{1}{2} r \rho$.  Ignoring differences in nuclear composition between the thin and quasi-spherical flows, the same correspondence must apply to the total internal energies integrated along the vertical column, $\tilde H_z \tilde\rho\tilde\epsilon \sim \frac{1}{2} r \rho\epsilon$, and thus, $\tilde\epsilon\sim \epsilon$ while $\tilde H_z \tilde P \sim \frac{1}{2} r P$.  Vertical force balance in the thin disk requires $\tilde P/\tilde H_z = \tilde \rho \tilde |g_z|$, where $\tilde g_z = -{\rm sgn}(z)\,g\,\tilde H_z/r$ is the gravitational acceleration in the $z$-direction, and $g\equiv (\mathbf{g}_{\rm BH}+\mathbf{g}_{\rm self})\cdot\hat{\mathbf{r}}$ is the radial gravitational acceleration (see Section \ref{sec:gravity}).   Enforcing that $\tilde H_z \leq H_z$, we obtain
\beq
\label{eq:scale_height}
\tilde H_z = {\rm min} \left[\left(-\frac{rP}{\rho g}\right)^{1/2},\frac{r}{2}\right] .
\eeq

\begin{figure*}
\begin{center}
\includegraphics[width=0.45\textwidth,clip]{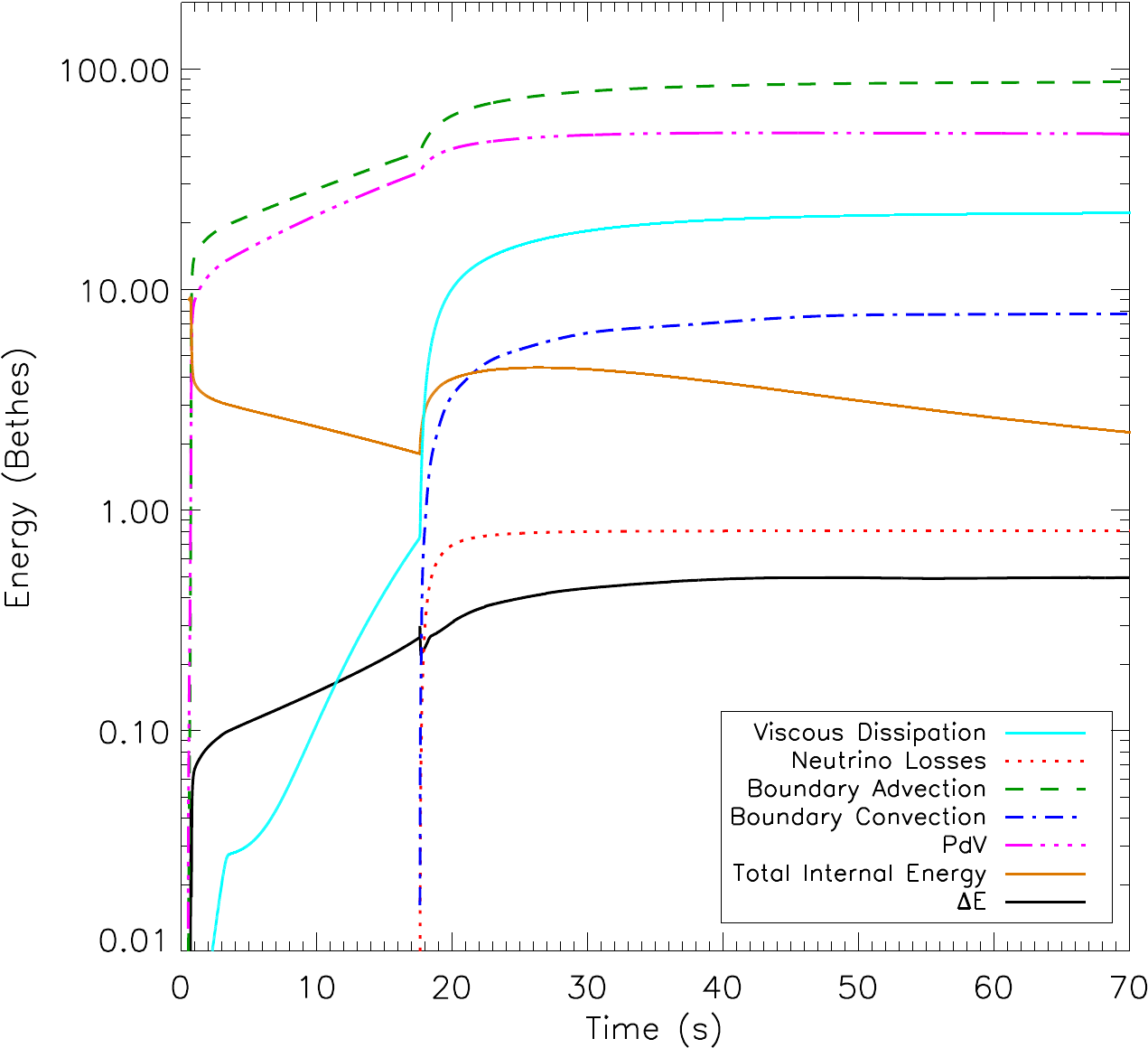}
\includegraphics[width=0.45\textwidth,clip]{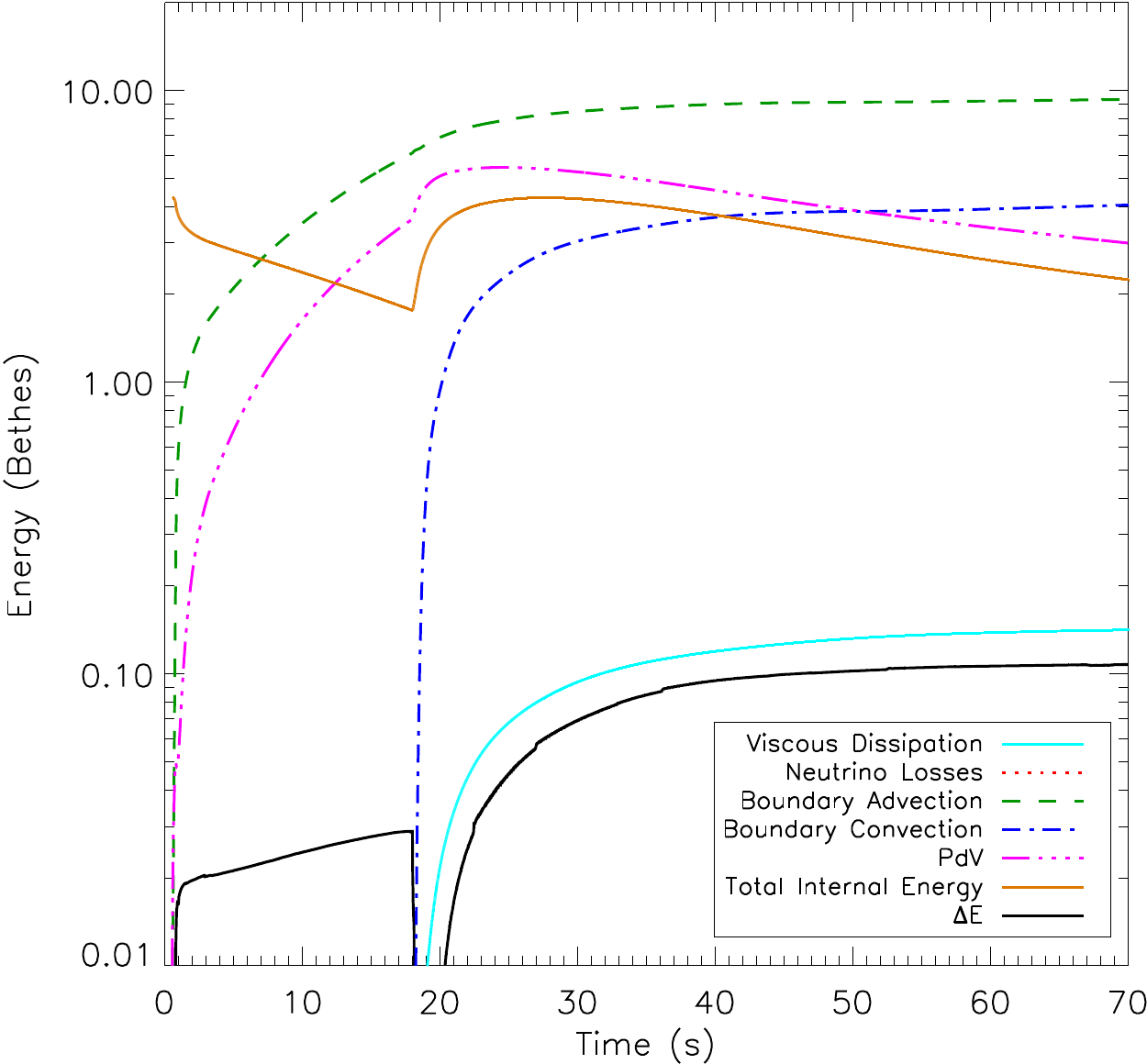}
\end{center}
\caption{Test of internal energy conservation in a run identical to Run 2 except with nuclear compositional change disabled and using a Newtonian gravitational potential.  Plotted are each of the terms in equation (\ref{eq:energy_conservation}) calculated at  $r_{\rm min} = 50 \,\textrm{km}$ (\textit{left}) and $r_{\rm min} = 1000 \,\textrm{km}$ (\textit{right}).  Note the difference in scales on the vertical axis.  The quantity $\Delta E$ represents the absolute value of the difference between the left and right hand side or equation (\ref{eq:energy_conservation}). A possibly dominant source of apparent error is inconsistent discretization of the various fluid variables and fluxes in PPM and in the post-processing and need not reflect an inaccuracy of the computation.  Neutrino losses are insignificant outside the inner $\sim1000\,\textrm{km}$.  Over the interval $0\,\textrm{s} \leq t \leq 70\,\textrm{s}$, we find that the total error is $<1\%$ of the largest term in equation (\ref{eq:energy_conservation}).
\label{fig:energy_conservation}}
\end{figure*}

To account for the higher density in the thin disk, we could pass $\tilde\rho$ to the EOS.  This, however, would result in a modified pressure $\tilde P$.  Out of a possibly unfounded concern that a modification of the pressure would introduce spurious dynamics in the spherically-averaged flow, we opted to modify our estimate of the disk midplane temperature in a manner not directly affecting the fluid pressure.  This corrected temperature then enters the calculation of the neutrino cooling rate and the NSE composition, both of which are highly sensitive to the midplane temperature.

We estimate the midplane temperature $\tilde T$ from the following extrapolation
\beq
\ln\left(\frac{\tilde T}{T}\right)\approx \left(\frac{\partial\ln T}{\partial\ln\rho}\right)_\epsilon \ln \left(\frac{\tilde\rho}{\rho}\right) ,
\eeq
where the partial derivative, which we denote with $\chi$, is evaluated at constant specific internal energy and can be expressed as
\bea
\label{eq:thin_disk_exponent}
\chi&\equiv& \left( \frac{\partial \ln T}{\partial \ln \rho}  \right)_\epsilon \nonumber\\
&=& -\left( \frac{\partial s}{\partial \ln \rho} \right)_T \left/ \left( \frac{\partial s}{\partial\ln T} \right)_\rho \right.- \frac{P}{\rho c_V T} 
\eea
where $c_V$ is the specific heat at constant volume.  The quantities on the right hand side of equation (\ref{eq:thin_disk_exponent}), are all provided by the Helmholtz EOS.

Since $\tilde\rho/\rho\sim r / (2\tilde H_z)$, the midplane temperature of the disk can be approximated via
\beq
\label{eq:midplane_temperature}
\tilde T = \left(\frac{r}{2\tilde H_z}\right)^\chi T \equiv \Xi \,T ,
\eeq
where the last equality defines the dimensionless temperature correction factor $\Xi$. 
To ensure continuity near the shock transition, we modify the correction factor to linearly approach unity at the shock transition by defining 
\beq
\label{eq:Xi_smooth}
\Xi_{\rm mod} = 1 + \left(1-\frac{r}{r_{\rm shock}}\right) (\Xi - 1 ) .
\eeq
For clarity of notation, we drop the subscript in $\Xi_{\rm  mod}$ in what follows.

The correction introduced in equation (\ref{eq:midplane_temperature}) affects both the temperature calculated from internal energy via the equation of state, as well as the NSE temperature calculated from the total internal and nuclear energy as described in Section \ref{sec:NSE} above.  Equation (\ref{eq:simple_eos}) is corrected to become 
\beq
\label{eq:simple_eos_corrected}
\epsilon_{\rm EOS}(\rho,\Xi^{-1}\tilde T,{\mathbf X}) = \epsilon ,
\eeq
while equation (\ref{eq:T_NSE}) is corrected to become 
\bea
\label{eq:T_NSE_corrected}
& &\epsilon_{\rm EOS}[\rho,\Xi^{-1}\tilde T_{\rm NSE},{\mathbf X}_{{\rm NSE}}( \tilde \rho,\tilde T_{\rm NSE},Y_e)] \nonumber\\ & &\ \ \ + \,\epsilon_{\rm nuc}[{\mathbf X}_{{\rm NSE}}(\tilde\rho,\tilde T_{\rm NSE},Y_e)]  \nonumber\\ 
& &= \epsilon(\rho,\Xi^{-1}\tilde T,{\bf X}) + \epsilon_{\rm nuc}(\mathbf{X}) .
\eea
Note that since $\Xi\geq 1$, the estimated midplane disk temperature is higher than the temperature calculated without this correction, but this allows the disk to cool faster than it would otherwise.  The rate of cooling by neutrino emission is then calculated from equations (\ref{eq:neutrino_cooling1}) and (\ref{eq:neutrino_cooling2}) but at density $\tilde\rho$ and temperature $\tilde T$.

\subsection{Initial Model and Boundary Conditions}
\label{sec:initial_model}

The initial model is the rotating $M_{\rm star}\approx 14\,M_\odot$ Wolf-Rayet star 16TI of \citet{Woosley:06a}, evolved to pre-core-collapse from a $16\,M_\odot$ main sequence progenitor.\footnote{\citet{LopezCamara:09} carried out SPH simulations of neutrino-cooled accretion during the first $0.5\,\textrm{s}$ of the collapse and \citet{Morsony:07} and \citet{Nagakura:11a} simulated the propagation of a relativistic jet using the same model star.  We discuss important caveats of using this model in Section \ref{sec:discussion}.} To prepare the model 16TI, Woosley \& Heger assumed that the rapidly rotating progenitor, which is near breakup at its surface at $r_{\rm star}\approx 4\times10^{5}\,\textrm{km}$, had low initial metallicity, $0.01\,Z_\odot$, and became a WR star shortly after central hydrogen depletion, which implied an unusually small amount of mass loss.  For illustration, the specific angular momentum at the three-quarters mass radius was $\ell_{3/4}\sim 8\times10^{17} \,\textrm{cm}^2\,\textrm{s}^{-1}$, implying circularization around a $5\,M_\odot$ black hole at $r\sim 2500 \,\textrm{km}$, much larger than ISCO.  The circularization radii of the outermost layers of the star are in the range $10^4-10^5 \,\textrm{km}$.  Woosley \& Heger provide a radius-dependent angular momentum profile $\ell_{\rm 16TI} (r)$.  We introduce the dimensionless parameter $\xi_{\ell}$ to scale the specific angular momentum $\ell(r)$ of our initial model relative to that of 16TI
\beq
\label{eq:ell}
\ell(r) = \xi_{\ell} \,\ell_{\rm 16TI} (r) .
\eeq
The plots of density, temperature, angular momentum, and composition in Section \ref{sec:results} show the initial conditions.  The angular momentum profile is specific to our fiducial Run 1 with $\xi_{\ell}=0.5$, half of the rotation rate of 16TI.  

The iron core of the model 16TI, with a mass $\sim 1\,M_\odot$, has mass too low to collapse directly into a black hole, but should instead first collapse into a neutron star. The latter could, but need not to, be driven to a successful explosion by the delayed neutrino mechanism.  A black hole can form by fallback.  We do not in any way account for the core bounce and its consequences, nor for the heating by the neutrinos emitted from the proto neutron star.  Our central compact object is a point mass from the outset equipped with, as we clarify below, an absorbing boundary condition.

Pseudo-logarithmic gridding is achieved by capping the adaptive resolution at radius $r$ with
$\Delta r > \frac{1}{8} \eta r$ where $\eta$ is a dimensionless parameter.  We choose $\eta=0.15$ for all but Run 2, where
$\eta = 0.075$.  Beyond the outer edge of the star we place a cold ($10^4\,\textrm{K}$), low-density, stellar-wind-like medium with density profile $\rho(r)=3\times10^{-7} \, (r/r_{\rm star})^{-2}\,\textrm{g cm}^{-3}$.

The simulation was carried out in the spherical domain $r_{\rm min}< r<r_{\rm max}$. We placed the inner boundary at  $r_{\rm min}\sim 25\,\textrm{km}$ and the outer boundary well outside the star at $r_{\rm max}=10^7\,\textrm{km}$.  In Table \ref{tab:simulations}, we summarize the main parameters of our simulations, and also present some of the key measurements, defined in Section \ref{sec:results}, characterizing the outcome of each simulation.  Each simulation was run for $\sim 10^7$ hydrodynamic time steps and required $\sim 5000$ CPU hours to complete.

The boundary condition at $r_{\rm min}$ was unidirectional ``outflow'' that allowed free flow from larger to smaller radii ($v_r<0$)  and disallowed flow from smaller to larger radii ($v_r>0$) by imposing a reflecting boundary condition.  We imposed the torque-free boundary condition via \citep[see, e.g.,][]{Zimmerman:05}
\beq
\label{eq:torque_free}
\frac{\partial}{\partial r}\left(\frac{\ell}{r^2}\right)_{r=r_{\rm min}} = 0 .
\eeq
As in other Eulerian codes, the boundary conditions in FLASH are set by assigning values to fluid variables in rows of ``guard'' cells just outside the boundary of the simulated domain.  Let $r_{1/2}$ denote the leftmost cell within the simulated domain, and let $r_{\mathcal G}$ where ${\mathcal G}=(-\frac{7}{2},-\frac{5}{2},-\frac{3}{2},-\frac{1}{2})$ be the four guard cells to the left of $r_{1/2}$ such that the grid separation corresponds to $\Delta {\mathcal G}=1$. The torque-free boundary condition, if assumed to apply for $r\leq r_{\rm min}$, implies $\ell_{\mathcal G}/r_{\mathcal G}^2=\ell_{1/2}/r_{1/2}^2$.  All other fluid variables $X$ were simply copied into the guard cells, $X_{\mathcal G}=X_{1/2}$, and were subsequently rendered thermodynamically consistent.  This simple prescription approximates free inflow (toward smaller $r$) across $r_{\rm min}$. The guard cell values for other fluid variables are assigned ignoring curvature of the coordinate mesh and formally violate conservation laws at $r<r_{\rm min}$.

The mass of the black hole $M_{\rm BH}$ was initialized with the mass of the initial stellar model contained within $r_{\rm min}$.  The black hole mass was evolved by integrating the mass  crossing the boundary at $r=r_{\rm min}$, 
\beq 
\frac{dM_{\rm BH}}{dt}=  (- 4 \pi r^2 \rho v_r)_{r=r_{\rm min}}.
\eeq
The sum of the mass of the black hole and the mass contained on the computational grid remains constant to a high level of precision throughout each simulation.

\subsection{Assessment and Tests of the Code}
\label{sec:tests}

We conducted tests of internal energy conservation, angular momentum transport, and spatial resolution convergence.  The time-integrated equation for the conservation of internal energy in absence of nuclear and thermal energy interconversion in spherical coordinates reads
\bea
\label{eq:energy_conservation}
& &E_{\rm int, tot}(t_{\rm max}) - E_{\rm int, tot}(t_{\rm min}) 
= 4 \pi  r_{\rm min,test}^2 \int_{t_{\rm min}}^{t_{\rm max}} (v_r \rho \epsilon + F_{\rm conv}) dt \nonumber \\
&-&4\pi  \int_{t_{\rm min}}^{t_{\rm max}} \int_{r_{\rm min,test}}^{r_{\rm max}}  \left[  \frac{P}{r^2}\frac{\partial (r^2v_r) }{\partial r}- Q_{\rm visc}+ Q_{\nu} \right] r^2 dr dt =0 ,
\eea
where 
\beq
E_{\rm int, tot} = \int_{r_{\rm min,test}}^{r_{\rm max}}{4 \pi r^2 \rho \epsilon dr}
\eeq
and $r_{\rm min,test}\geq r_{\rm min}$ is a reference radius defining the inner boundary of the spherical annulus in which we test energy conservation.
We have ignored any flow of energy through $r_{\rm max}$, since stellar material does not reach this radius in the course of any simulation.

\begin{figure}
\begin{center}
\includegraphics[width=.45\textwidth,clip]{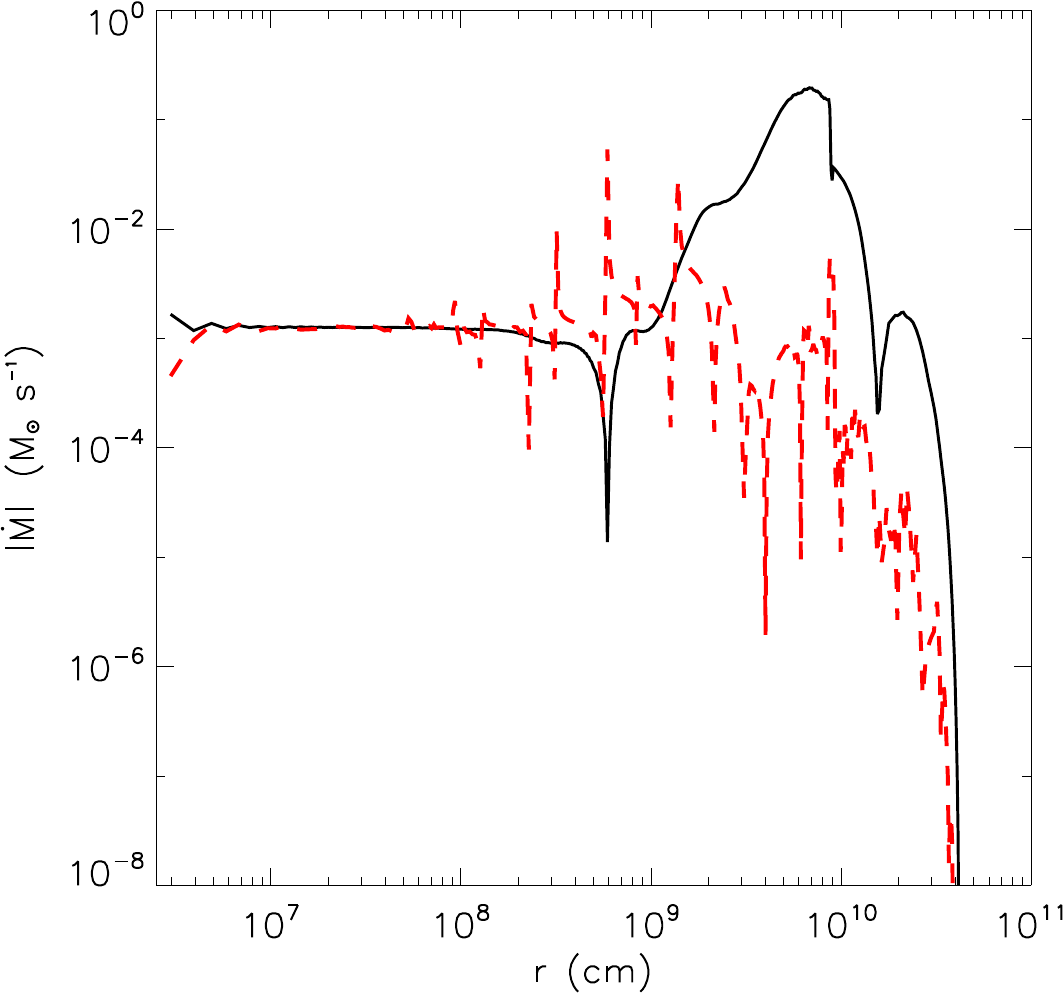}
\caption{The absolute value of the actual (\emph{black, solid}) and steady state analytic (\emph{red, dotted}, see equation [\ref{eq:mdot}]) mass flow, $\dot{M}$, as a function of radius in Run 1 at $t=50\,\textrm{s}$.  The deviation from the analytical value is $\lesssim 5\%$ at radii $100\,\textrm{km} \lesssim r \lesssim 1000\,\textrm{km}$.  At this time, $r_{\rm shock} \sim 7 \times 10^4\,\textrm{km}$. \label{fig:mass_analytic}}
\end{center}
\end{figure}

In Figure \ref{fig:energy_conservation}, we utilize equation (\ref{eq:energy_conservation}) to test the global conservation of internal energy in a run with identical test parameters to Run 2, except that we disabled nuclear compositional change and used a Newtonian gravitational potential without relativistic corrections. In the legend, the apparent error $\Delta E$ is defined as the absolute value of the difference between the left and right hand side or equation (\ref{eq:energy_conservation}).  The evaluation of the various terms in equation (\ref{eq:energy_conservation}) was carried out in post-processing from cell-centered data recorded in $\Delta t =0.01\,\textrm{s}$ intervals, which, in retrospect, is prone to the introduction of various spatial and temporal discretization artifacts not present in the actual simulation. 

We find that the apparent error is $<1\%$ of the largest term in equation (\ref{eq:energy_conservation}) when calculated for $r_{\rm min,test} = 50 \,\textrm{km}$ for the time interval $0\,\textrm{s} \leq t \leq 70\,\textrm{s}$.  The apparent error is most significant, $\Delta E\sim 4\times10^{50}\,\textrm{ergs}$, prior to and during the first few seconds after shock passage. The apparent error that accrues after the first few seconds following shock passage is less than $10^{50}\,\textrm{ergs}$.  This can be compared to the total binding energy change on the simulation grid, which if sufficiently large can imply a supernova. We calculate this energy in Section \ref{sec:explosion} below and find that it increases by $\sim (1.5-2)\times10^{51}\,\textrm{ergs}$ following shock formation, and a significant fraction ($\sim 5\times10^{50}\,\textrm{ergs}$) of the increase is accrued later than a few seconds after shock formation, when the change in the cumulative apparent error is very small. Therefore, it does not seem that the apparent energy conservation error at the levels seen in the simulations should significantly impact the prospects for explosion.  We note that in our calculations we explicitly transport specific internal energy rather than the total energy, by setting the parameter {\tt eintSwitch} to a very large value.  We would like to reiterate that it is likely that the apparent error is an artifact of post processing and the true energy conservation is better.  To demonstrate the latter, however, one would have to reconstruct the diagnostic energy fluxes using the very same interpolation procedure as is performed within the PPM in FLASH.  We anticipate carrying out such a test in an extension of this work.

In steady state accretion, mass accretion associated with viscous angular momentum transport should occur at the rate
\beq
\label{eq:mdot}
\dot{M}_{\rm s.s.} = - 4 \pi \left( \frac{\partial \ell}{\partial r} \right)^{-1} \frac{\partial}{\partial r} \left( r^4 \nu \rho \frac{\partial \Omega}{\partial r} \right) .
\eeq
In Figure \ref{fig:mass_analytic} we show $\dot{M}(r)$ for Run 2 at $t=50 \,\textrm{s}$ along with the analytic steady-state estimate of equation (\ref{eq:mdot}).  In the rotationally supported region $100\,\textrm{km} \lesssim r \lesssim 1000\,\textrm{km}$, the deviation from the analytical value is $\lesssim 5\%$, which lends credence to the accuracy of our angular momentum transport scheme.

Our spatial resolution was chosen such that we resolve the innermost, neutrino-cooled region of the disk over several zones.  One caveat is that we do not resolve the sonic radius of the flow, an issue discussed in \cite{McKinney:02}.  Because we use a torque-free boundary condition, the calculation is not subject to spurious viscous dissipation at the inner boundary.  However, our boundary condition may still influence the fluid flow, and it therefore may be more apt to consider values of $\dot{M}$, as opposed to, for example, $\alpha$ or $\ell$, when comparing our work to other simulations.

\begin{deluxetable}{cccccc}
\label{tab:simulations}
\tablecolumns{6}
\tablewidth{3.4in}
\tablecaption{Summary of Simulation Parameters\label{tab:simulations}}
\tablehead{ 
  \colhead{Run} &  
  \colhead{$\Delta r_{\rm min} (\textrm{km})$ \tablenotemark{a}} &  
  \colhead{$\alpha$ \tablenotemark{b}} &
  \colhead{$\xi_{\ell}$ \tablenotemark{c}} &
  \colhead{$\xi_{\rm conv}$ \tablenotemark{d}} &
  \colhead{$\xi_{\rm conv,mix}$ \tablenotemark{e}} }
\startdata
1 & $10.0$ & $0.1$ & $0.5$           & $2.0$          & $6.0$      \\
2\tablenotemark{f} & $\mathbf{0.5} $  & $0.1$            & $0.5$           & $2.0$          & $6.0$        \\
3 & $10.0$ & $\mathbf{0.2}$   & $0.5$           & $2.0$          & $6.0$        \\
4 & $10.0$ & $\mathbf{0.025}$ & $0.5$           & $2.0$          & $6.0$     \\
5 & $10.0$ & $0.1$            & $0.5$           & $\mathbf{5.0}$ & $\mathbf{15.0}$  \\
6 & $10.0$ & $0.1$            & $0.5$           & $\mathbf{0.5}$ & $6.0$        \\
7 & $10.0$ & $0.1$            & $0.5$           & $\mathbf{1.0}$ & $6.0$       \\
8 & $10.0$ & $0.1$            & $\mathbf{0.25}$ & $2.0$          & $6.0$         \\
9 & $10.0$ & $0.1$            & $0.5$           & $2.0$          & $\mathbf{3.0}$ 
\enddata
\tablenotetext{a}{The minimum resolution element size}
\tablenotetext{b}{The dimensionless viscous stress-to-pressure ratio}
\tablenotetext{c}{Rotational profile parameter (see equation [\ref{eq:ell}])}
\tablenotetext{d}{Convective efficiency parameter (see equation [\ref{eq:F_C}])}
\tablenotetext{e}{Convective compositional mixing efficiency parameter (see equation [\ref{eq:mixing}])}
\tablenotetext{f}{This run also had additional angular resolution (see Section \ref{sec:initial_model})}
\end{deluxetable}

\begin{figure*}
\begin{center}
\includegraphics[width=0.35\textwidth,clip]{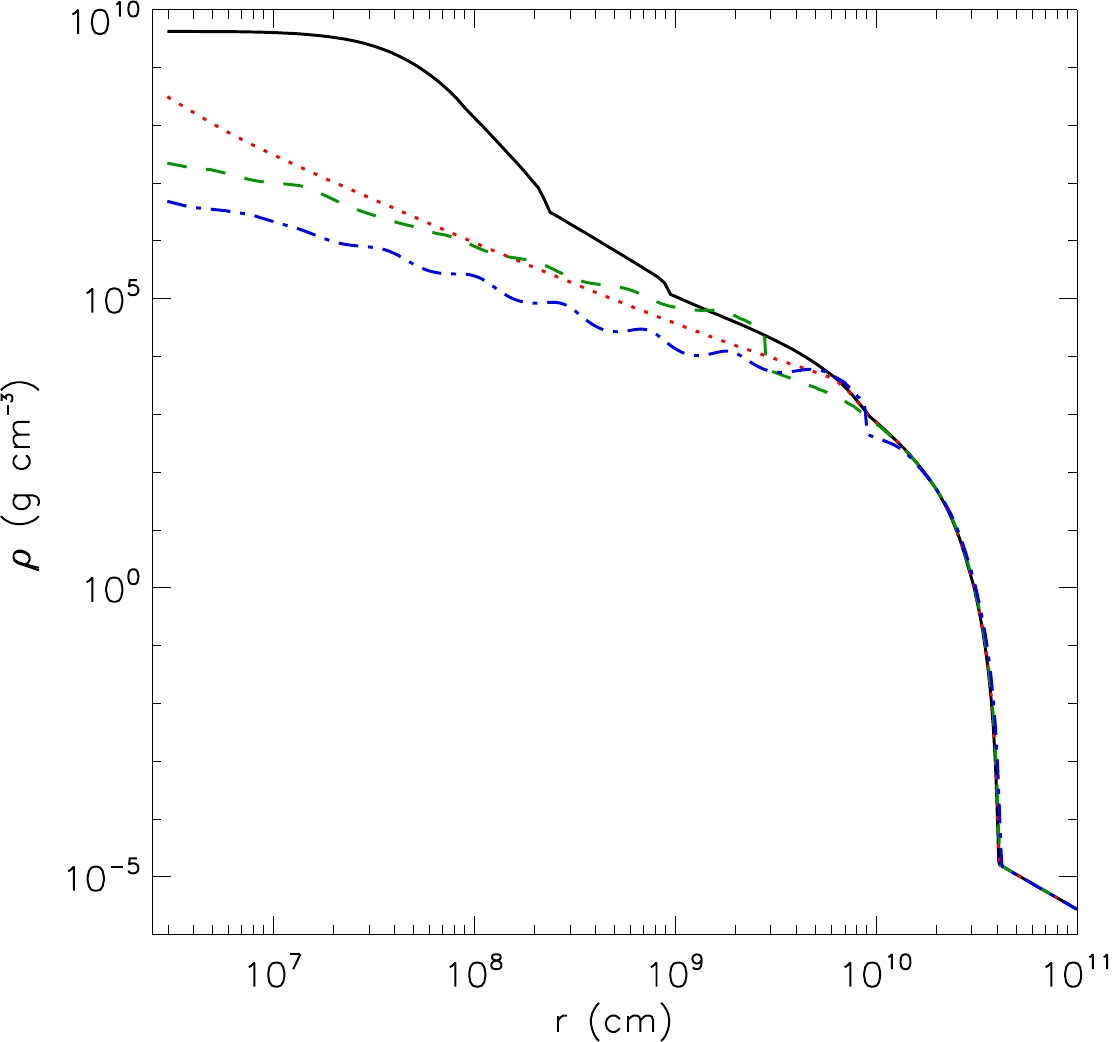}
\includegraphics[width=0.35\textwidth,clip]{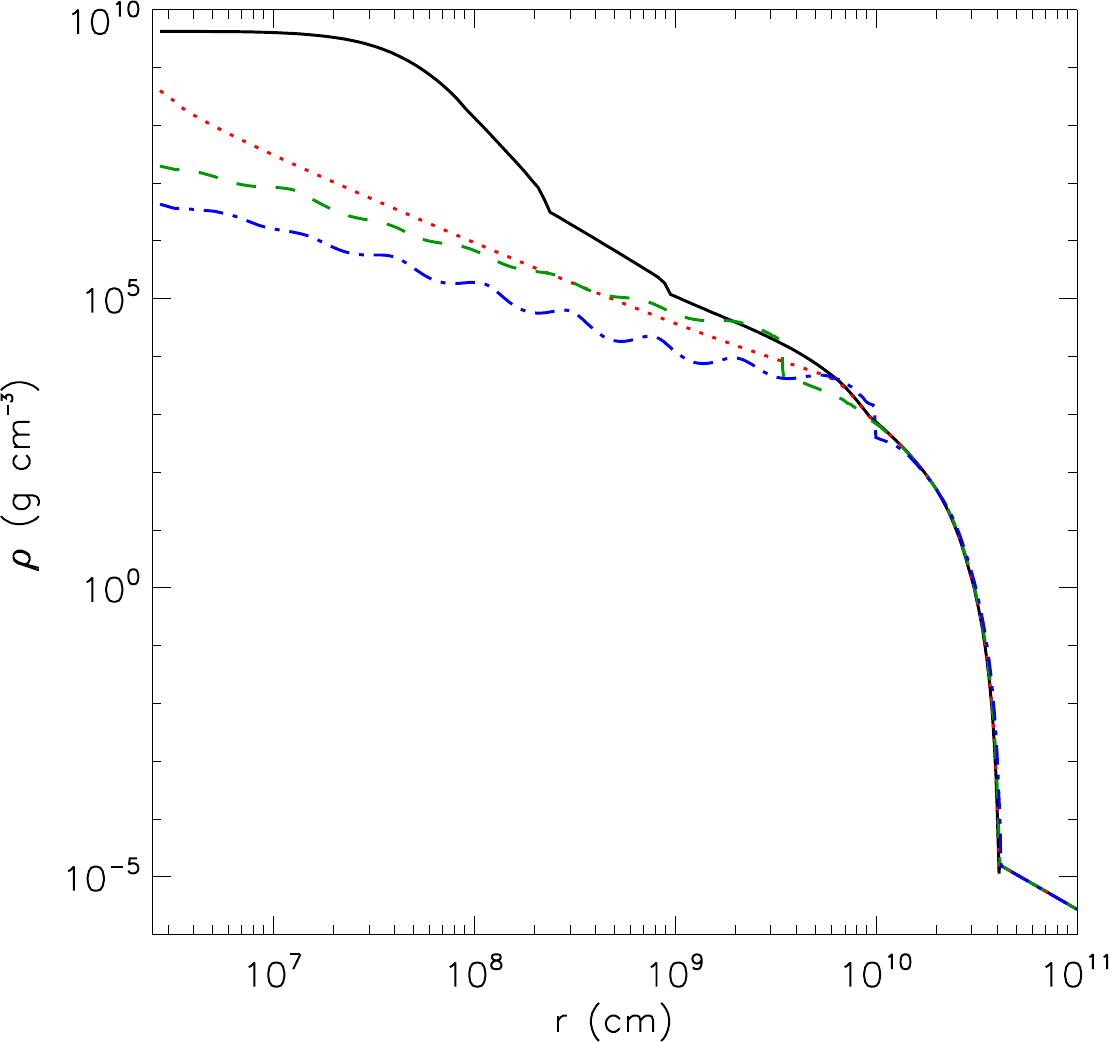}
\end{center}
\caption{Density in Run 1 (\emph{left}) and the higher resolution Run 2 (\emph{right}) at $t=0 \,\textrm{s}$ (\emph{black,solid}), $t=15 \,\textrm{s}$ (\emph{red,dotted}), $t=25 \,\textrm{s}$ (\emph{green,dashed}), and $t=50 \,\textrm{s}$ (\emph{blue,dash-dotted}).  In the convective region behind the shock front, some waves form due to an instability, developing at late times, that is likely an artifact of including mixing length theory convection as an explicit term in the transport equations.  The density jump across the shock front is approximately an order of magnitude. \label{fig:density}}
\end{figure*}

\begin{figure*}
\begin{center}
\includegraphics[width=0.35\textwidth,clip]{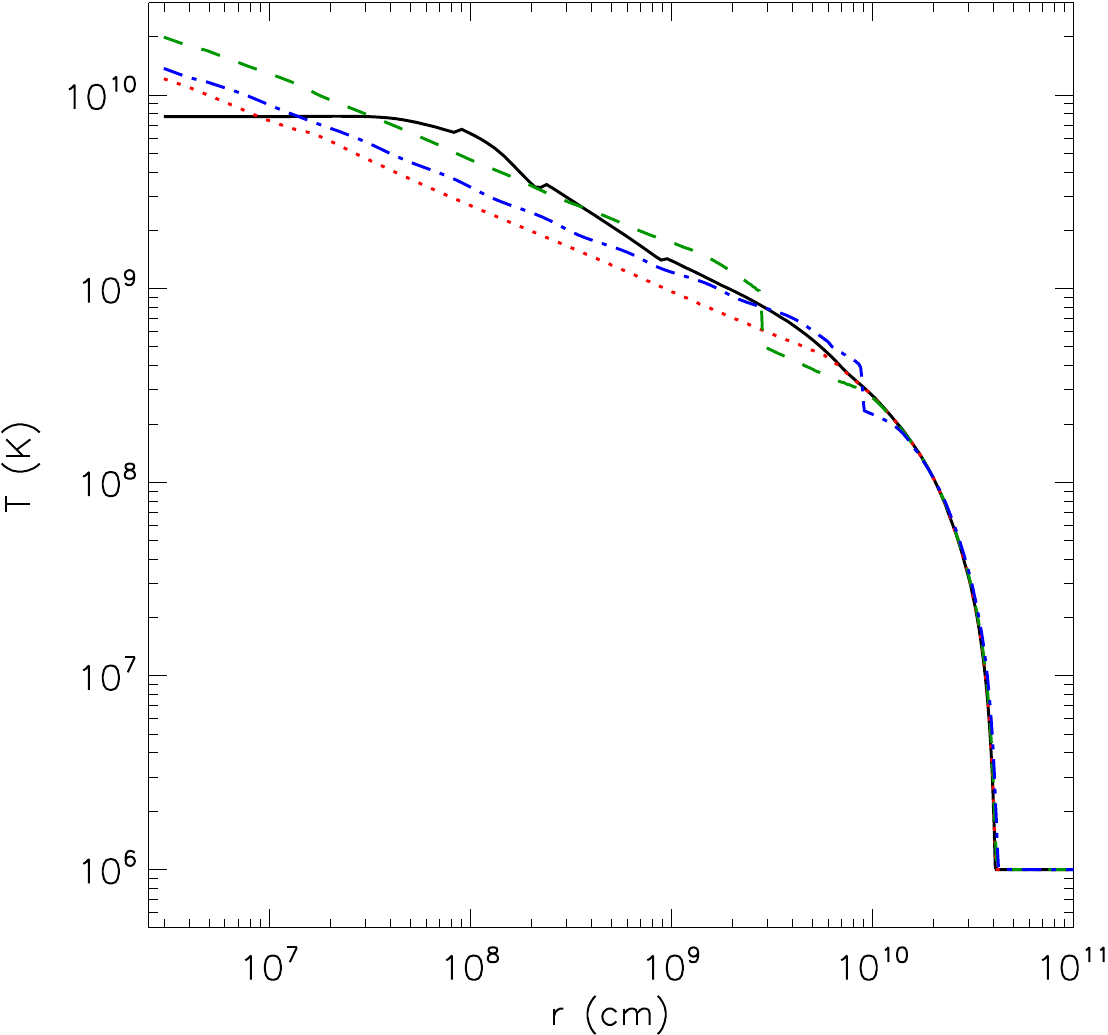}
\includegraphics[width=0.35\textwidth,clip]{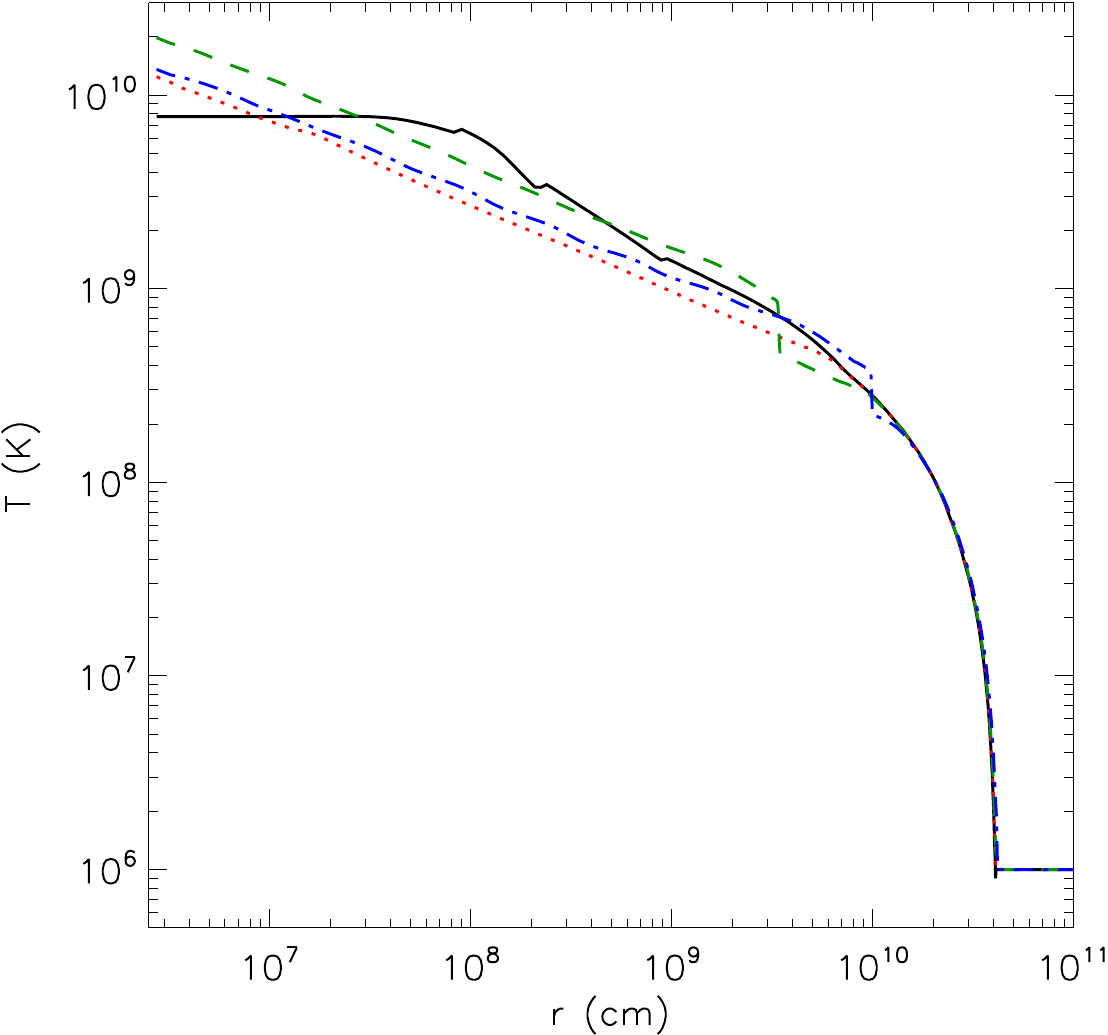}
\end{center}
\caption{Temperature in Run 1 (\emph{left}) and the higher resolution Run 2 (\emph{right}) at $t=0 \,\textrm{s}$ (\emph{black,solid}), $t=15 \,\textrm{s}$ (\emph{red,dotted}), $t=25 \,\textrm{s}$ (\emph{green,dashed}), and $t=50 \,\textrm{s}$ (\emph{blue,dash-dotted}).  Photodisintegration and neutrino emission cool the innermost disk, while nuclear fusion provides additional heating in the post-shock region (see Section \ref{sec:nucleosynthesis}). \label{fig:temperature}}
\end{figure*}

\begin{figure*}
\begin{center}
\includegraphics[width=0.35\textwidth,clip]{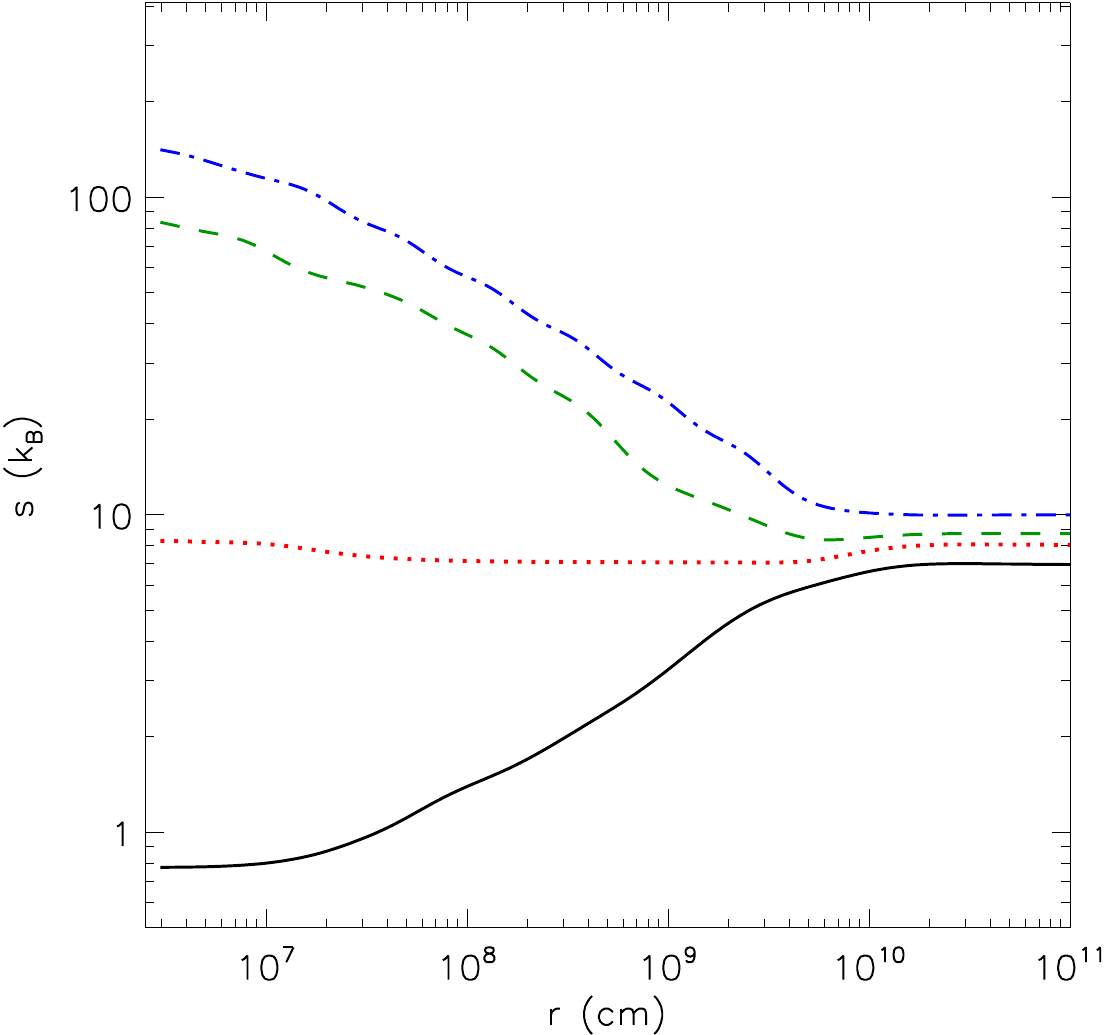}
\includegraphics[width=0.35\textwidth,clip]{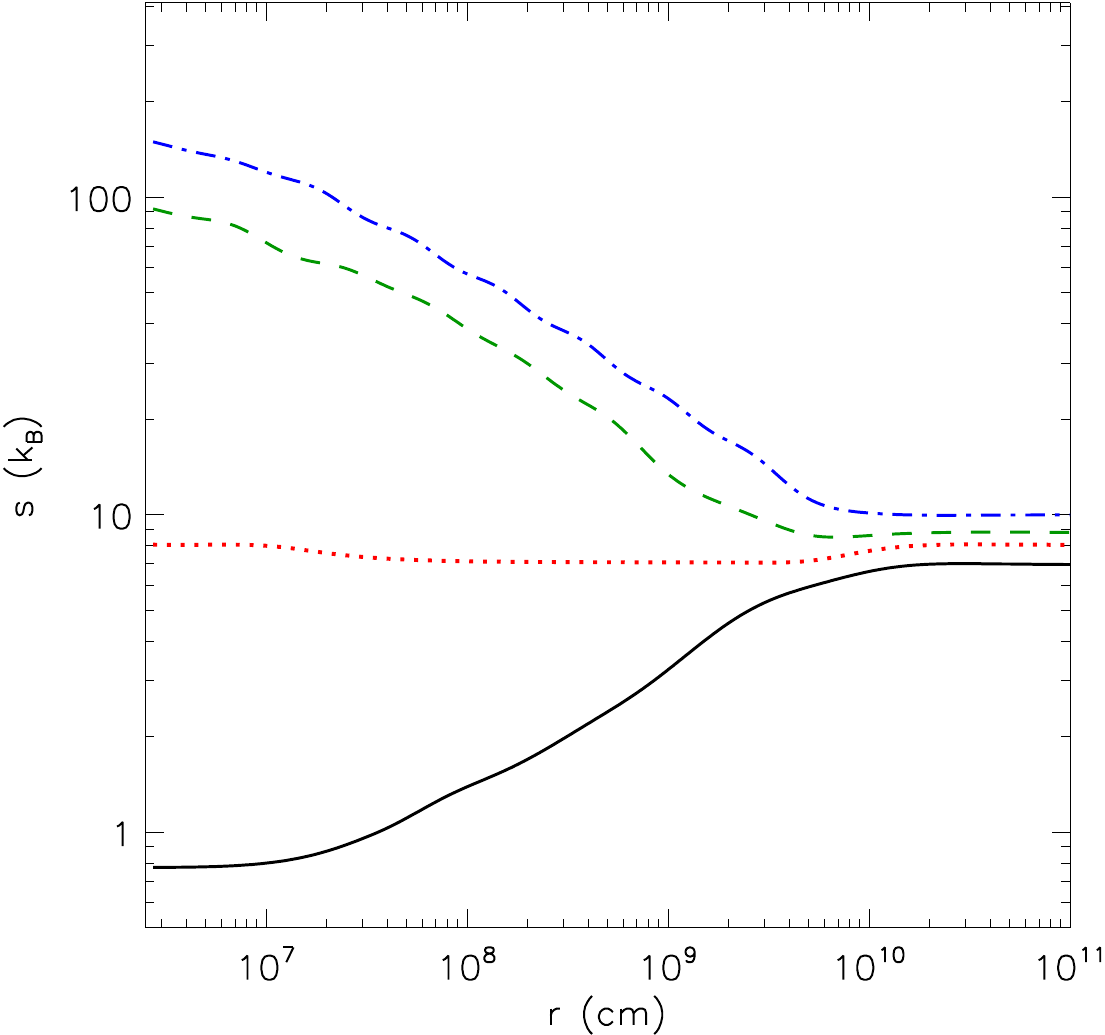}
\end{center}
\caption{The smoothed entropy $s_{\rm smooth}$ per baryon in Run 1 (\emph{left}) and the higher resolution Run 2 (\emph{right}) at $t=0 \,\textrm{s}$ (\emph{black, solid}), $t=15 \,\textrm{s}$ (\emph{red, dotted}), $t=25 \,\textrm{s}$ (\emph{green, dashed}), and $t=50 \,\textrm{s}$ (\emph{blue, dash-dotted}) in units of the Boltzmann constant ($k_{\rm B}$).  After fluid comes into radial force balance, a strong entropy inversion is observed, giving rise to convection.\label{fig:entropy}}
\end{figure*}

\begin{figure*}
\begin{center}
\includegraphics[width=0.35\textwidth,clip]{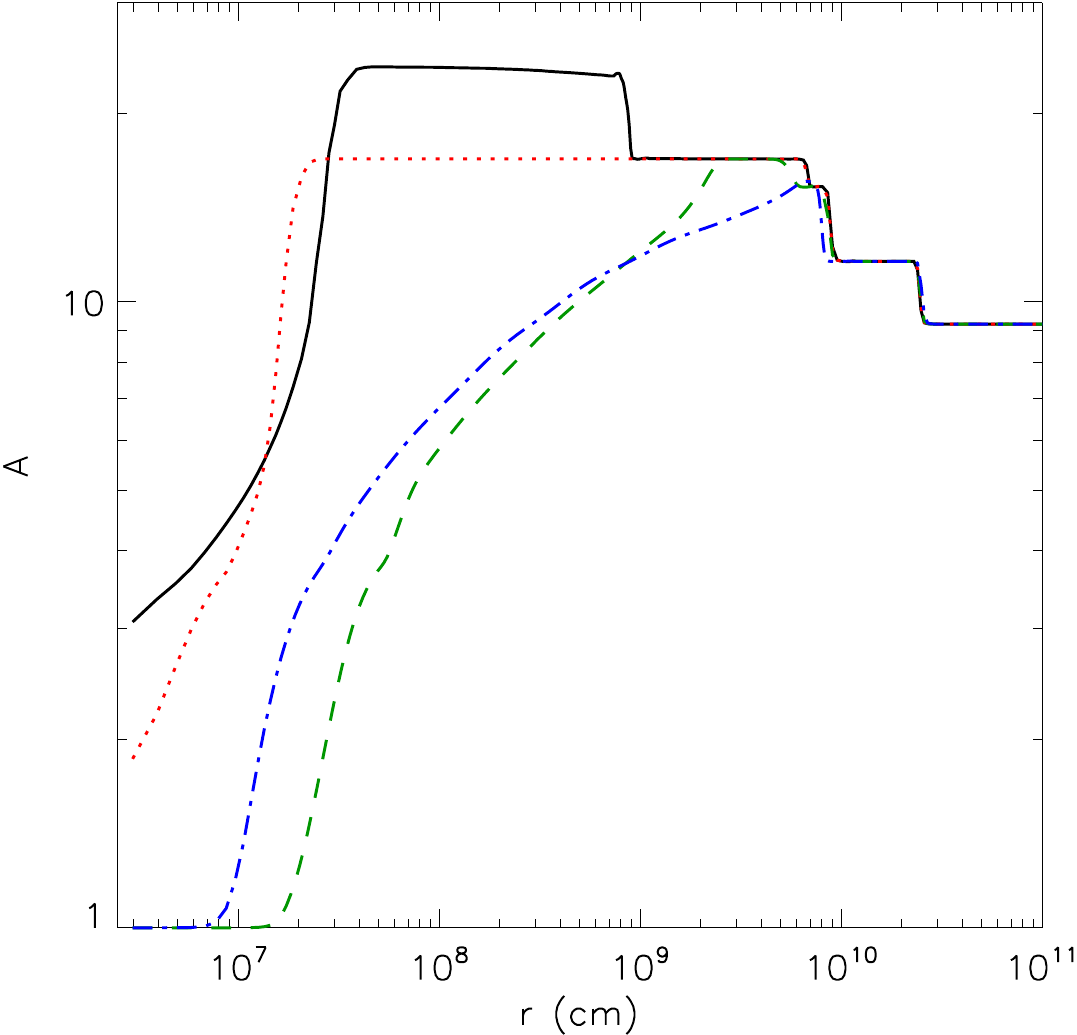}
\includegraphics[width=0.35\textwidth,clip]{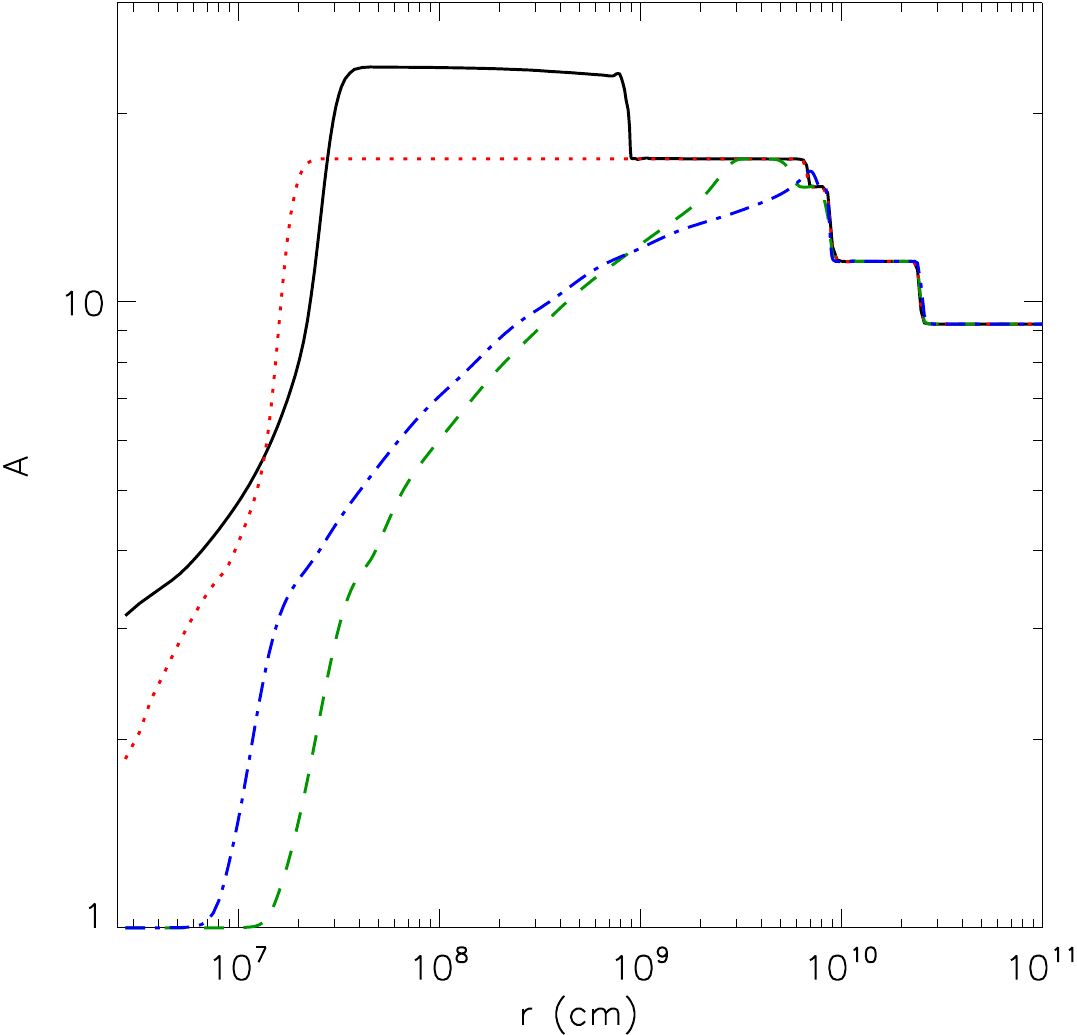}
\end{center}
\caption{Mass-weighted average of the atomic mass $\bar{A}$ in Run 1 (\emph{left}) and Run 2 (\emph{right}) at $t=1\,\textrm{s}$ (\emph{black,solid}), $t=15\,\textrm{s}$ (\emph{red,dotted}), $t=25\,\textrm{s}$ (\emph{green,dashed}), and $t=50\,\textrm{s}$ (\emph{blue,dot-dashed}).  Note that at $t=1\,\textrm{s}$, the iron core has already accreted onto the central point mass.  At late times, photodisintegration in the hottest inner regions behind the shock front reduces the value of $\bar{A}$.  Convective mixing is able to dredge up lighter elements; our scheme for nuclear compositional transformation does not correctly model the subsequent recombination and freezeout well outside NSE. \label{fig:abar}}
\end{figure*}

Run 1 and Run 2 contained identical hydrodynamic parameters and differed only in spatial resolution.  Run 2 was capable of one additional level of resolution refinement over Run 1 and the parameter $\eta$ described in Section \ref{sec:initial_model} was set to one-half the value in Run 1, allowing for significantly higher resolution as a function of radius.  Figures \ref{fig:density}, \ref{fig:temperature}, \ref{fig:entropy}, and \ref{fig:abar} show the density, temperature, specific entropy, and the mean atomic weight in Run 1, and also in the higher-resolution Run 2 at different times.  Substantial agreement is seen between the low and high resolution simulations.

\subsection{Limitations of the Method}
\label{sec:limitations}

The primary limitations of the model of collapsar accretion that we have presented here include: (1) a very approximate one-dimensional treatment of the intrinsically two and three dimensional flow structure; (2) limited adequacy of the Navier-Stokes viscous stress as a model for the magnetic stress arising in the nonlinear development of the magnetorotational instability; (3) no a priori knowledge of the expected efficiency of convective energy transport $\xi_{\rm conv}$ and of compositional transport $\xi_{\rm conv,mix}$; (4) treatment of nuclear compositional transformation through relaxation to NSE rather than by integrating the nuclear reaction network that would have allowed us to make predictions about the nucleosynthetic output; (5) neglect of ambient compositional stratification and nuclear compositional transformation inside convective cells in the calculation of the convective heat flux; (6) the lack of modeling of the axial relativistic jet and its enveloping cocoon that are thought to be present in LRGB sources; and (7) the use of a relatively low mass progenitor star, which may or may not be able to yield a black hole and an explosion with an energy as high as has been inferred in supernovae associated with LGRBs.  Overcoming limitations (1) through (6) will require much more computationally expensive multidimensional hydrodynamic and magnetohydrodynamic simulations.  Limitation (7) can be addressed by applying our current method to other, more massive stellar models; here, we speculate what collapsars in higher mass progenitors may behave like in Section \ref{sec:discussion} below.

It would be tempting in view of limitation (5) to try to incorporate the effects of compositional stratification ambient to convective cells in the convective energy flux, which is normally achieved by multiplying the energy flux in equation (\ref{eq:F_C}) with an factor
\beq
\left[1-\left(\frac{dT}{d\mu}\right)_{P,\rho}\frac{d\mu}{dr} \left/ \left(\frac{T}{c_p} \frac{ds}{dr}\right) \right.\right]^{1/2} ,
\eeq
where $\mu$ is the mean nuclear mass, and utilizing the Ledoux instead of the Schwarzschild criterion \citep[see, e.g.,][]{Bisnovatyi-Kogan:01}.  This would be meaningful as long as the convective eddy turnover time $\tau_{\rm conv}\sim \lambda_{\rm conv}/v_{\rm conv}$ were shorter than the nuclear time scale $\tau_{\rm nuc}\lesssim \tau_{\rm NSE}$, so that the convective cells can be treated as adiabatic before they mix.  However at radii where Ledoux convection would differ most from Schwarzschild convection, namely where the photodissociation into helium nuclei and free nucleons is substantial, the convective time scale is much longer than the nuclear time scale, $\tau_{\rm conv}\gg \tau_{\rm nuc}$.  The internal composition of a convective cell evolves as it rises, and the associated entropy change is a much stronger effect than the variation of the ambient composition treated in Ledoux convection. Magnetization of the medium may play a role in this regime but its effects are poorly understood. Cognizant of the ongoing research on the interplay of convection and nuclear burning \citep[see, e.g.,][]{Arnett:11}, we adopt the Schwarzschild model and consider it but a parametrization of a complex, still to be explored physics.

\section{Results}
\label{sec:results}

Nine simulations were carried out to explore sensitivity to the resolution of the simulation $\Delta r_{\rm min}$, the viscous stress-to-pressure ratio $\alpha$, the stellar rotation $\xi_\ell$, the efficiency of convective energy transport $\xi_{\rm conv}$, and the efficiency of convective mixing $\xi_{\rm conv,mix}$.  The values of these parameters in each of the simulations are summarized in Table \ref{tab:simulations}.  Among these, Run 1 can be considered the fiducial model.  Each simulation was run for $100 \,\textrm{s}$, except for Runs 4 and 5, where strong numerical instabilities associated with our convection scheme prevented us from simulating for more than $40\,\textrm{s}$ and $50\,\textrm{s}$, respectively.  In what follows, we present the results.  In Section \ref{sec:accretion}, we address the evolution of the rate with which mass accretes onto the central black hole.  In Section \ref{sec:envelope}, we discuss the nature of radial force balance in the fraction of the stellar material that has been traversed by the outgoing shock wave, but has not accreted onto the black hole and also discuss the mass and angular momentum transport in the system. In Section \ref{sec:energy_transport}, we address energy transport.  In Section \ref{sec:nucleosynthesis}, we address the nuclear composition of the flow and discuss the limitations inherent in our simplified treatment of nuclear compositional transformation.  In Section \ref{sec:explosion}, we discuss the global energetics and check whether sufficient energy may be transported into a portion of the stellar envelope to produce a supernova. 

\subsection{Central Accretion Rate and Black Hole Mass}
\label{sec:accretion}

\begin{figure*}
\begin{center}
\includegraphics[width=0.28\textwidth,clip]{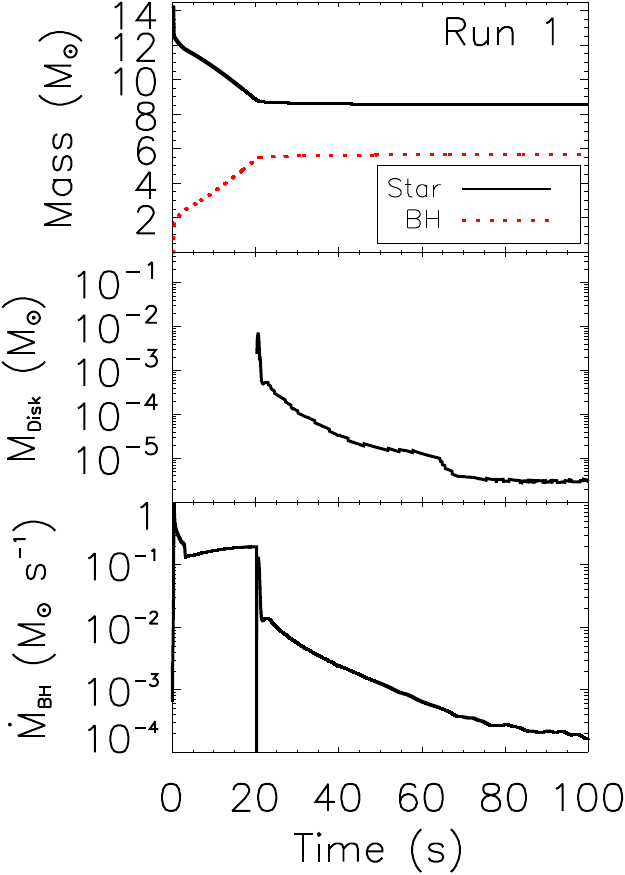}
\includegraphics[width=0.28\textwidth,clip]{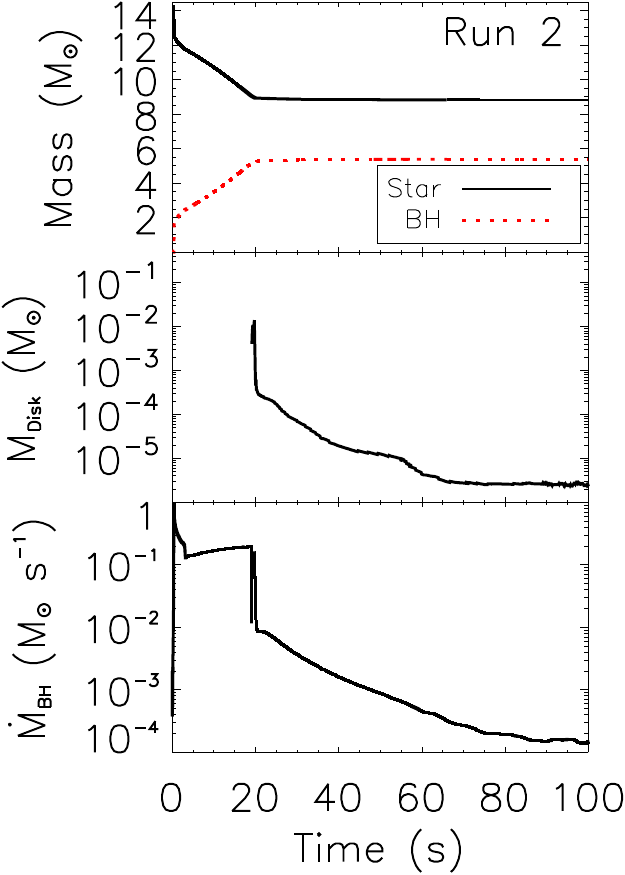}
\includegraphics[width=0.28\textwidth,clip]{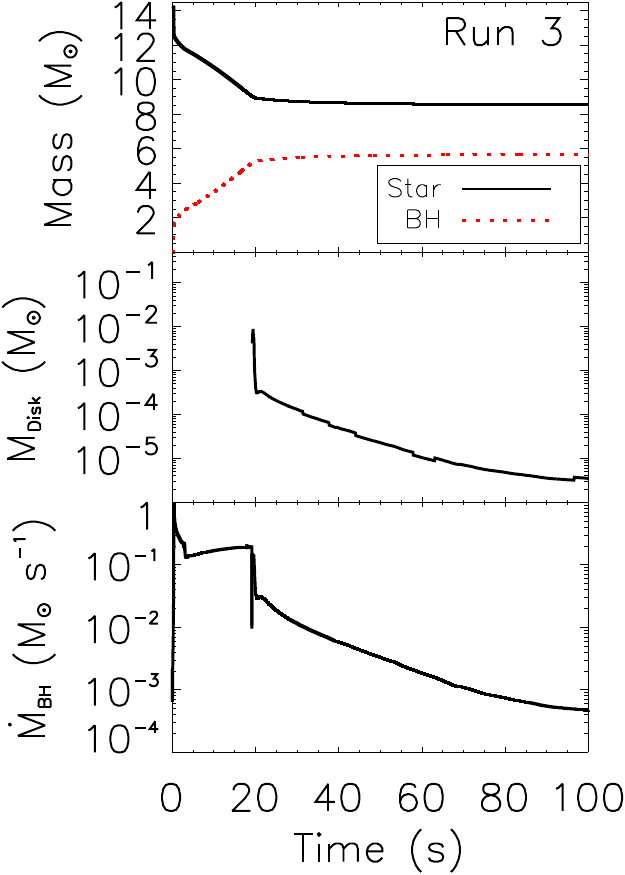} \\
\includegraphics[width=0.28\textwidth,clip]{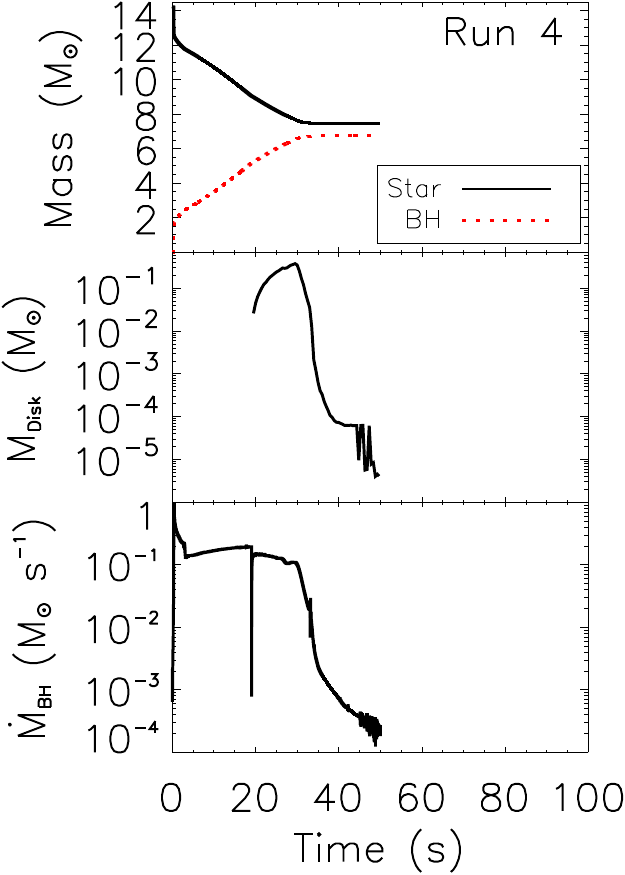}
\includegraphics[width=0.28\textwidth,clip]{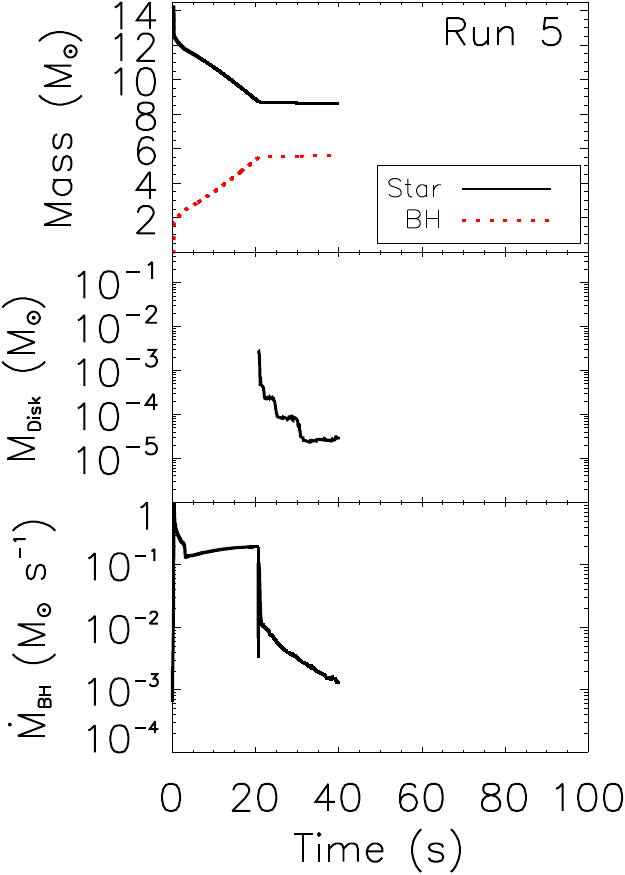}
\includegraphics[width=0.28\textwidth,clip]{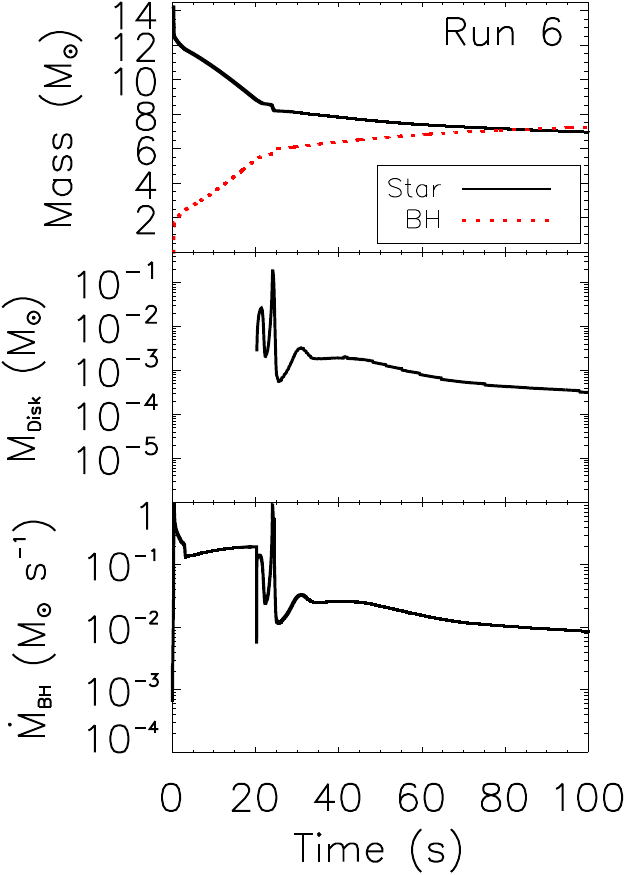} \\
\includegraphics[width=0.28\textwidth,clip]{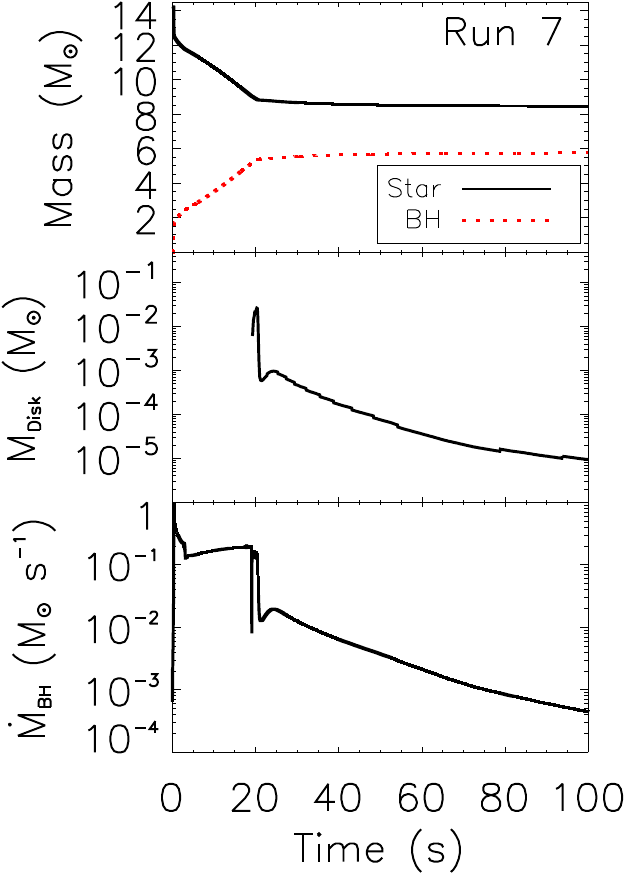}
\includegraphics[width=0.28\textwidth,clip]{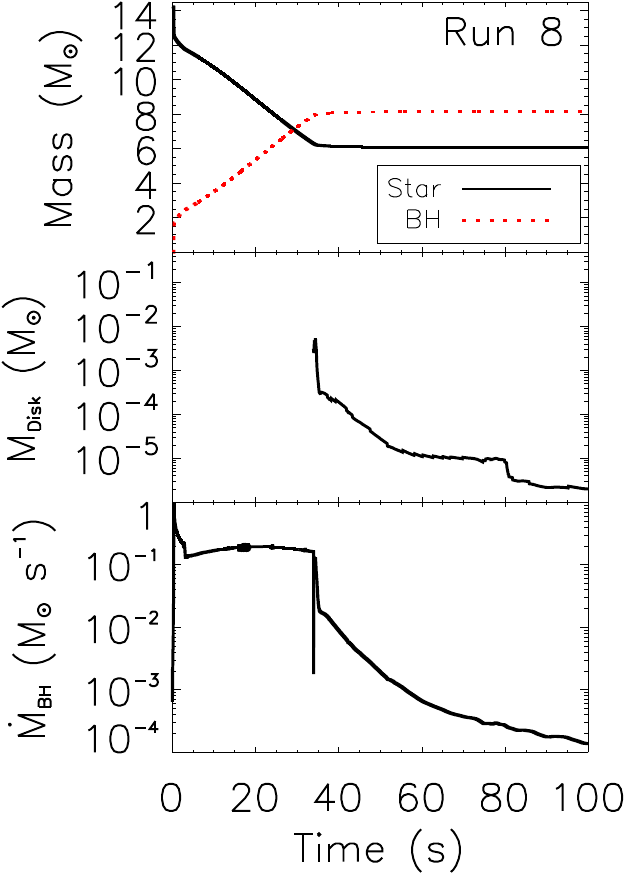}
\includegraphics[width=0.28\textwidth,clip]{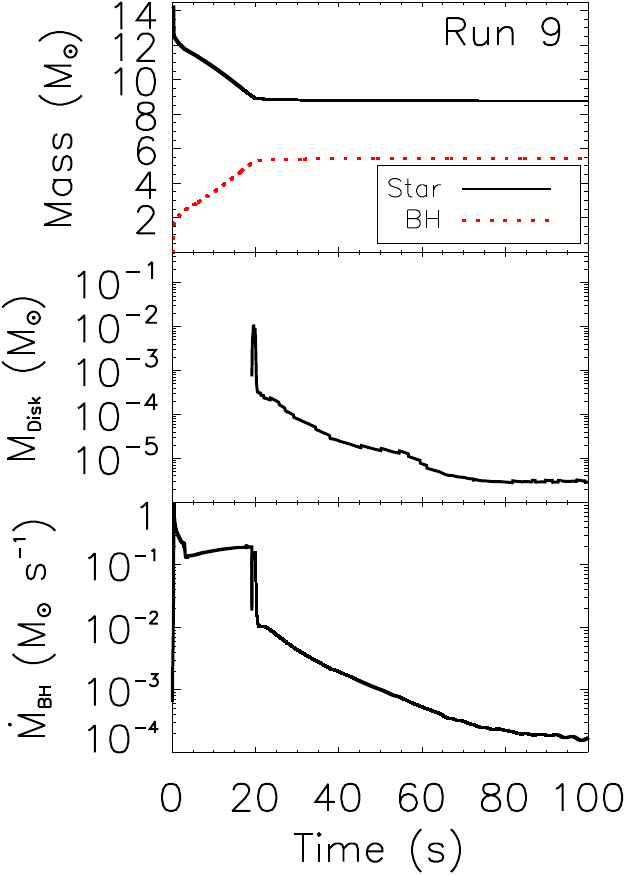}
\end{center}
\caption{Mass of the stellar envelope (\emph{top, black, solid}), mass of the central object (\emph{top, red, dashed}), mass of the disk (\emph{middle}) as defined in Section \ref{sec:accretion}, and mass accretion rate through the inner boundary (\emph{bottom}) in each of the runs.  Most of the mass is accreted onto the central object in the first $\sim 20 \textrm{s}$ in most runs.  The disk mass makes up only a small portion of the total remaining, while the rest of the mass exists in a pressure supported atmosphere that may continue to feed the disk, or may be potentially unbound by the accretion shock.  Plots of the disk mass begin when $t=t_{\rm shock}$.  The quick drop in accretion rate seen in most of the simulations occurs around the time of shock formation. \label{fig:fourplot}}
\end{figure*}

Each simulation exhibits unshocked radial accretion of the inner, low-angular momentum mass shells of the progenitor star through the inner boundary lasting $\sim (20-30) \,\textrm{s}$ at relatively steady accretion rates of $\sim (0.1 - 0.2) \,M_\odot \,\textrm{s}^{-1}$. The central mass accretion rate, black hole mass, and total mass on the computational grid as a function of time in each simulation are shown in Figure \ref{fig:fourplot}.  The abrupt drop of the central accretion rate at $\sim(20-30)\,\textrm{s}$ is associated with the appearance of an accretion shock precipitated by the arrival of the mass shells with specific angular momentum sufficient to lead to circularization around the black hole.

\begin{figure*}
\begin{center}
\includegraphics[width=0.30\textwidth,clip]{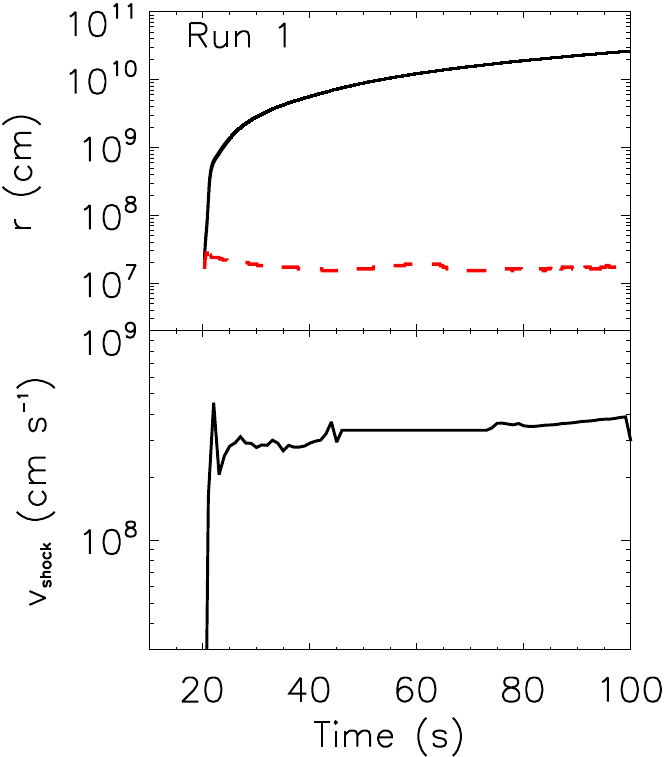}
\includegraphics[width=0.30\textwidth,clip]{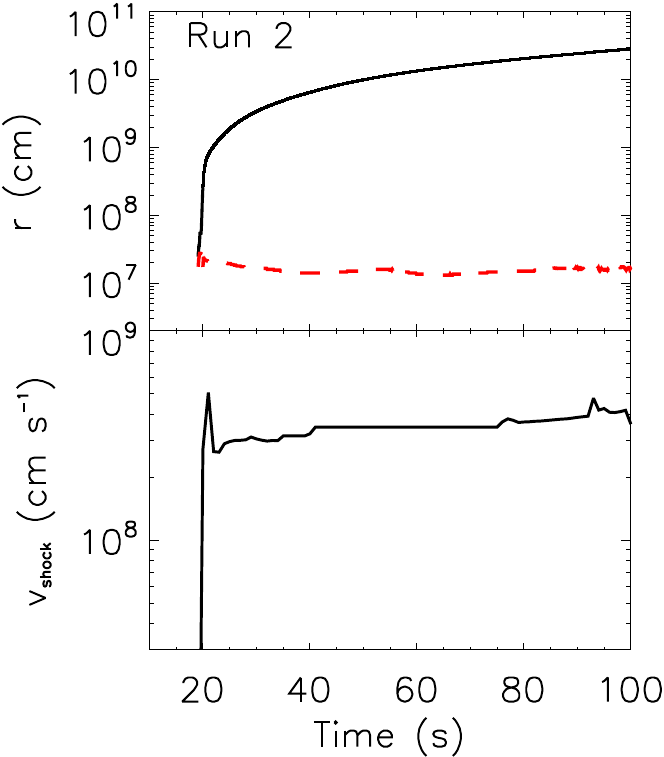}
\includegraphics[width=0.30\textwidth,clip]{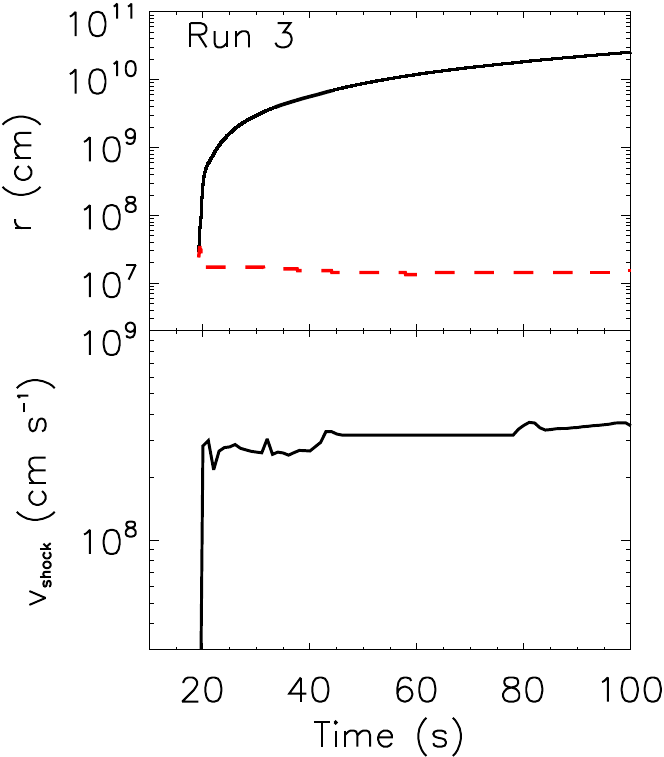} \\
\includegraphics[width=0.30\textwidth,clip]{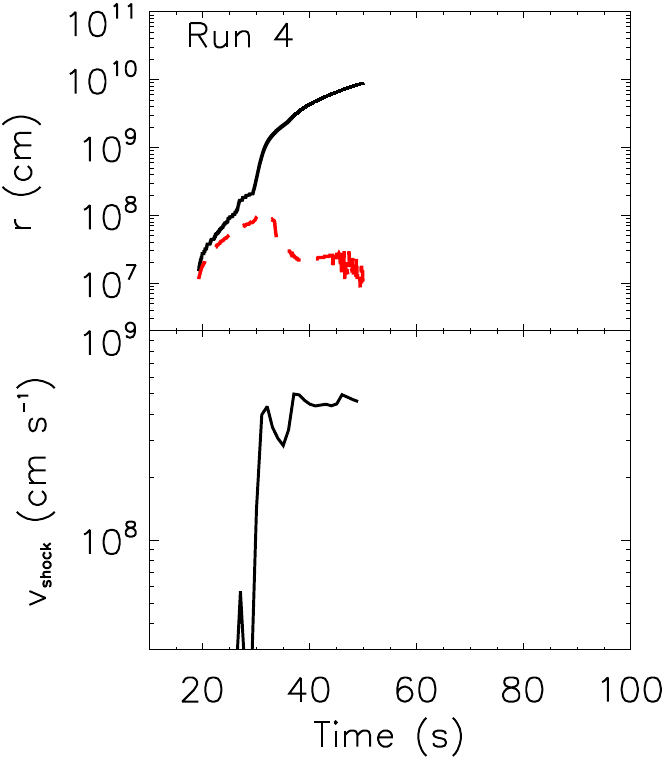}
\includegraphics[width=0.30\textwidth,clip]{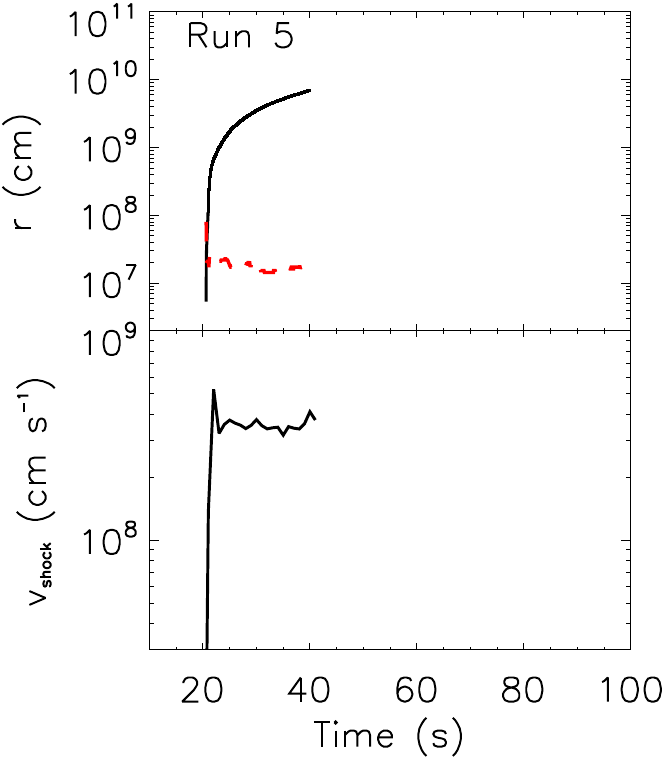} 
\includegraphics[width=0.30\textwidth,clip]{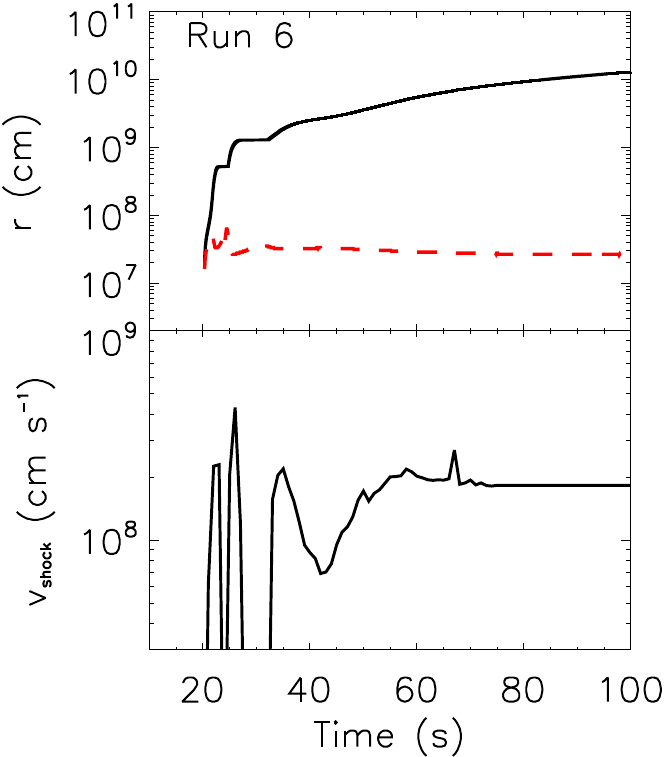} \\
\includegraphics[width=0.30\textwidth,clip]{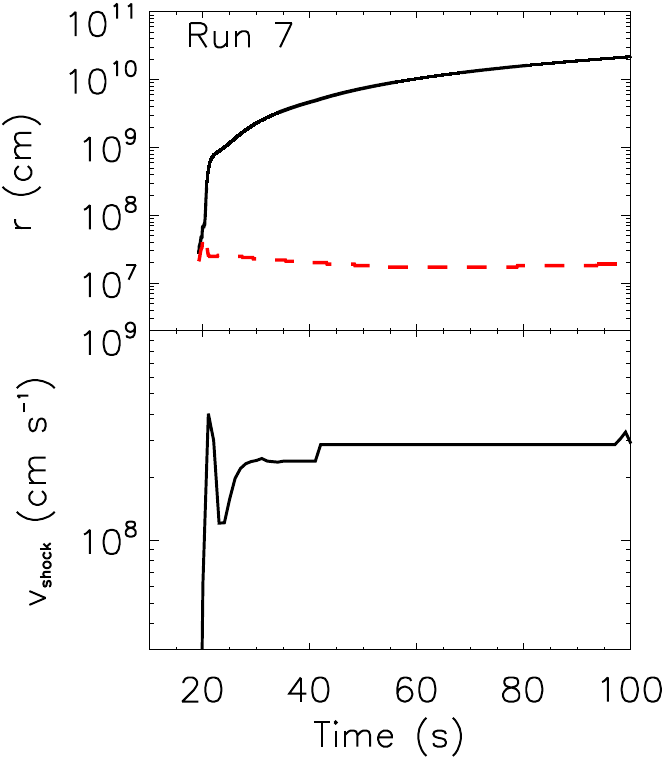}
\includegraphics[width=0.30\textwidth,clip]{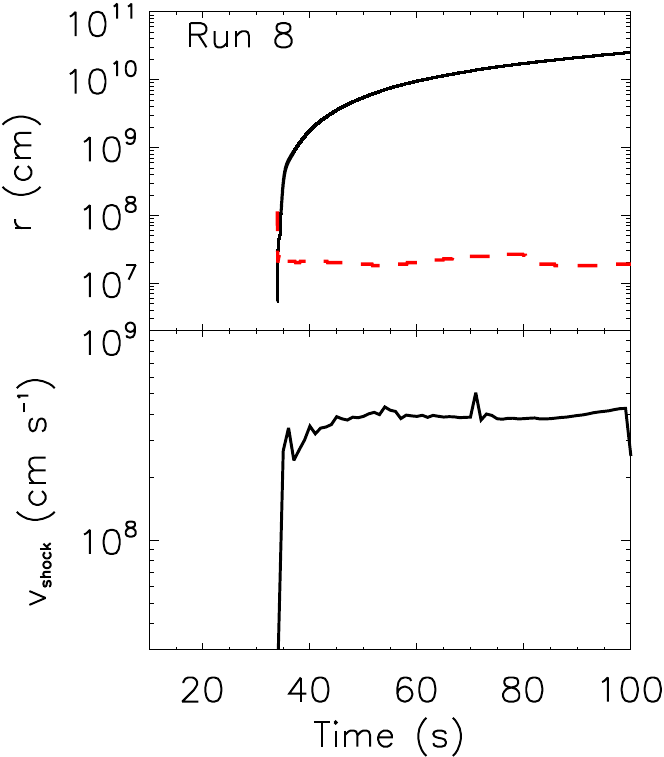}
\includegraphics[width=0.30\textwidth,clip]{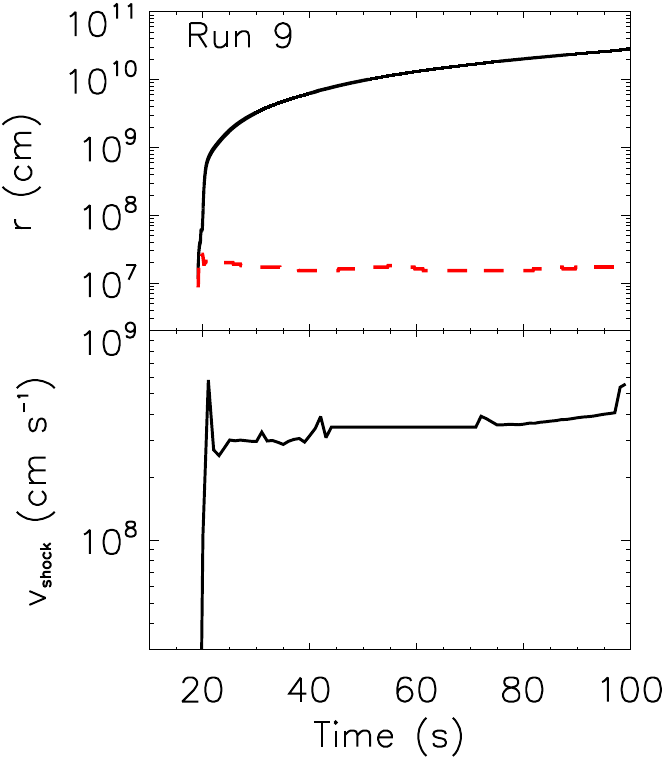}
\end{center}
\caption{Shock location (\emph{top}) and velocity (\emph{bottom}) in each of the runs.  The red dashed line shows $r_{\rm disk}$, the outermost radius where the acceleration due to the centrifugal force is at least $50\%$ the acceleration due to the pressure gradient.  In Run 4 and Run 6, the shock stalls and is reinvigorated.  Shock velocities were typically $2000-4000 \,\textrm{km}\,\textrm{s}^{-1}$. The small fluctuations in the shock velocity are numerical artifacts of the discreteness in our shock detection algorithm.
\label{fig:twoplot}}
\end{figure*}

In Figure \ref{fig:twoplot} we show the location of the shock $r_{\rm shock}$ and its velocity $v_{\rm shock}\equiv dr_{\rm shock}/dt$ as a function of time.  For each of the runs, we identify the time when the shock first reaches radius $10\,r_{\rm min}=250\,\textrm{km}$ as the shock formation time $t_{\rm shock}$ and list the shock formation times in column 2 of Table \ref{tab:results}.  We also provide the mass of the black hole at this point, $M_{\rm BH}(t_{\rm shock})$, in column 8.  The black hole mass at the time of shock formation was $M_{\rm BH}(t_{\rm shock})\sim (5.2-5.5)\,M_\odot$ in Runs $1-7$ and 9.  In Run 8, which was initiated with reduced initial angular momentum, the accretion shock appeared later and the black hole mass is correspondingly larger.

After the formation of the shock, the fluid nearest the inner boundary is rotationally supported and accretes as a result of angular momentum transport driven by the viscous shear stress.  Subsequent to shock formation, the accretion rate declines rapidly either promptly or following a short delay.   The typical rapid drop of the accretion rate is by a factor $\sim 5-10$ (Runs 1, 2, 3, 5, 7, and 9), and this is followed by a continued power-law-like decline.  By the end of each simulation at $100\,\textrm{s}$, the accretion rate has typically declined to $\sim (10^{-3}-10^{-4})\,M_\odot\,\textrm{s}^{-1}$ (Runs 1, 2, 3, 4, 7, 8, and 9), or a factor of $100-1000$ of the pre-shock value.  Final black hole masses were $\sim(6 - 7) \,M_\odot$ in the simulations with $\xi_{\ell} = 0.5$, and $\sim10\,M_\odot$ in Run 8 with reduced initial angular momentum $\xi_\ell = 0.25$.

In Run 4, with a low value of the viscosity parameter $\alpha=0.025$, the shock first made a very slow progress from $300\,\textrm{km}$ to $2000\,\textrm{km}$ during the first $10\,\textrm{s}$ from its appearance.  Then, at $30\,\textrm{s}$, the shock suddenly accelerated to $v_{\rm shock} \sim 5000\,\textrm{km s}^{-1}$.  The near-stagnation of the shock can be understood by noticing that during the $10\,\textrm{s}$, the neutrino cooling rate matches the viscous heating rate; the rapid cooling prevents the central entropy rise and convection seen in all other runs (see Section \ref{sec:energy_transport} below).  In Section 4.2 of \cite{Lindner:10}, we discussed the scenario in which the shock stagnation brought about by efficient neutrino cooling prolongs the LGRB central engine activity resulting in a longer prompt emission.  

In Run 6, which had convective efficiency $\xi_{\rm conv}=0.5$, the shock stalled at the radius $\sim 10^4\,\textrm{km}$ for $\sim 5\,\textrm{s}$ before proceeding outward.  Note that the reinvigoration of the shock is solely driven by the convective energy transport, as we do not simulate the negligible neutrino energy and momentum deposition. The stalling and restarting of the shock was reflected in a strong variability of the central accretion rate.

\begin{deluxetable}{ccccccc}
\label{tab:simulations}
\tablecolumns{7}
\tablewidth{3.2in}
\tablecaption{Summary of Key Measurements\label{tab:results}}
\tablehead{ 
  \colhead{Run } &  
  \colhead{$t_{\rm shock}$ \tablenotemark{a}} &
  \colhead{$M_{\rm BH}(t_{\rm shock})$ \tablenotemark{b}} &
  \colhead{$M_{\rm unbound}$ \tablenotemark{c}} &
  \colhead{$E_{\rm bind}$ \tablenotemark{d}} &
  \colhead{$E_{\rm kin}$ \tablenotemark{e}} &
  \colhead{$M_{\rm Fe}$ \tablenotemark{f}} }  
\startdata
1      &  $ 20.3 $   & $ 5.4 $    & $6.0$      & $~0.40$   & $0.31$ & $0.06$\\
2\tablenotemark{g} &  $ 19.1 $   & $ 5.2 $    & $6.4$      & $~0.44$   & $0.36$ & $0.04$\\
3 & $ 19.2 $   & $ 5.2 $    & $5.7$      & $~0.34$   & $0.28$ & $0.06$ \\
4 &  $ 19.8 $   & $ 5.3 $    & $4.4$      & $~0.62$   & $0.29$ & $0.07$\\
5 &$ 20.6 $   & $ 5.5 $    & $3.1$      & $~0.40$   & $0.18$ & $0.04$\\
6 & $ 20.3 $   & $ 5.4 $    & $0.0$      & $-0.43$   & $0.16$ & $0.03$\\
7 & $ 19.2 $   & $ 5.2 $    & $4.4$      & $~0.08$   & $0.17$ & $0.09$\\
8 & $ 34.0 $   & $ 7.9 $    & $5.9$      & $~0.54$   & $0.29$ & $0.02$\\
9 &  $ 19.2 $   & $ 5.2 $    & $6.8$      & $~0.46$   & $0.37$ & $0.03$
\enddata
\tablenotetext{a}{Time at which shock reaches $r=250\,\textrm{km}$ ($\textrm{s}$)}
\tablenotetext{b}{Black hole mass at when the shock reaches $r=250 \,\textrm{km}s$ ($M_\odot$)}
\tablenotetext{c}{Unbound mass at the end of the simulation ($M_\odot$) }
\tablenotetext{d}{Total energy in the stellar material at the end of the simulation ($10^{51}\,\textrm{ergs}\,\textrm{s}^{-1}$; see Section \ref{sec:explosion} and Figure \ref{fig:energy})}
\tablenotetext{e}{Total kinetic energy of outbound material ($10^{51}\,\textrm{ergs}\,\textrm{s}^{-1}$; see Section \ref{sec:explosion})}
\tablenotetext{f}{Total mass of newly synthesized Fe-group elements at the end of the simulation ($M_\odot$; see Section \ref{sec:nucleosynthesis})} 
\tablenotetext{g}{This run also had additional angular resolution (see Section \ref{sec:initial_model})}
\end{deluxetable}

\subsection{The Shocked Envelope and Angular Momentum}
\label{sec:envelope}

Shock passage leaves a shock- and convection-heated, pressure supported envelope which contains much more mass than the disk, consistent with what we saw in \cite{Lindner:10}.  Figure \ref{fig:density} shows that the density in the envelope is an approximate power-law of radius $\rho\propto r^{-0.9}$.  Figure \ref{fig:temperature} indicates that the temperature is also a power-law $T\propto r^{-0.4}$.  The pressure (not shown) is an approximate power law $P\propto r^{-1.8}$. The profiles extend inward into the regime in which rotational support dominates pressure support.  The mass of the rotationally supported material in the grid, where $a_{\rm cent} > \frac{1}{2} |g_{\rm self} +g_{\rm BH}|$, promptly following disk formation was typically $\lesssim 5\%$ of the total mass on the grid.  Most of the mass on the grid was in the pressure supported atmosphere seamlessly connecting to the disk. The mass of the disk in each simulation is shown in Figure \ref{fig:fourplot}.    In some of the runs, certain variability is seen in the disk mass over the first few seconds of disk formation.  Afterwards, the disk mass in each simulation declines monotonically. In most runs ($1-4$, $7-9$) the disk mass declines to $M_{\rm disk}\lesssim 10^{-5}\,M_\odot$ by the end of the simulation, while in Run 6, the mass at the end of the simulation is somewhat larger but still very small, $M_{\rm disk}\sim 3\times10^{-4}\,M_\odot$.

\begin{figure}
\begin{center}
\includegraphics[width=.42\textwidth,clip]{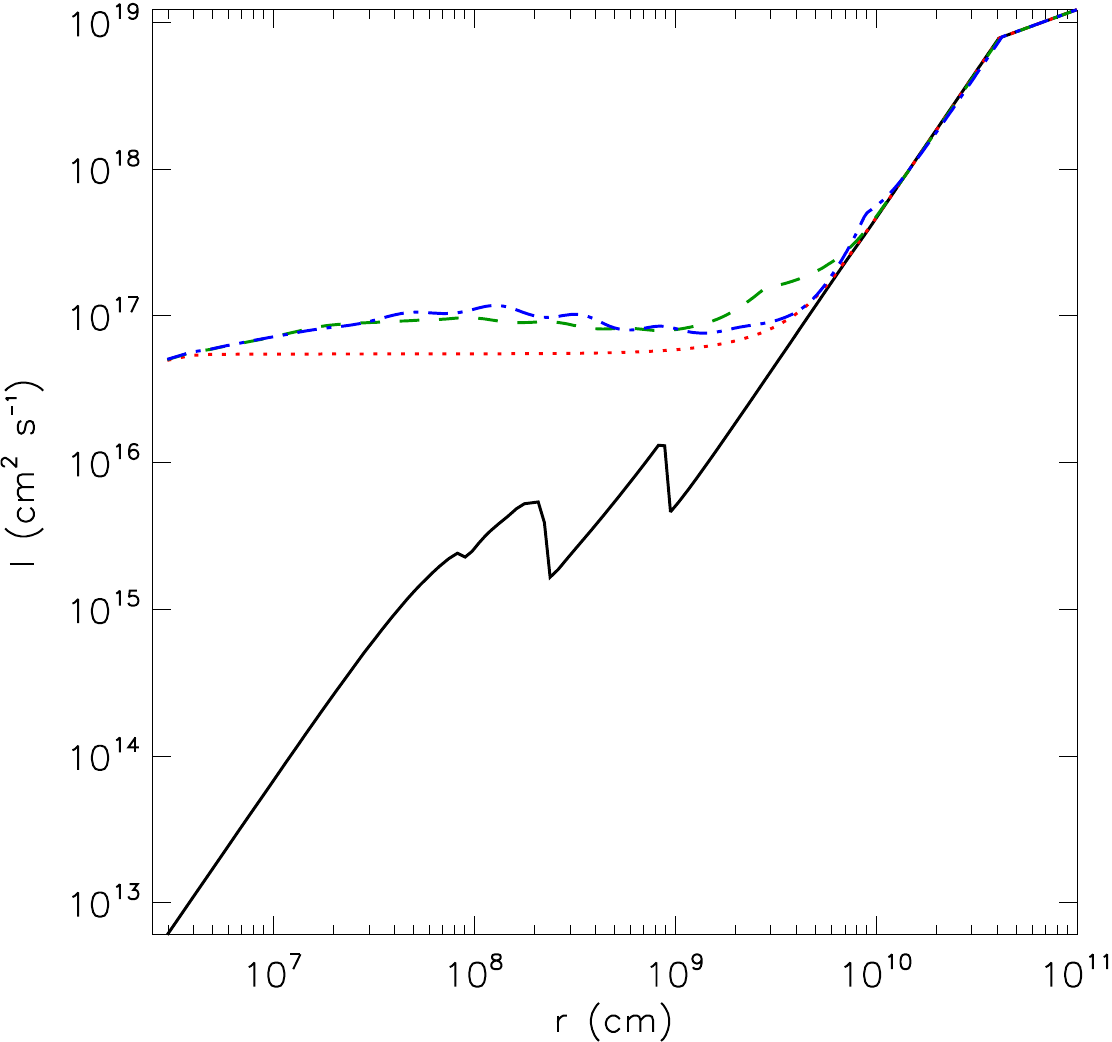}
\caption{Specific angular momentum, $\ell$, as a function of radius in Run 1 at $t=0\,\textrm{s}$ (\emph{black, solid}), $t=15\,\textrm{s}$ (\emph{red, dotted}), $t=25\,\textrm{s}$ (\emph{green, dashed}), and $t=50\,\textrm{s}$ (\emph{blue,dot-dashed}).  The initial rotational profile for the stellar model shows large spikes at compositional boundaries.  Early in the simulation, low angular momentum material is quickly accreted. \label{fig:angm}}
\end{center}
\end{figure}

Specific angular momentum as a function of radius is shown in Figure \ref{fig:angm}.  In the initial angular momentum profile of the model, compositional boundaries coincide with discontinuities in the profile, but in 16TI these occur only at mass coordinates that are accreted directly onto the black hole, prior to the initial circularization.  The angular momentum profile of the mass shells remaining at initial circularization is monotonically increasing and most of the remaining mass has nearly the same angular momentum, $\sim (1-2)\times10^{17}\,\textrm{cm}^2\,\textrm{s}^{-1}$.\footnote{Stellar models exist in which nonmonotonicity is pronounced.  This can produce an interesting variability of the central accretion rate \citep[e.g.,][]{LopezCamara:10,Perna:10}.}  This implies that the shocked atmosphere has nearly uniform specific angular momentum everywhere except at the radii where the time scale on which the viscous torque transport angular momentum is shorter than the time since circularization.  At $(25-50)\,\textrm{s}$, there is a mild, sub-Keplerian inward downturn in $\ell(r)$ at $r\lesssim 1000\,\textrm{km}$.  Angular momentum transport is too slow within the initial $\sim100\,\textrm{s}$ to affect the radii $\gtrsim 10^4\,\textrm{km}$.

\begin{figure*}
\begin{center}
\includegraphics[width=.45\textwidth,clip]{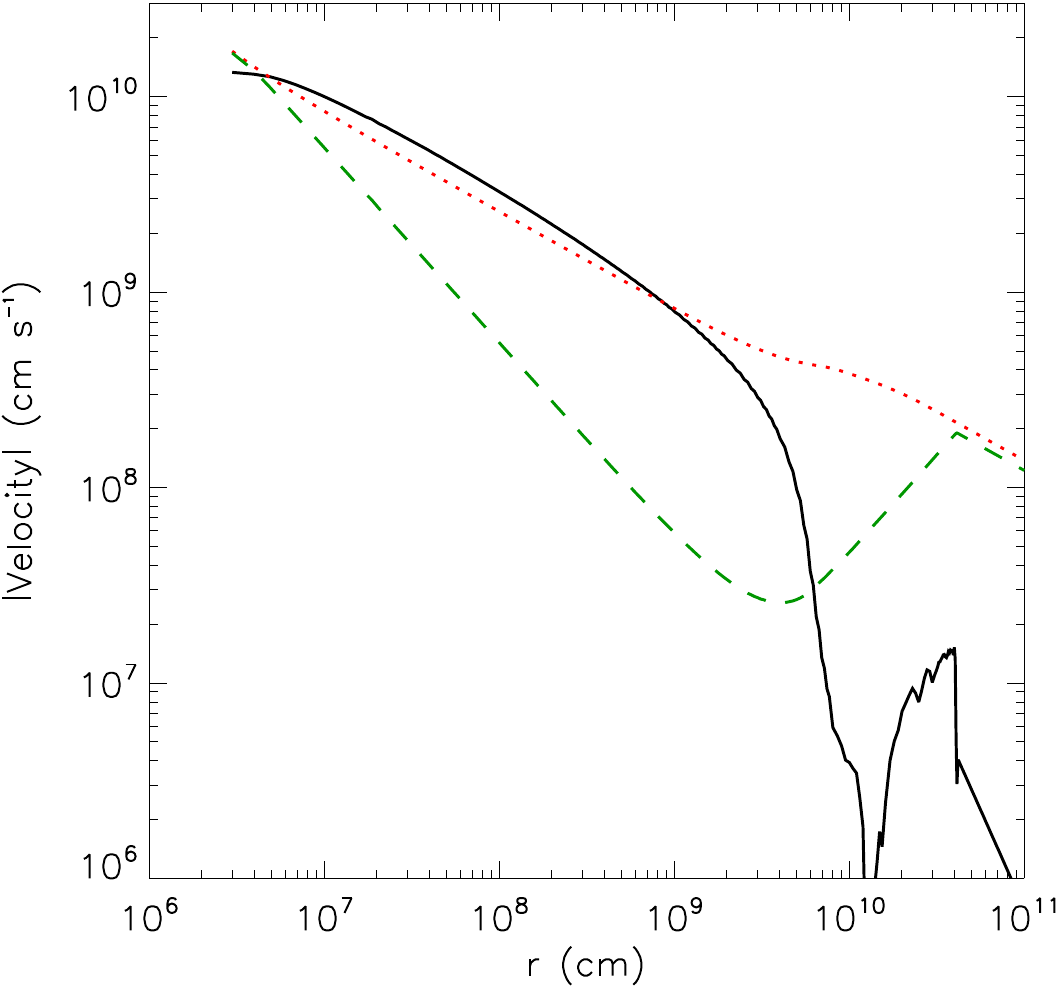}
\includegraphics[width=.45\textwidth,clip]{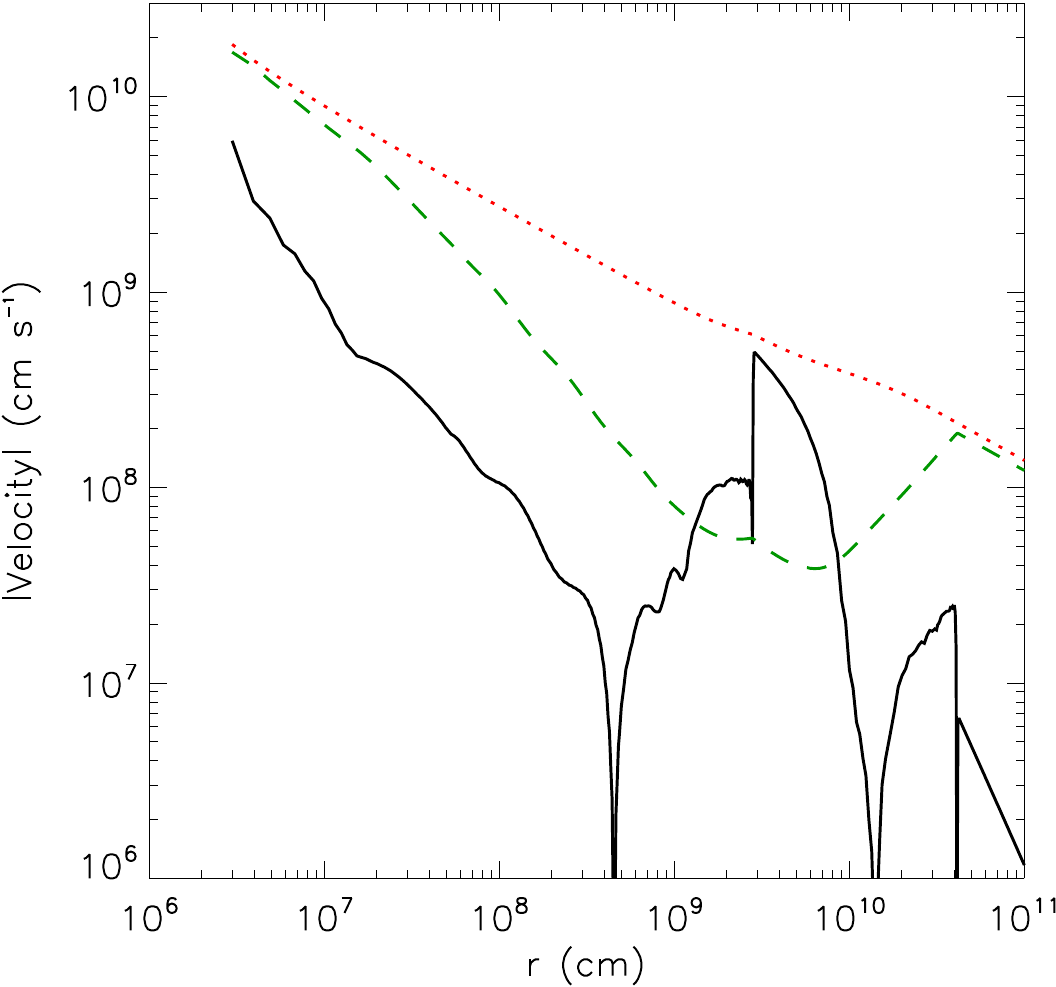}
\caption{The absolute value of the radial velocity, $v_r$ (\emph{black, solid}), Keplerian velocity including pseudo-relativistic corrections and self gravity (\emph{red, dotted}), and $v_\phi$ (\emph{green, dashed}) as a function of radius in Run 1 at $t=18\,\textrm{s}$ (\textit{left}) and at $t=30\,\textrm{s}$ (\textit{right}), just as material begins to circularize.  Notice that the rotational velocity is approaching the Keplerian velocity at the inner radii.  Once material has become rotationally supported, there is a dramatic drop in $v_r$.  At radii $4000\,\textrm{km} \lesssim r < r_{\rm shock}$, the radial velocity is positive, indicating an outflow.   \label{fig:velx}}
\end{center}
\end{figure*}

The mass accretion rate as a function of radius in Run 1 at $t=50\,\textrm{s}$ is shown in Figure \ref{fig:mass_analytic}. The accretion rate is independent of radius for $r \lesssim 2000\,\textrm{km}$, which is the radii where the angular momentum profile has relaxed to a viscous quasi-equilibrium.  The analytic expectation, given in equation (\ref{eq:mdot}), is shown as well. Figure \ref{fig:velx} shows the radial velocity $v_r$, angular velocity $v_{\phi}$ and Keplerian velocity as a function of radius at $t=18\,\textrm{s}$, just as material begins to circularize outside of the black hole, and at $t=30\,\textrm{s}$, after an accretion disk has formed.  At $t=18\,\textrm{s}$, the velocity $v_{\phi}$ reaches the Keplerian value at the innermost radii.

\begin{figure*}
\begin{center}
\includegraphics[width=0.30\textwidth,clip]{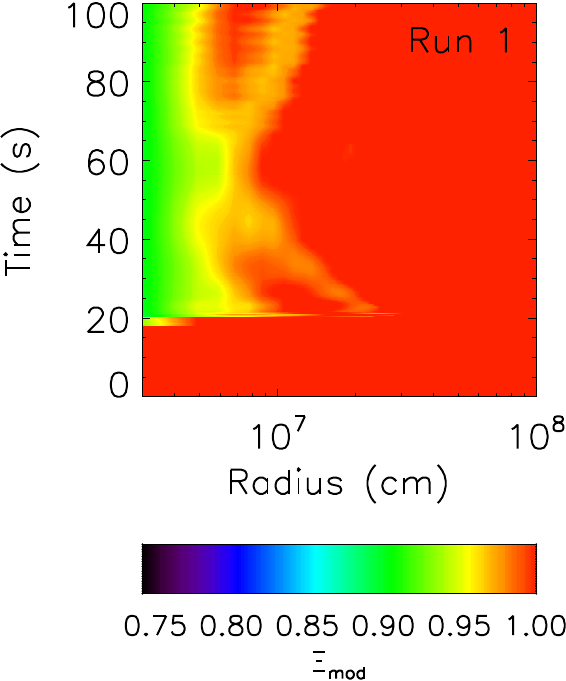}
\includegraphics[width=0.30\textwidth,clip]{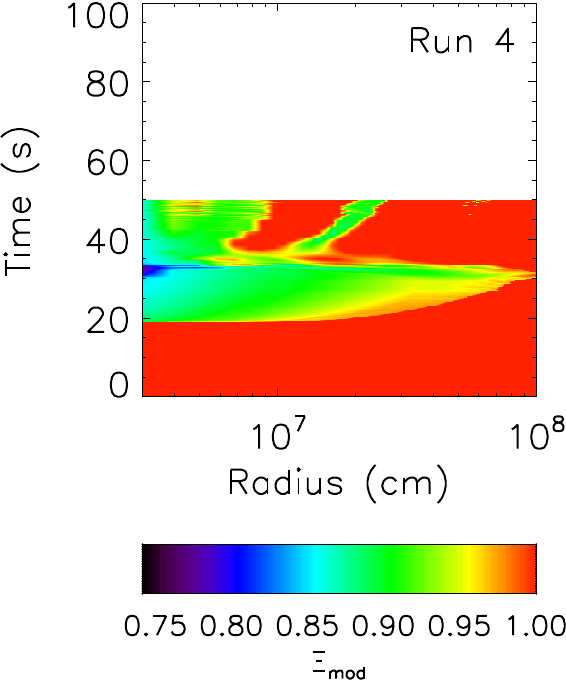}
\includegraphics[width=0.30\textwidth,clip]{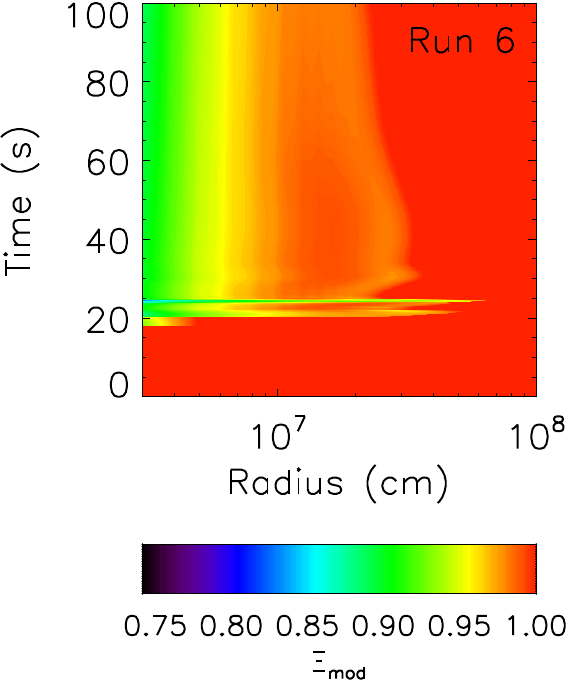}
\end{center}
\caption{The value of the correction factor  $\Xi_{\rm mod}$ applied to the temperature to account for the possibility of a reduced vertical scale height in, from left to right, Runs 1, 4, and 6  (see Section \ref{sec:thin_disk}).  This correction factor does not drop below $\sim 0.77$ in any of the simulations, indicating that geometric thinning of the accretion flow is a relatively weak effect.  The correction is only applied in regions where $\mathbf{a}_{\rm cent} > 0.5\,\mathbf{a}_{\rm pres}$, which occurs only for $r < 1000\,{\rm km}$; thus $\Xi_{\rm mod}=1.0$ for $r > 1000\,{\rm km}$.
\label{fig:tfac}}
\end{figure*}

Throughout the simulations, we tracked the value of our estimate of the vertical ($z$-directed) pressure scale height-to-radius ratio $\tilde H_z/r$, as described in Section \ref{sec:thin_disk}.  When the estimated ratio is below one-half, this indicates that in two dimensions, the flow should be disk-like, and when $\tilde H_z/r\ll 0.5$, the flow is a geometrically thin disk.   We found that  $\tilde H_z/r$ is below $0.5$ but is still always above a minimum of $0.3$ everywhere, except in Run 4, which had the lowest viscosity.  The disk-like radii where the vertical pressure scale height-to-radius ratio is below one-half are $r \lesssim 200 \,\textrm{km}$ immediately following circularization and shrink to $r\lesssim 100\,\textrm{km}$ by the end of the simulation.  In Run 4 with a reduced viscous stress-to-pressure ratio $\alpha$, neutrino cooling drove the disk to be geometrically thin, where $H_z/r \lesssim 0.3$ in the inner $r\lesssim 500\,\textrm{km}$.  In the innermost zone in Run 4, $H_z/r = 0.1$ at $t=20\,\textrm{s}$, the lowest seen in any simulation.  By $t=35\,\textrm{s}$, no thin disk is present.  In Figure \ref{fig:tfac} we show the value of $\Xi_{\rm mod}$ defined in equation (\ref{eq:Xi_smooth}) throughout the simulation in Runs 1, 4, and 6; it does not drop below $\sim 0.77$.  Only in Run 4 is a genuinely thin accretion disk present, and there it is limited to small radii.  The outer radius of the thin disk decreases as the neutrino luminosity drops (see Section \ref{sec:energy_transport}).  We attribute the observed moderate thinning of the accretion flow to the cooling of the flow by the photodisintegration of helium nuclei into free nucleons, and in Run 4, the additional contribution of neutrino cooling is also significant.

\subsection{Energy Transport}
\label{sec:energy_transport}

To understand the energetics of the accretion flow in a collapsar, we need to consider the transport of mechanical, thermal, and nuclear binding energy, as well as the loss to neutrino emission.  Before turning to energy transport, we discuss the neutrino losses, which turn out to be not significant in the regime we consider.

\begin{figure}
\begin{center}
\includegraphics[width=0.45\textwidth,clip]{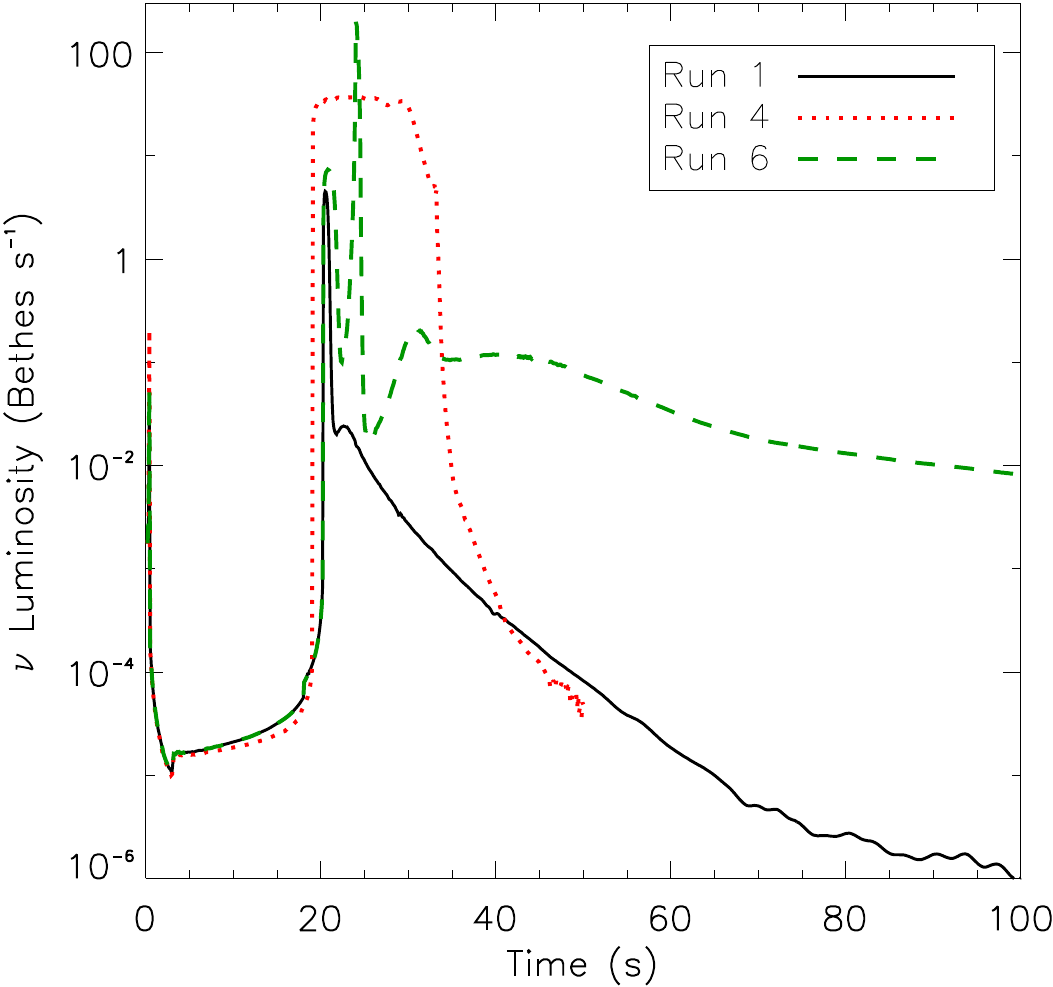}
\end{center}
\caption{Neutrino luminosity integrated over the entire computational domain in representative Runs 1, 4, and 6 (see Section \ref{sec:neutrino_cooling}).  The peak in luminosity occurs shortly after the formation of the accretion shock.  Note that we do not capture neutrino emission from the region $r < 25 \,\textrm{km}$, where much neutronization and peak neutrino luminosity is expected to occur in the first few seconds after the formation and collapse of the iron core.  The total neutrino luminosities integrated over the entire simulation in Runs 1, 4, and 6 were $2.7$, $419.3$, and $83.9\times10^{51}\,\textrm{ergs}$, respectively.
\label{fig:n_cl}}
\end{figure}

The integrated neutrino luminosity is dominated by the emission from the inner $\sim 100\,\textrm{km}$.  The luminosity as a function of time in the representative Runs 1, 4, and 6 is shown in Figure \ref{fig:n_cl}. In simulations with $\alpha=0.1$, neutrino luminosities integrated over the entire computational domain peaked immediately following shock formation at $\sim (1-200)\times10^{51} \,\textrm{ergs}\,\textrm{s}^{-1}$.  The peak luminosity lasted anywhere from less than a second in the runs with high peak luminosities to a few seconds in the runs with low peak luminosities. After the peak, the luminosity decays first very rapidly until it has dropped to $\sim10^{50}\,\textrm{ergs}\,\textrm{s}^{-1}$, and then continues to decay approximately exponentially by several orders of magnitude to settle at $\sim (10^{-6}-10^{-5})\,\textrm{ergs}\,\textrm{s}^{-1}$ after $\sim50\,\textrm{s}$.  The sharp luminosity peak is an artifact of the abrupt nature of shock formation in our 1.5D dimensional treatment and is probably not physically significant.  The total energy that would be deposited by an absorption of the emitted neutrinos, which we do not calculate, is negligible.

Now turning to energy transport, we examine the radial transport of all forms of energy, the thermal and kinetic energies, the nuclear binding energy, and the gravitational potential energy.  The gravitational potential energy is a nonlocal functional of the mass distribution.  However, ignoring relativistic effects, one can define the gravitational potential energy per unit volume to be $\rho(\Phi_{\rm BH}+\frac{1}{2}\Phi_{\rm s})$, where $\Phi_{\rm BH}$ is the gravitational potential of the black hole which we define via $\Phi_{\rm BH}(r)\equiv \int_r^\infty g_{\rm BH}(r') dr' $ with $g_{\rm BH}$ given in equation (\ref{eq:Artemova}) and $\Phi_{\rm self}$ is that of the self-gravity of the star. Then, $\rho v_r\Phi$, where $\Phi=\Phi_{\rm BH}+\Phi_{\rm self}$, can be interpreted as the flux of  gravitational energy advected by the fluid, but one must additionally include the flux of gravitational energy transported by self gravity \citep[see, e.g.,][Appendix F]{Binney:08}, which equals
\beq
F_{\rm grav,self}=\frac{1}{8\pi G} \left(\Phi_{\rm self}\nabla\frac{\partial\Phi_{\rm self}}{\partial t}-\frac{\partial\Phi_{\rm self}}{\partial t}\nabla\Phi_{\rm self}\right) .
\eeq 
This term is significant only in the outer envelope of the star.
The rate with which the sum total of these energies is transported radially is given by
\bea
\label{eq:edot}
 \dot{E} &=& 4 \pi r^2\left[\rho v_r \left(\epsilon +  \epsilon_{\rm nuc}+ \frac{P}{\rho} + \frac{1}{2}   v_r^2 + \frac{1}{2}\frac{\ell^2}{r^2} + \Phi\right) + F_{\rm grav,self}\right. \nonumber\\ 
&-&\left. \rho \nu \ell \frac{\partial \Omega}{\partial r} + F_{\rm conv}+F_{\rm nuc,mix}\right] ,
\eea
where the convention is such that $\dot E>0$ implies the transport of positive energy outward, opposite from the convection employed in the definition of the mass accretion rate $\dot M$.  Here, $-\rho\nu\ell \partial\Omega/\partial r$ is flux of energy transported by the viscous stress.

\begin{figure}
\begin{center}
\includegraphics[width=0.45\textwidth,clip]{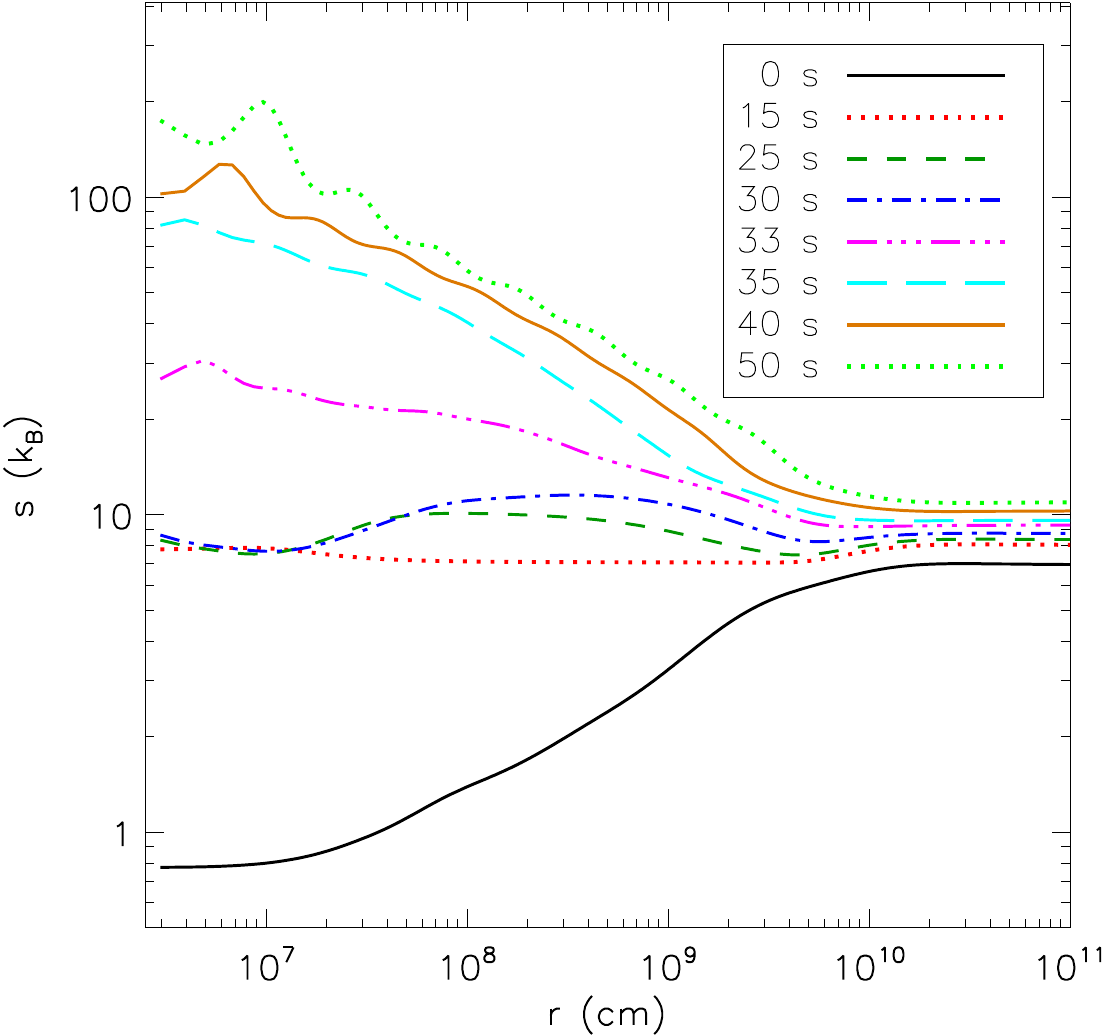}
\end{center}
\caption{The smoothed entropy $s_{\rm smooth}$ per baryon in units of the Boltzmann constant ($k_{\rm B}$) at various times in the low-viscosity Run 4 (cf.\ Fig.\ \ref{fig:entropy}).  Unlike in the other runs, here a negative specific entropy gradient is not seen immediately after the fluid comes into radial force balance.  Even by $t=30\,\textrm{s}$ the fluid is still stable against convection; neutrino cooling prevents the early rise of convective instability.  Once the neutrino luminosity begins to drop around $t=33\,\textrm{s}$, entropy in the post-shock region begins to rise, bringing about strong entropy inversion.  By the end of the simulation, Run 4 has the largest value of entropy seen in any of the simulations.\label{fig:entropy4}}
\end{figure}

Specific entropy as a function of radius at several times in Runs 1 and 2 is shown in Figure \ref{fig:entropy}.  After the formation of the accretion shock and the rotationally supported flow, viscous dissipation heats the fluid, thus producing a negative entropy gradient in the shock downstream.  The negative specific entropy gradient extends almost to the shock front, and thus the energy injected at small radii can travel to raise the entropy of the entire post-shock region.  Figure \ref{fig:entropy4} shows the specific entropy in Run 4.  Here, the high neutrino luminosity after the accretion shock has formed keeps the entropy in the post shock region relatively low.  For the first $\sim 10 \,\textrm{s}$ after shock formation, no specific entropy inversion is seen, and the fluid is stable against convection.  When the neutrino luminosity begins to drop around $t\approx 33\,\textrm{s}$, the entropy rises, a negative specific entropy gradient appears in the post-shock region, and convection starts transporting the viscously dissipated energy outward.

\begin{figure*}
\begin{center}
\includegraphics[width=0.75\textwidth,clip]{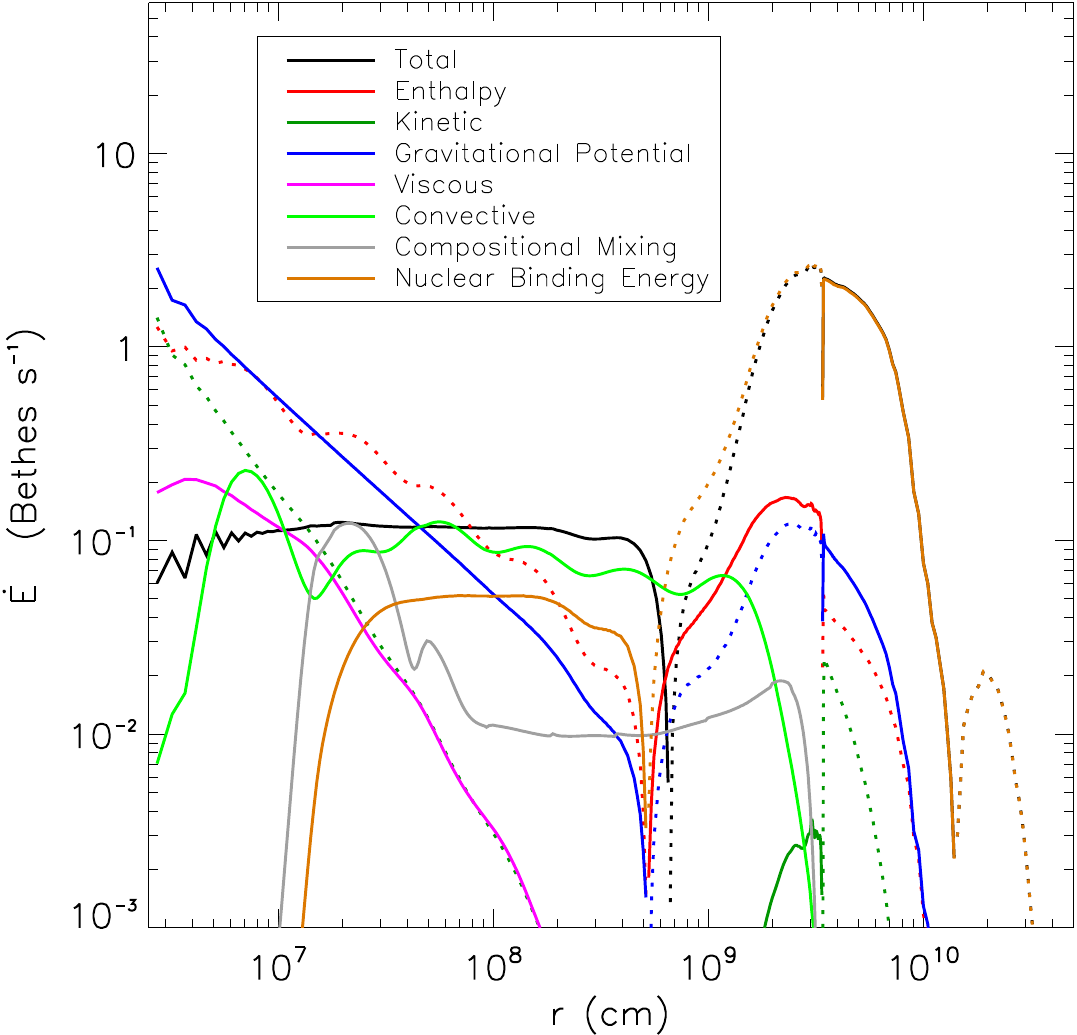}
\caption{The total and partial energy transport rates in Run 2 at $t=30\,\textrm{s}$ (see Section \ref{sec:energy_transport} and equation [\ref{eq:edot}]).  The curves show $\dot E$ (\emph{black}), the enthalpy advection rate $4\pi r^2v_r(\rho\epsilon+P)$ (\emph{red}), the kinetic energy advection rate $2\pi r^2v_r\rho(v_r^2+\ell^2/r^2)$ (\emph{green}), the gravitational potential energy transport rate $4\pi r^2(\rho v_r\Phi+F_{\rm grav,self})$ (\emph{dark blue}), the rate of energy transport by the viscous stress $-4\pi r^2\rho\nu\ell\partial\Omega/\partial r$ (\emph{pink}), the rate of thermal energy transport by convection $4\pi r^2 F_{\rm conv}$ (\emph{green}), the nuclear energy transport rate associated with convective compositional mixing $4\pi r^2 F_{\rm nuc,mix}$ (\emph{gray}), and the nuclear binding energy advection rate $4\pi r^2v_r \rho \epsilon_{\rm nuc}$ (\emph{orange}).  Negative values are indicated by dotted lines. \label{fig:edot}}
\end{center}
\end{figure*}

Figure \ref{fig:edot} plots the net transport rate $\dot E$ and the various constituent terms in Run 2 at $t=30\,\textrm{s}$; the radii and other observables quoted in the remainder of this section will be specific to this particular simulation snapshot and will vary across different simulations and different times within a simulation.  Approximate radial independence of the energy transport rate, $\partial \dot E/\partial r\approx 0$ for $200\,\textrm{km}\,\lesssim r \lesssim 4000\,\textrm{km}$, where the transport rate is positive $\dot E\approx 10^{50}\,\textrm{ergs}\,\textrm{s}^{-1}>0$, is indicative of quasi-steady-state accretion.  At larger radii, $r\gtrsim 5000\,\textrm{km}$, where the inner inflow gives way to an outer outflow---a precursor of the brewing explosion---no quasi-steady state is present and the fluid variables evolve on the dynamical time in the wake of the expanding shock.  At small radii, $r\lesssim 100\,\textrm{km}$, where one expects a steady state, the curve $\dot E(r)$ exhibits a small positive gradient, as well as a sawtooth consistent with that seen in the accretion rate $\dot M(r)$.  The constancy of the plotted energy transport rate is contingent on an accurate cancellation of the other transport terms.  We suspect that the observed nonconstancy is arising from relatively small inconsistencies in the discretization or gravitational source terms in FLASH and in the calculation of the gravitational energy during post-processing.

\begin{figure*}
\begin{center}
\includegraphics[width=1.0\textwidth,clip]{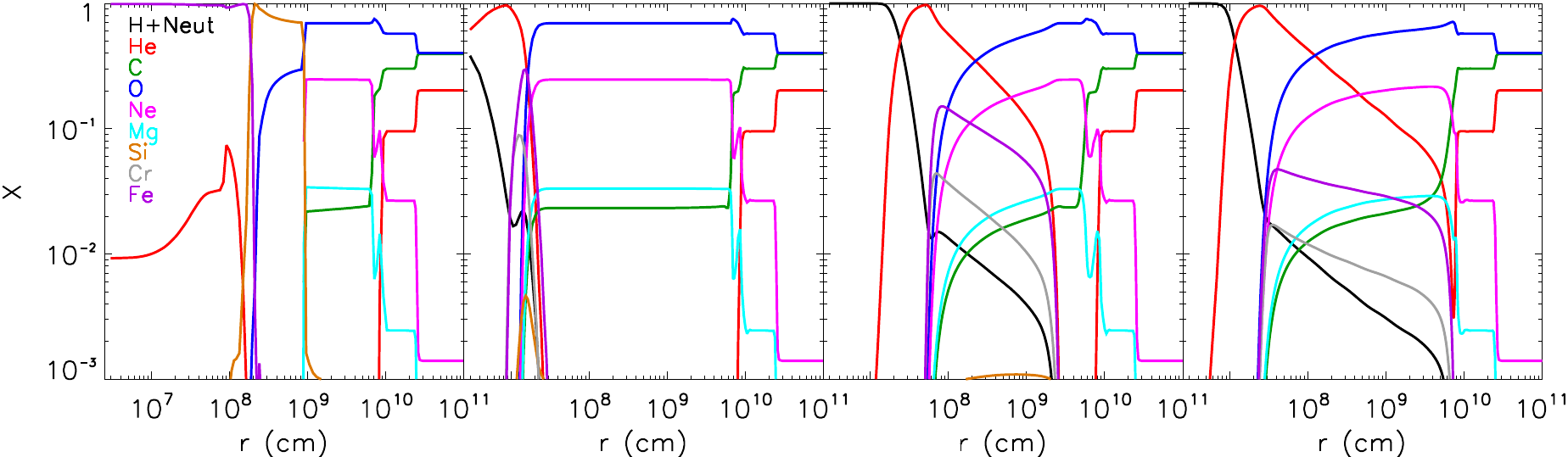}
\end{center}
\caption{The abundances of the common elements summed over all isotopic species at (\emph{left to right}) $t=0,15,25,50\,\textrm{s}$ in Run 1. \label{fig:elements}}
\end{figure*}

At $r\lesssim 1000\,\textrm{km}$, the inward advection of thermal and kinetic energy dominates over the outward transport by convection and the viscous stress. Therefore, the innermost flow is an advection-dominated accretion flow (ADAF; \citealt{Narayan:94,Narayan:95,Blandford:99}). At $r\gtrsim1000\,\textrm{km}$, the outward transport of thermal energy by convection dominates the inward transport by advection and this region is thus an convection-dominated accretion flow (CDAF, see, e.g., \citealt{Stone:99,Igumenshchev:00b,Blandford:04}).  Inward nuclear binding energy advection and nuclear compositional mixing both act to transport the total energy outward if one counts the negative nuclear binding energy in the total energy budget.  Convection transports energy from the ADAF-CDAF transition radius where the magnitude of the enthalpy advection flux $\sim |v_r (\rho\epsilon+P)|$ equals the convection flux $F_{\rm conv}$ to the shock radius $r_{\rm shock}$.  In Figure \ref{fig:edot} the former is at $r_{\rm ADAF}\approx 1000\,\textrm{km}$ and the latter is at $r_{\rm shock}\approx 3.5\times10^4\,\textrm{km}$.  In Run 1 and similar runs, $r_{\rm ADAF}$ increases very slowly from $\sim 1000\,\textrm{km}$ to $\sim 2000\,\textrm{km}$ from shock formation until $t=100\,\textrm{s}$.  In Run 4, the radius is $\sim 200\,\textrm{km}$ throughout the simulation, and in Run 6, the radius grows from $\sim 5000\,\textrm{km}$ to over $\sim 10^4\,\textrm{km}$ in the course of the simulation.

We suspect that the location of $r_{\rm ADAF}$ determines the amount of energy that can be carried to the shock front, and that in turn, the ADAF-CDAF transition radius is primarily a function of the convective efficiency $\xi_{\rm conv}$ and the viscous stress-to-pressure ratio $\alpha$.  Simulations with larger values of $\xi_{\rm conv}$ resulted in stronger shocks and larger amounts of unbound material. The convective compositional mixing parameter $\xi_{\rm conv, mix}$ had little effect on the final outcome of our simulations.  Turning to the viscosity parameter, Run 3 with a large values of $\alpha$ produced a somewhat less energetic shock with less unbound material at the end of the simulation.  Run 4, the simulation with the lowest viscosity, produced the most energetic explosion, even in the presence of more pervasive cooling by neutrino emission and by photodisintegration in the low-$\alpha$ regime; these flows are denser and hotter \citep[see, e.g.,][]{Popham:99,Chen:07}.  In \cite{Milosavljevic:12} we show that $r_{\rm ADAF}$ is expected to be smaller under low viscosity conditions because the advective luminosity is proportional to $v_r$, which is proportional to $\alpha$.  This trend is reproduced in the present simulations.

\subsection{Nuclear Composition of the Flow}
\label{sec:nucleosynthesis}

Our simplified treatment of nuclear compositional transformation, which entails relaxation to NSE on a temperature-dependent time scale, is designed to track the impact of nuclear photodisintegration and recombination on the thermodynamics of the flow.  However, it does not allow the computation of the ultimate nucleosynthetic output in the presence of out-of-NSE burning.  Thus, the results presented here can only be understood in the light of the limitations of the method.\footnote{\cite{Metzger:11} modeled accretion disks associated with the mergers of white dwarfs and neutron stars or black holes.  He found that nuclear processes taking place at temperatures $T\lesssim 4\times10^9\,\textrm{K}$ may lead to significant heating in the resulting outflows.  The relaxation to NSE we employ underestimates the heating due to out-of-NSE nuclear recombination.}  It is also worth recalling that we do not calculate neutronization that could modify the proton-to-nucleon ratio $Y_e$.  

In the hottest, innermost accretion flow, photodisintegration of heavy nuclei saps energy that could otherwise be transported by convection to larger radii to energize the shocked envelope.  However, once the nuclei are broken down, convective mixing can dredge up free nucleons to larger and cooler radii, where they can recombine and heat the fluid locally.  Figure \ref{fig:abar} shows the mean atomic mass $\bar{A}$ as a function of radius at various times in Run 1 and Run 2.   The mean atomic mass drops below $4$ in the innermost $(200-300)\,\textrm{km}$.  The positive gradient in $\bar{A}$ seen in portions of the convective region would in the Ledoux picture enhance the convective energy flux, but our Schwarzchild treatment of convection does not capture this effect.  We argue in Section \ref{sec:limitations} that since the nuclear time scale is shorter than the convective time scale at radii where Ledoux convection implies an enhanced energy transport, the nuclear compositional transformation \emph{inside} convective cells, not consider in the Ledoux treatment, should dominate. Lacking a theory of convection in this regime we adhere to the simpler Schwarzschild parametrization.

In Figure \ref{fig:elements}, we show the mass-weighted abundances of the most common elements in our simulations in Run 1 at various times.  By $t=30\,\textrm{s}$, again, the inner $200\,\textrm{km}$ is made up almost entirely of free nucleons in nearly equal portions, as $Y_e\approx0.5$ everywhere. The effect of convective compositional transport of the reprocessed nuclear species---the free nucleons, helium, and iron---from the hot innermost accretion flow is seen in the power law tails extending to near the location of the shock in the right panels. 

Although in our calculations we do not allow the evolution of $Y_e$, we can still speculate about the effects of neutronization.  \cite{Chen:07} computed the structure of time-independent accretion disks around Kerr black holes including the effects of pair capture and neutronization.  In their models with $\alpha=0.1$, the same as our fiducial viscous-stress-to-pressure ratio, at the radii where $\rho \sim 10^7\,\textrm{g}\,\textrm{cm}^{-3}$, corresponding to the density in the innermost disk in our simulations, they find $Y_e\approx0.5$, the same as in our non-neutronizing treatment.  In their models with $\alpha=0.01$, however, \cite{Chen:07} find that at densities corresponding to the innermost disk in our simulations, significant neutronization was in effect and $Y_e$ dropped well below neutron-proton equality.  It is therefore possible that in the very innermost regions of the disks in our Run 4 with a low viscosity $\alpha=0.025$, the true value of $Y_e$ should be lower than we assume. This would modify the abundances and thermodynamics of the portion of the flow in NSE.  The key question of consequence for the viability of the mechanism we propose for the production of luminous supernovae is, will the neutron-rich material pollute larger radii and drive a tendency toward the synthesis of iron instead of $^{56}$Ni?  Because this neutronization only seems to be most significant in the hottest innermost regions, where neutrino cooling is efficient and the flow is predominantly rotationally supported, it is possible that most of the neutron-enhanced material would be advected into the black hole.  This is a quantitative question that can be answered only with multidimensional simulations.

\begin{figure}
\begin{center}
\includegraphics[width=0.45\textwidth,clip]{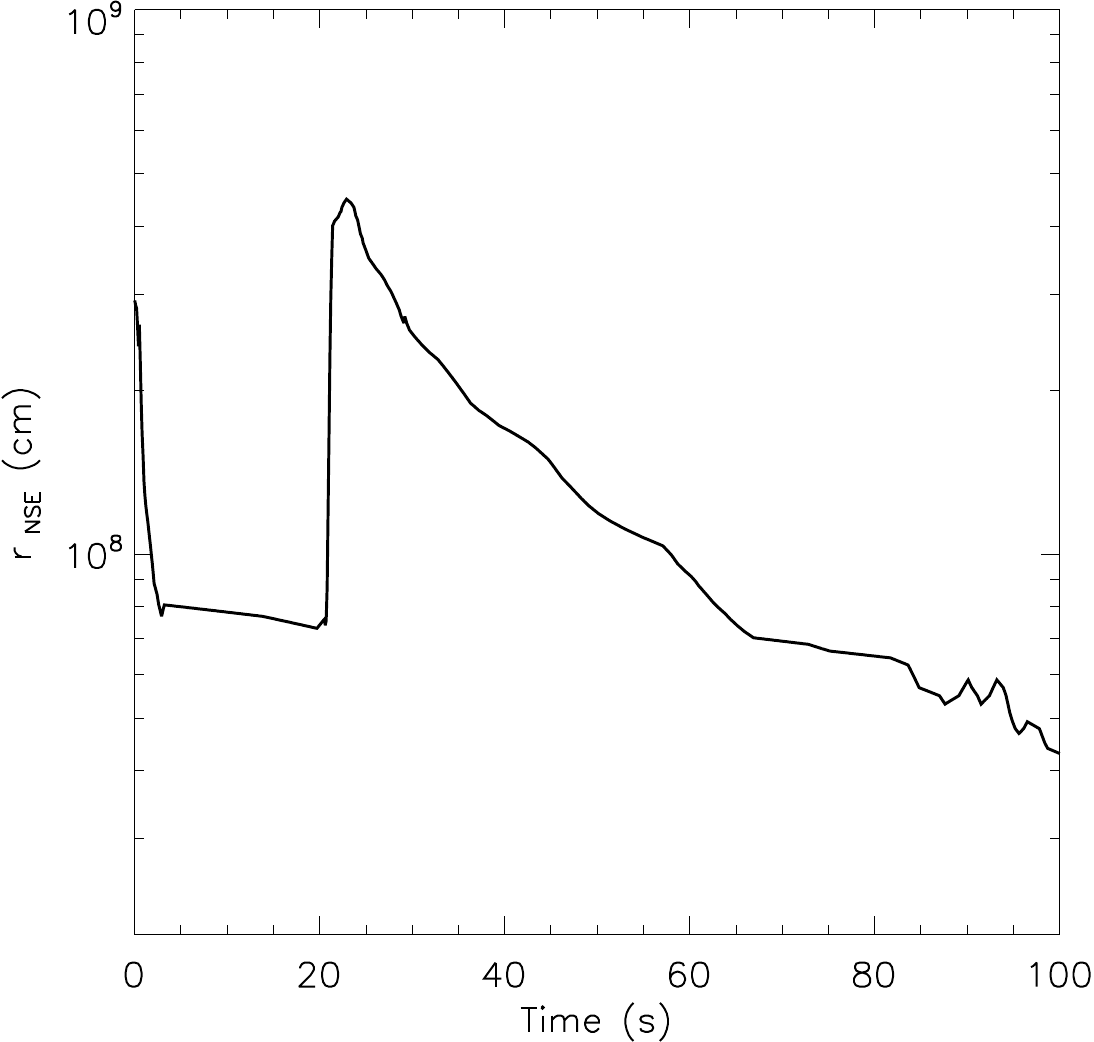}
\caption{Location of $r_{\rm NSE}$, the largest radius at which nuclear compositional transformation, in the form of gradual convergence toward NSE, is taking place in our Run 1; this radius is closely approximated with the radius where the temperature crosses $\approx3\times10^9\,\textrm{K}$.\label{fig:rnse}}
\end{center}
\end{figure}

In Figure \ref{fig:rnse}, we show the location of $r_{\rm NSE}$, the largest radius at which nuclear compositional transformation, in the form of gradual convergence toward NSE, is taking place in our Run 1.  This is the only region in which our calculations will capture nucleosynthesis and photodisintegration.  Effectively, it is the radius at which $T > 3\times10^9 \,\textrm{K}$.  Early in the simulation, the hot stellar core accretes into the black hole, and $r_{\rm NSE}$ quickly drops from $\approx 3000 \,\textrm{km}$ to $\approx 800 \,\textrm{km}$.  After the shock forms, additional heating in the shock and by viscous dissipation rapidly increases $r_{\rm NSE}$ and it peaks at $\approx 4000\, \textrm{km}$.  As the density and temperature drop and the viscous heating rate decreases, the inner regions of the star begin to cool, and $r_{\rm NSE}$ declines again. 

\begin{figure}
\begin{center}
\includegraphics[width=0.465\textwidth,clip]{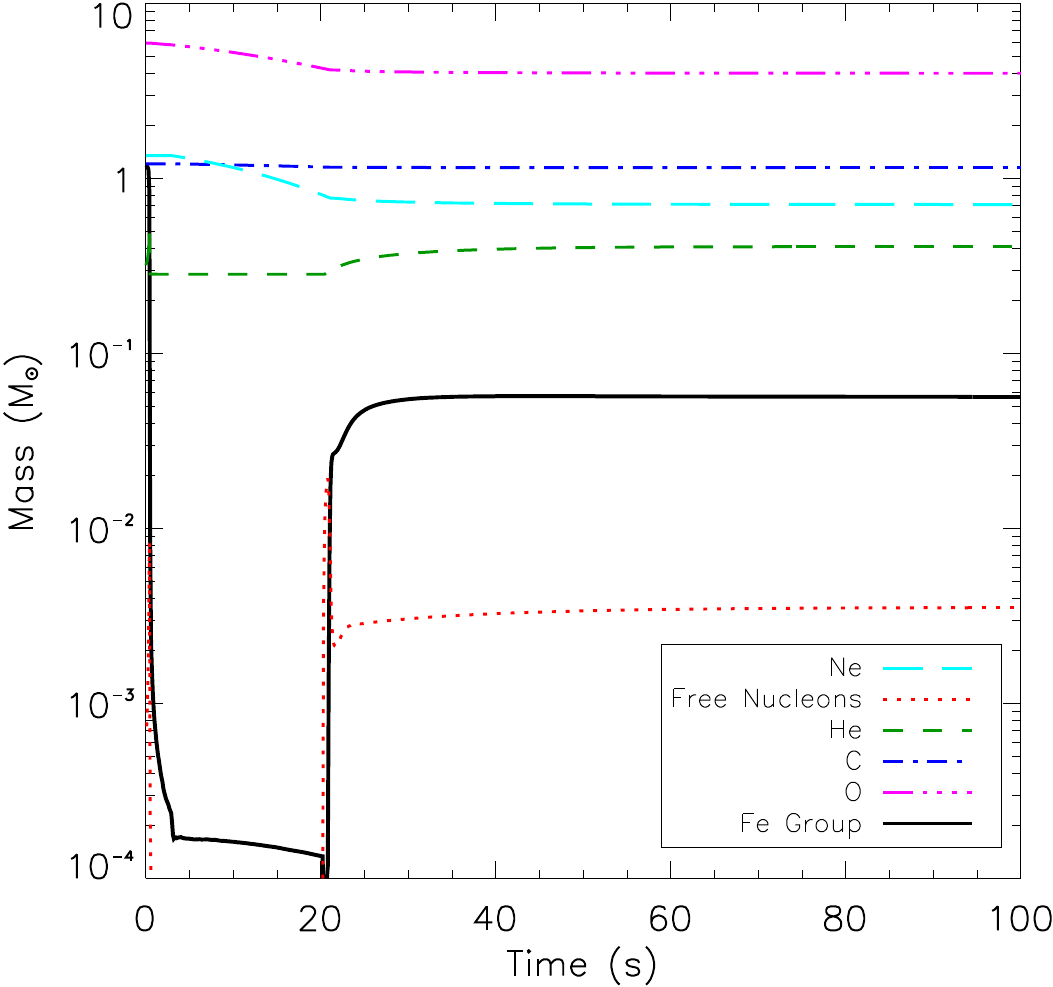}
\caption{Mass of the most common elements summed over isotopic species in Run 1 as a function of time.  Here, Fe represent all the isotopes of the iron group.  At the start of the simulation, there is a large dip in the amount of iron group elements, due to the accretion of the iron core.  After shock formation, $\sim 0.05 \,M_\odot$ of iron group elements are synthesized. \label{fig:element_mass}}
\end{center}
\end{figure}

In Figure \ref{fig:element_mass}, we show the total mass of various nuclear species in the entire simulation domain as a function of time.  The most notable change is the dip in the mass of iron-group elements. The initial decrease in in the iron group mass is the accretion of the core of the star onto the black hole.  After shock formation, a rapid increase in the amount of free nucleons is seen, in addition to production of additional helium and iron group elements.  In Table \ref{tab:results}, we show the total amount of newly fused Fe-group elements present at the end of the simulation, which fall in the range of $0.02\,M_\odot- 0.09\,M_\odot$.  Since we do not calculate the out-of-NSE burning in convectively dredged up material, at least a fraction of the extended helium tail seen in Figure \ref{fig:elements} could be expected to burn into iron and thus the iron group mass in Figure \ref{fig:element_mass} and Table \ref{tab:results} can be interpreted as a lower limit.

\subsection{Prospects for Explosion}
\label{sec:explosion}

\begin{figure}
\begin{center}
\includegraphics[width=0.475\textwidth,clip]{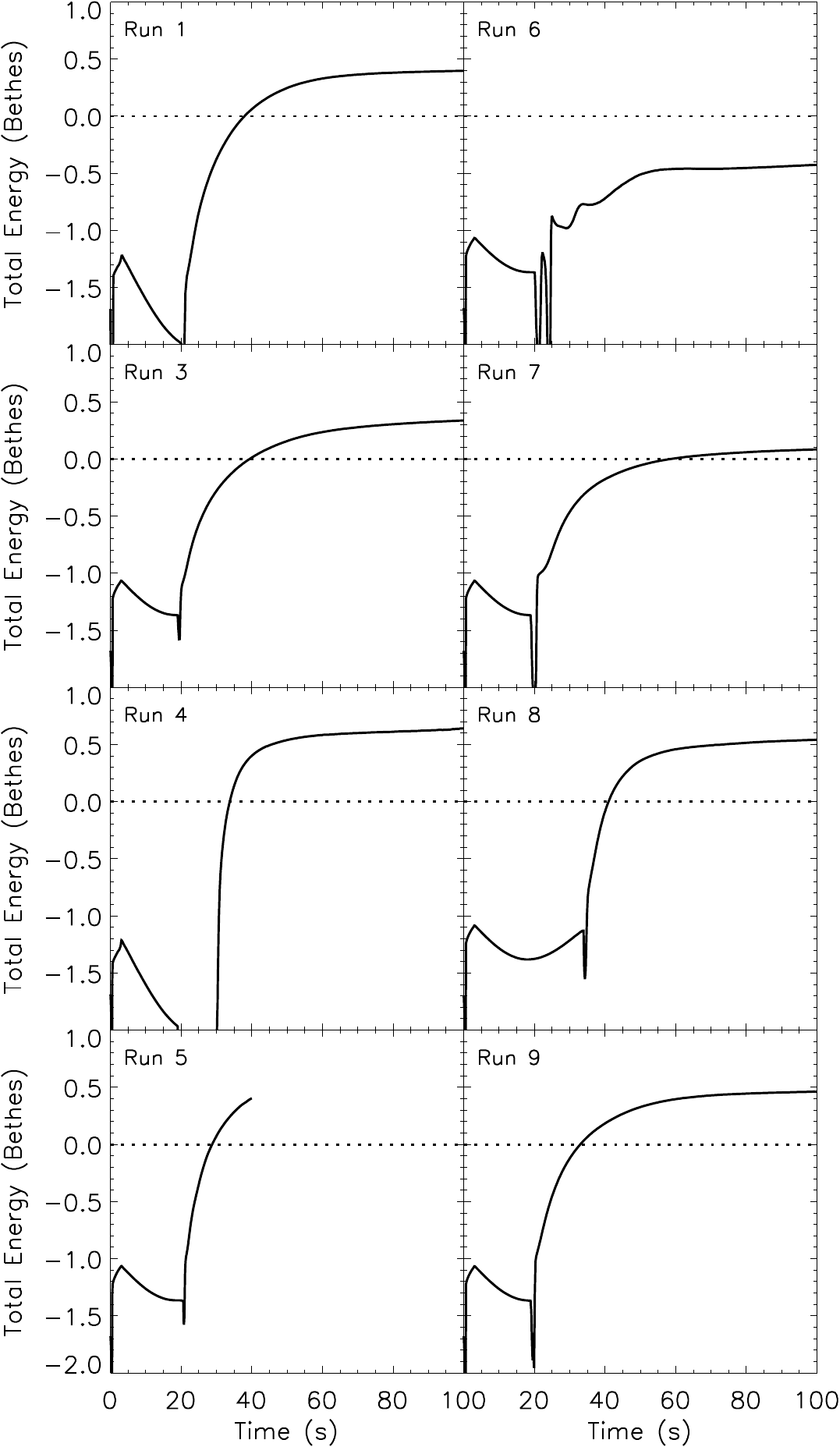}
\end{center}
\caption{The total binding energy on the simulation grid in Runs $1$, and 2$-$9.  Simulations in which the convective efficiency factor, $\xi_{\rm conv} \geq 0.5$ reached a positive total binding energy by the end of the simulation.  The large, brief dips in energy visible in some simulations around the time of shock formation are primarily due to cooling by neutrino emission.  The large dip seen in Run 4 is also predominantly due to neutrino cooling and has a minimum of $E_\textrm{bind}=-1.5\,\times10^{52}\,\textrm{ergs}$.
\label{fig:energy}}
\end{figure}

\begin{figure}
\begin{center}
\includegraphics[width=0.45\textwidth,clip]{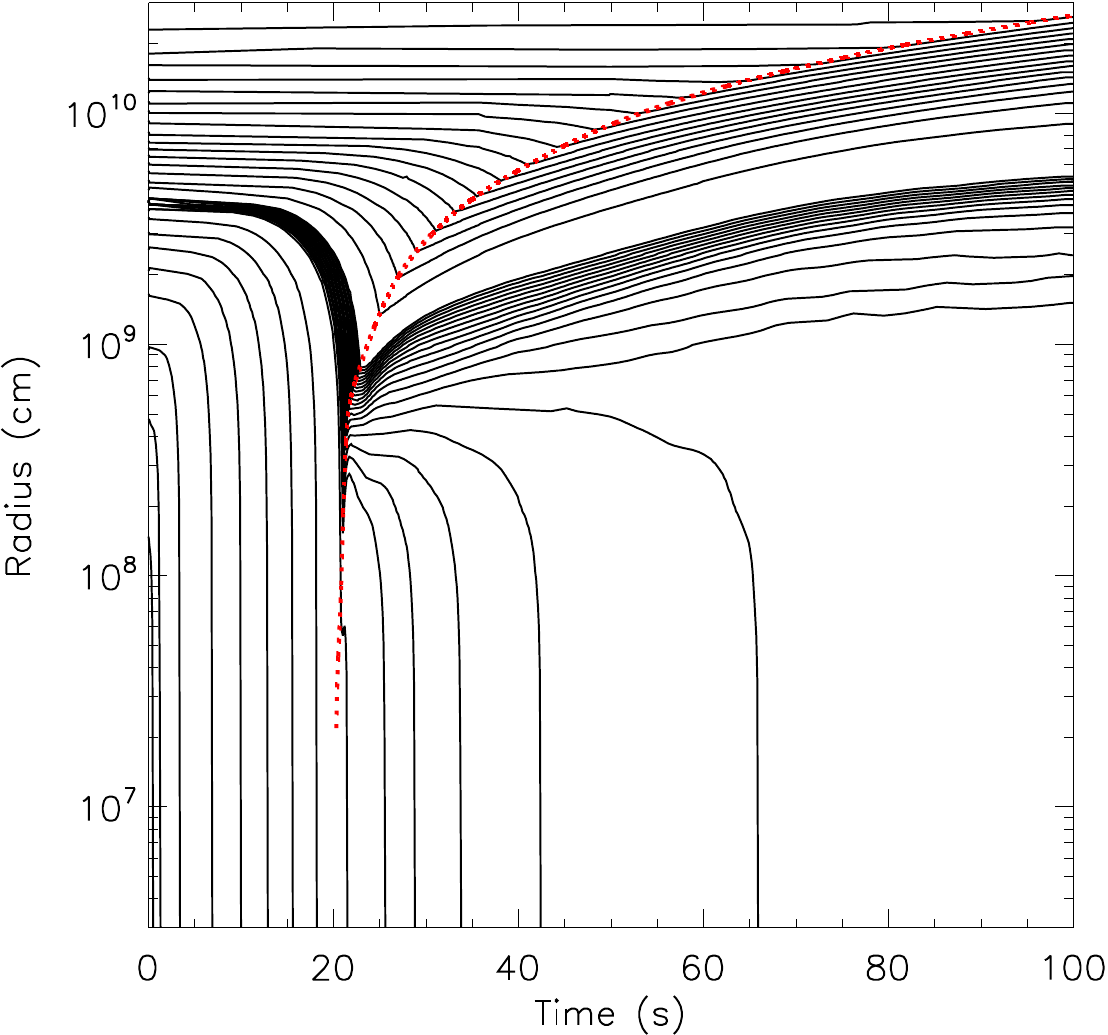}
\caption{The evolution Lagrangian mass coordinates in Run 1 (\emph{black, solid}).  The location of $r_{\rm shock}$ is also shown (\emph{red, dashed}).  The bifurcation between an inner inflow and an outer outflow occurs at $r\sim 4000\,\textrm{km}$.  This corresponds to a mass coordinate of $\sim 5.6\,M_\odot$.\label{fig:mass_coords}}
\end{center}
\end{figure}

The total thermal and mechanical energy, which we also refer to as the total binding energy, present on the grid was computed for each simulation via
\beq
E_{\rm bind} = \int_{r_{\rm min}}^{r_{\rm max}}{ \rho\left(\epsilon + \frac{1}{2}v_r^2 + \frac{1}{2}\frac{\ell^2}{r^2}  +\Phi_{\rm BH} +\frac{1}{2}\Phi_{\rm self}\right) 4\pi r^2dr} 
\eeq
and is shown in Figure \ref{fig:energy}.  A positive binding energy indicates a potential for explosion.  Runs 1, 2, 3, 4, 5, 8, and 9 acquired a positive total binding energy $E_{\rm bind}\sim (3.5-6.2)\times10^{50}\,\textrm{ergs}$ by the end of the simulation, indicating the potential for explosion.  These runs have $\xi_{\rm conv}\geq 2$ in common.  Run 7 with $\xi_{\rm conv}=1$ reached marginally unbound condition with $E_{\rm bind}\sim (0.5-1)\times10^{50}\,\textrm{ergs}$.  Run 6 with $\xi_{\rm conv}=0.5$ remained gravitationally bound throughout the entire simulation. 

The total unbound mass in each simulation was calculated by summing the masses of any fluid element with a positive value of $E_{\rm bind}$ and a positive radial velocity.  The total unbound masses in each simulation defined by this diagnostic are shown in Table \ref{tab:results}.  This criterion does not take into account any interaction that unbound and bound material might have subsequent to the measurement.  This criterion also neglects future energy gain or loss from nuclear processes.  The location and velocity of the outward moving accretion shock is shown in Figure \ref{fig:twoplot}.  In exploding models, the typical shock velocities were $(2000-4000) \,\textrm{km}\,\textrm{s}^{-1}$.  In Run 4 and Run 6, the shock stalled or slowed, and later restarted once or several times.  In Figure \ref{fig:mass_coords}, we show the evolution of various Lagrangian mass coordinates in Run 1 throughout the simulation.  Once the shock moves beyond $\sim 4000\,\textrm{km}$, the infalling material obtains a positive velocity once it reaches $r_{\rm shock}$.  

These results suggest that a high convective efficiency is required for sufficient transfer of energy from the inner accretion flow to the envelope to unbind the envelope.  Simulations with higher values of $\xi_{\rm conv}$ had relatively larger shock velocities and amounts of unbound mass at the ends of their simulations, and Run 4 with a lower $\alpha$ had the largest value of $E_{\rm bind}$.  Run 8 with reduced initial specific angular momentum produced an explosion comparable to that seen in the fiducial model.

\section{Observational Signatures and Progenitor Types}
\label{sec:discussion}

Modeling of light curves and spectra of supernovae associated with LGRBs has yielded information about the nature of these explosions.  The mass of $^{56}$Ni present in the supernova ejecta is easily estimated from the light curve by fitting simple radioactive decay models. The velocity of the ejecta is  inferred from observed line widths. 
Then, in a standard approach, the kinetic energy and mass of the ejecta are derived by comparing the inferred $^{56}$Ni yield with that implied by one-dimensional hydrodynamic models in which a spherical shock wave is introduced by hand (by the action of a hypothetical ``piston'') into the progenitor's envelope.\footnote{In ``piston'' and ``thermal bomb'' models, an explosion is mimicked by injecting a large kinetic or thermal energy into a narrow shell over a relatively short $\lesssim 1\,\textrm{s}$ time interval \citep[e.g.,][]{Jones:81,Woosley:86,Arnett:87,Shigeyama:87,Woosley:88,Thielemann:90,Woosley:95,Limongi:03,Chieffi:04,Young:07,Freyer:08,Kasen:09,Maeda:09,Joggerst:10, Dessart:11}.}  This approach relies on the hypothesis that $^{56}$Ni is produced at the shock front, for which the shock must be very strong.  

Our simplified simulations suggest an alternative scenario in which nucleosynthesis takes place in the accretion flow in the interior of the star, similar to the wind-nucleosynthesis models \citep[e.g.,][]{Beloborodov:03,Pruet:03,Pruet:04,Nagataki:06,Surman:06,Maeda:09,Metzger:11}.  Our results, however, suggest that the accretion flow is long lived, lasting tens or hundreds of seconds or longer, and so the nucleosynthesis can be sustained at lower densities than in the wind models, and its products can be delivered to the envelope by vigorous convection.  

We also find that a supernova-like shock wave may be powered by the sustained input of accretion energy, without energization by neutrino energy deposition.  The dynamics of the accretion-energy-powered shock wave is fundamentally different from that powered by a piston. It remains to be explored whether the accretion scenario will call for a modification of the standard approach to modeling the light curves and spectra of the supernovae that could be yielding black holes.  We are thus somewhat reluctant to compare our results directly with observational inferences obtained with existing supernova models.  Previous work has attributed kinetic energies of $\sim(2 - 50)\times10^{51}\,\textrm{ergs}$ to supernovae associated with LGRBs \citep[see, e.g., ][and references therein]{Woosley:06b,Hjorth:11}.  Our models come short of these energies, but they are consistent with the low energy end among the more typical Type Ib and Ic supernovae (see, e.g., Table 4 in \citealt{Hamuy:03}).  Unfortunately the simplified treatment of nuclear compositional transformation does not allow us to predict the $^{56}$Ni synthesized in our models.  We can only say that supernovae powered by collapsar accretion should exhibit a high degree of mixing of hydrostatic and explosive elements.

The shock velocities at $t=100\,\textrm{s}$, when the shock is still in the interior of the envelope, in the models that achieve explosion, are $v_{\rm shock}\approx 4000 \,\textrm{km}\,\textrm{s}^{-1}$.  This is a half or smaller fraction of the commonly cited values for shock velocities measured in the observed supernovae. Of course, leading to the breakout of the stellar surface, the shock accelerates as it traverses the negative density gradient.  The mass-weighted rms free expansion velocity inferred from the total energy at the end of the simulation is $v_{\rm FE}\sim (2E_{\rm bind}/M_{\rm unbound})^{1/2}\sim 3000\,(E_{\rm bind}/0.5\times10^{51}\,\textrm{ergs})^{1/2}(M_{\rm unbound}/5M_\odot)^{-1/2}\,\textrm{km}\,\textrm{s}^{-1}$, again lower than usually quoted for the observed supernovae.

Our initial model of choice was the $M_{\rm star}\approx 14\,M_\odot$ Wolf-Rayet model star 16TI of \citet{Woosley:06a}, evolved to pre-core-collapse from a $16\,M_\odot$ main sequence progenitor.  The model 16TI is commonly used in LGRB investigations \citep[e.g.,][]{Morsony:07,Morsony:10,LopezCamara:09,LopezCamara:10,Lazzati:10,Lazzati:11a,Lazzati:11b,Lindner:10,Nagakura:11a,Nagakura:11b}, but it has been suggested that the progenitors of supernovae with confirmed association with LGRBs must be associated with the core collapse of more massive stars, perhaps with masses in the range of $(25 - 60)\,M_\odot$ or higher \citep[e.g.,][and references therein]{Podsiadlowski:04, Smartt:09}.  However, predictions regarding the nature of the final remnant in such explosions are sensitive to the highly-debated details of the explosion mechanisms in core collapse supernovae.  Observational studies of the spatial distribution of Type Ic supernovae and GRBs in galaxies suggest that the respective progenitors should be at least $\sim25\,M_\odot$ and $\sim 43\,M_\odot$ \citep[][see, also, \citealt{Larsson:07}]{Raskin:08}.  It is of interest to note that simulations of neutron-star-powered explosions have been successful only in the lowest mass progenitors.  The accretion powered mechanism we propose will operate in more massive progenitors that produce black holes.   It is reasonable to speculate that in progenitors more massive than in our model, or with different internal structure, the explosion energies would be much higher than we find, more in line with the high energies of the LGRB supernovae.  The long term accretion in massive collapsar progenitors deserves further study.

\section{Conclusions}
\label{sec:conclusions}

We have conducted a series of hydrodynamic simulations of the viscous post-core-collapse accretion of a rapidly rotating $\sim 14 \,M_\odot$ Wolf-Rayet star of \citet{Woosley:06a} onto the central black hole that we assumed was in place at the beginning of the simulation.  The spherically-symmetric simulations with rotation were carried out for $100\,\textrm{s}$ and resolved the radii down to $25\,\textrm{km}$, including the range of radii where the collapsing stellar material circularizes around the black hole.  The simulations tracked cooling by neutrino emission and the relaxation to nuclear statistical equilibrium in the hot inner accretion flow. The simulations also tracked convection and convective compositional mixing in the mixing length theory approximation.  Finally, the simulations tracked viscous angular momentum transport and the associated heating in the flow.  To explore the sensitivity to model parameters, we varied the initial angular momentum profile, convective energy transport and compositional mixing efficiencies, and the viscous stress-to-pressure ratio $\alpha$.  Our main findings are as follows.

Lacking sufficient angular momentum to be rotationally supported around the black hole, the inner mass coordinates of the stellar envelope accrete unshocked onto the black hole.  At $t\sim20\,\textrm{s}$, or later with reduced initial angular momentum, the first mass shell able to circularize around the black hole arrives.  Once material becomes circularized, an accretion shock forms as the mass shells in near free fall interact with the rotationally supported material.

This shock front travels outwards, leaving behind a mostly pressure supported, shock heated, convective stellar envelope.  Only a very small fraction of the mass is predominantly rotationally supported; the rotationally supported, geometrically thick disk connects smoothly to the pressure supported, shock-heated envelope.  The structure and energetics of the flow are governed by accretion mechanics.  The energy dissipated by the viscous stress at the innermost radii, the radii smaller than some critical $r_{\rm ADAF}$, is advected into the black hole. The innermost flow is thus an ADAF. At larger radii, convection transports the dissipated energy outward, into the stellar envelope and toward the expanding shock front, implying that the outer flow is a CDAF.   The delivery of energy from $\sim r_{\rm ADAF}$ to the envelope proceeds for tens of seconds, and the total energies delivered are sufficient to produce supernovae, albeit not as energetic as the ones inferred in association with LRGBs.

We found that the final energy deposited into the envelope strongly depended on the efficiency of convective energy transport and depended somewhat on the viscous stress-to-pressure ratio $\alpha$.  These two parameters strongly influence the location of the ADAF/CDAF transition, as we have explored with crude analytical arguments in \citet{Milosavljevic:12}.  The low-$\alpha$ model has a hotter disk with more pervasive cooling by neutrino emission and nuclear photodisintegration, sapping energy from the final explosion.  However, the low-$\alpha$ disk also has an ADAF/CDAF transition at a smaller radius, potentially allowing for a higher convective luminosity.

For sufficiently high convective efficiencies, the stellar envelope was capable of obtaining positive total thermal and mechanical energies $\sim 0.5\times10^{51}\,\textrm{ergs}$, shock velocities $\sim 4000\,\textrm{km}\,\textrm{s}^{-1}$, and unbound masses $\sim 6\,M_\odot$.  We suggest that the accretion powered mechanism, which is distinct from and possibly mutually exclusive with the standard neutron-star-powered ``delayed-neutrino'' mechanism, could explain low luminosity Type Ib and Ic supernovae, but multidimensional study is needed to pin down the true efficiency of convective energy transport and to estimate the expected $^{56}$Ni yield.

\acknowledgements

We would like to thank J. Craig Wheeler for helpful comments on a draft of this manuscript, and Rodolfo Barniol Duran, Lars Bildsten, Adam Burrows, Emmanouil Chatzopoulos, and Sean Couch for valuable conversations.  The software used in this work was in part developed  by the DOE-supported ASC/Alliance Center for Astrophysical  Thermonuclear Flashes at the University of Chicago.   The authors acknowledge the Texas Advanced Computing Center  (TACC) at the University of Texas at Austin for providing high-performance computing resources that have contributed to this research.   C.~C.~L.\ acknowledges support from a National Science Foundation Graduate Research Fellowship.  M.~M.\ acknowledges support from NSF grants AST-0708795 and AST-1009928.  P.~K.\ acknowledges support from NSF grant AST-0909110.

\end{document}